\UseRawInputEncoding
\documentclass[final,3p,times,onecolumn]{elsarticle}
\usepackage[hyphens]{url}
\usepackage[breaklinks=false]{hyperref}
\hypersetup{breaklinks=false}
\usepackage[spaces,hyphens]{xurl} 
\usepackage{natbib}
\onecolumn
\usepackage{pifont}
\usepackage{color}
\usepackage{tabularx, booktabs}
\usepackage{amsmath}
\usepackage{algorithmic}
\usepackage{float}
\usepackage{graphicx}
\usepackage{array} 
\newcolumntype{C}[1]{>{\centering\arraybackslash}p{#1}}
\usepackage{subcaption}

\usepackage{xcolor}
\usepackage{multirow}
\usepackage{algorithm}
\usepackage{caption}
\usepackage{fancyhdr}
\usepackage{tabularx, booktabs} 
\usepackage{color}
\usepackage{tabularx, booktabs}
\usepackage{amsmath}
\usepackage{algorithmic}
\usepackage{float}

\usepackage{xcolor}
\usepackage{multirow}
\usepackage{algorithm}
\usepackage{caption}
\usepackage{fancyhdr}
\usepackage{tabularx, booktabs} 
\usepackage{graphicx}
\usepackage{tikz}
\usetikzlibrary{shapes.geometric, arrows.meta, positioning}
\usepackage[superscript]{cite}
\begin{document}

\begin{frontmatter}

\title{Improving the Clinical Generalizability of Machine Learning-Based Alzheimer's Disease Detection Across Domains and Modalities: A Comprehensive Review}

\author{Yingchao Huang\corref{mycorrespondingauthor}}
\ead{yingcha1@ualberta.ca}
\address{Faculty of Engineering, Electrical \& Computer Engineering Department, University of Alberta, Edmonton AB T6G 2H5, Canada}
\author{Yuhan Su}
\address{School of Basic Medical Sciences, Hebei University, Baoding 071000, China}
\author{Xin Wang}
\address{Faculty of Digital Innovation, Arts \& Sciences, Saskatchewan Polytechnic, Regina SK S4S 5X1, Canada}
\author{Shanshan Yao}
\address{Department of Civil \& Environmental Engineering and School of Mining \& Petroleum Engineering, University of Alberta, Edmonton AB T6G 2H5, Canada}

\cortext[mycorrespondingauthor]{Corresponding author}

\begin{abstract}

\noindent\textit{Background}: Machine learning (ML)-based diagnostic systems have shown promising performance for Alzheimer’s disease (AD) detection. However, their clinical translation remains limited by reduced performance across different cohorts, imaging protocols, acquisition settings, and heterogeneous participant and clinical populations. This lack of generalizability remains a major barrier to the reliable deployment of AI tools in real-world clinical settings.

\noindent\textit{Objective}: This review aims to provide a comprehensive synthesis of domain adaptation (DA) and domain generalization (DG) strategies for AD detection. It focuses on their potential to improve the robustness, generalizability, and clinical reliability of ML-based diagnostic systems.

\noindent\textit{Methods}: We conduct a comprehensive scoping review of existing DA and DG approaches across multiple data modalities. The reviewed studies are organized according to data modality, algorithmic framework, adaptation strategy, clinical relevance, and reported cross-domain performance. Key methodological trends, performance gains, limitations, and remaining challenges are analyzed to identify future research directions.

\noindent\textit{Results}: DA and DG approaches have demonstrated encouraging improvements in cross-domain performance and model robustness. Emerging modalities, such as speech-based approaches, EEG and eye-tracking data, show great potential for early, accessible, and non-invasive cognitive assessments. However, important limitations remain, including insufficient external validation, limited interpretability, computational complexity, privacy concerns, and challenges in scaling to routine clinical practice.

\noindent\textit{Conclusion}: DA and DG provide important methodological pathways for developing more generalizable ML-based systems for AD detection. Future research should prioritize explainable, computationally efficient, privacy-preserving, and clinically validated DA/DG techniques. Strengthening these aspects will be essential for translating robust AI-based AD detection tools into clinically deployable solutions.

\end{abstract}

\end{frontmatter}

\section{Introduction}
\label{sec:intro}

Alzheimer's disease (AD) is the leading cause of dementia in elderly adults, characterized by progressive degeneration of memory and cognitive skills \cite{kumar2024alzheimer}. It accounts for approximately 60\%-70\% of all dementia cases globally \cite{reitz2011epidemiology, who2017dementia}. As the World Health Organization (WHO) reported, 12.7 million people over 65 years old will be diagnosed with AD by 2050 \cite{who2017dementia}. AD typically progresses through three stages: Healthy Control (HC) or Cognitively Normal (CN), Mild Cognitive Impairment (MCI), and Alzheimer's Disease (AD) itself \cite{10179900}. Although there are currently no cures for AD, early diagnosis, particularly during the MCI stage, is critical, which can help slow disease progression through timely therapeutic interventions and improve the quality of life for people with AD \cite{crous-bou2017alzheimer}. 

Traditionally, AD diagnosis relies on clinical evaluations, neuropsychological assessments, and advanced neuroimaging techniques such as Magnetic Resonance Imaging (MRI) and Positron Emission Tomography (PET) scans \cite{c2bc6ca63e9047f185981d4b59b748b9}. Structural changes in the brain, including early enlargement of the lateral ventricles and degeneration of the hippocampus and amygdala, often precede noticeable cognitive decline \cite{10182253}. Neuroimaging is therefore a vital tool, offering high spatial resolution and excellent contrast for studying brain tissue. Accurate interpretation of neuroimaging data enables clinicians to distinguish between diseased and healthy regions of the brain \cite{pei2022alzheimers}.

In recent years, technological advancements have introduced non-invasive and cost-effective approaches that serve as viable alternatives to traditional diagnostic techniques. These include multimodal signals capturing different aspects of brain and cognitive function, which are capable of remote, continuous, and scalable monitoring. Among them, electroencephalography (EEG) has shown strong potential due to its high temporal resolution, affordability \cite{kim2018detection, amini2021diagnosis}, and ability to provide real-time insights into brain activity \cite{sudharsan2022recognition, safi2021early}. EEG abnormalities in people with AD typically include slower brainwave activity, reduced signal complexity, and disrupted synchronization between brain regions \cite{safi2021early, alsharabi2023eeg}.

Beyond EEG, behavioral biomarkers such as eye-tracking (ET), facial expressions, and speech patterns have attracted increasing attention in recent AD research. ET technology offers a non-intrusive way to evaluate visual and cognitive functions by capturing abnormalities in eye movements such as saccades, fixation times, and attention patterns that are often linked with impairments in memory, spatial orientation, and cognitive processing speed \cite{tokushige2023early}. Its ease of use and adaptability across clinical and home environments make ET particularly valuable for longitudinal monitoring \cite{medicina59081361}. Facial expression analysis also contributes to AD detection, as neurodegenerative processes can impair the physiological basis of emotional expression, leading to reduced facial muscle activity and diminished affective responses \cite{doi:10.1176/appi.neuropsych.21070186, zheng2023detecting}. Similarly, changes in both acoustic features (e.g., pitch, pause duration) and linguistic features (e.g., coherence, lexical diversity) reflect the progressive deterioration of cognitive and language abilities in people with AD \cite{konig2025novel}. Recent developments in large language models (LLMs) have enhanced the analysis of speech by enabling the identification of intricate linguistic features, such as grammatical structure, semantic flow, and lexical retrieval that support earlier and more accurate detection of cognitive decline \cite{AlsuhaibaniMuath2025ARoM}.

In recent years, machine learning (ML) has significantly advanced the AD detection, particularly through applications in computer vision (CV) and natural language processing (NLP). CV techniques have demonstrated strong potential in identifying AD by analyzing neuroimaging data, particularly MRI and PET scans, to identify subtle structural changes in brain regions such as the hippocampus, an area commonly affected in the early stages of neurodegeneration \cite{ren2023mri, lu2022two, el2022two, https://doi.org/10.1002/alz.14600, aberathne2023detection, beach2012accuracy}. Beyond neuroimaging, CV methods have also enabled the automated extraction and analysis of facial features, offering a non-invasive means to detect AD-related alterations in emotional expression and facial muscle activity \cite{CHEN201777, zheng2023detecting, 11031459}. CV has also been applied to EEG signals to enhance AD detection by transforming brain signals into spectral images, improving classification accuracy by capturing abnormalities in brain wave patterns associated with AD \cite{jpm15010027, XIA2023e14858, alsharabi2023eeg}. Additionally, CV-based analysis of ET data has shown promise in identifying impairments in visual attention, fixation, and eye movement patterns in individuals with AD, highlighting another promising non-invasive biomarker \cite{10.1145/3577190.3614149, sun2022novel, 10433663, math13111867}. In parallel, NLP techniques, especially those based on LLMs, have emerged as powerful tools for detecting cognitive decline through speech-based data. These models analyze both acoustic features and linguistic patterns such as coherence, lexical richness, and syntactic complexity, which are known to deteriorate with AD progression \cite{k2023alzheimersdiseasedetectionspontaneous, mortensen2025early, Han_Lam_Li_Cheung_2025, Mo2024.08.22.24312463, 10495546}.

While ML has demonstrated strong potential in AD detection across various data modalities, its deployment in real-world settings remains limited without the integration of domain adaptation (DA) and domain generalization (DG) techniques. Many ML models are developed under the assumption that training and testing data share the same underlying distribution, even when sourced from different datasets \cite{ijcai2021p628}. This assumption undermines model reliability in practical applications, particularly when DG and DA are not considered. In reality, this condition is rarely met, especially in medical domains like AD detection, where significant variability exists across modalities such as MRI, PET, and EEG. Even within a single modality, differences in imaging equipment, acquisition protocols, and patient demographics can introduce domain shifts that degrade model performance when applied beyond the training environment \cite{10.1145/3654664}. To overcome these challenges, DA focuses on adapting models trained on a source domain to perform well on a different target domain \cite{10.1007/978-981-19-7867-8_47}, while DG technique aims to develop models that are inherently robust to unseen domains \cite{9847099}. The DA/DG techniques are critical for building ML systems that are both generalizable, adaptable, and effective in real-world applications for AD detection.

Formally, a domain can be represented as $\mathcal{D}=\{\mathcal{X},P(X),P(Y|X)\}$, where $\mathcal{X}$ is the feature space, $P(X)$ is the marginal feature distribution, and $P(Y|X)$ is the conditional relationship between features and diagnostic labels. Domain shift occurs when at least one of these components differs between a source domain $\mathcal{D}_s$ and a target domain $\mathcal{D}_t$. In AD detection, covariate shift commonly arises from scanner protocols, electrode layouts, microphones, languages, or ET devices that alter $P(X)$; label shift occurs when the prevalence of CN, MCI, AD, or other cognitive conditions differs between cohorts; and concept shift occurs when the relationship between biomarkers and labels changes because of diagnostic criteria, comorbidities, disease stage, or demographic composition. DA assumes that some target-domain information is available during model development and therefore seeks to reduce the discrepancy between $\mathcal{D}_s$ and $\mathcal{D}_t$. DG is more restrictive: it trains on one or more source domains and aims to perform well on unseen domains without accessing the target domain during training. This distinction is important clinically because DA is suitable when a deployment site can provide unlabeled or limited labeled calibration data, whereas DG is more relevant when models must be deployed in new clinics, languages, or devices without local adaptation data.

\begin{figure}[htbp]
\centering
\begin{tikzpicture}[
  dim/.style={rectangle, draw, rounded corners, align=center, text width=3.5cm, minimum height=1.4cm, inner sep=3pt, font=\scriptsize, fill=red!5},
  node distance=0.35cm and 0.35cm
]
\node[dim] (a) {\textbf{Technical}\\ scanners, vendors, protocols, electrodes, microphones, ET devices\\ \emph{(covariate shift)}};
\node[dim, right=of a] (b) {\textbf{Population}\\ age, sex, education, language, ethnicity, comorbidities\\ \emph{(label/concept shift)}};
\node[dim, right=of b] (c) {\textbf{Site / geographic}\\ institutions, regions, clinic vs.\ community/remote};
\node[dim, below=of a] (d) {\textbf{Clinical-task}\\ HC--AD, CN--MCI, MCI--AD, MCI conversion, multi-class staging};
\node[dim, below=of b] (e) {\textbf{Temporal}\\ dataset drift, evolving criteria, longitudinal follow-up};
\node[dim, below=of c] (f) {\textbf{Workflow}\\ pathway integration, missing modalities, operator variability};
\end{tikzpicture}
\caption{A framework for clinical generalizability in ML-based AD detection. The six dimensions organize the sources of domain shift addressed (or left unaddressed) by DA/DG studies; most reviewed studies evaluate primarily technical and, less often, site/geographic generalization.}
\label{fig:generalizability}
\end{figure}
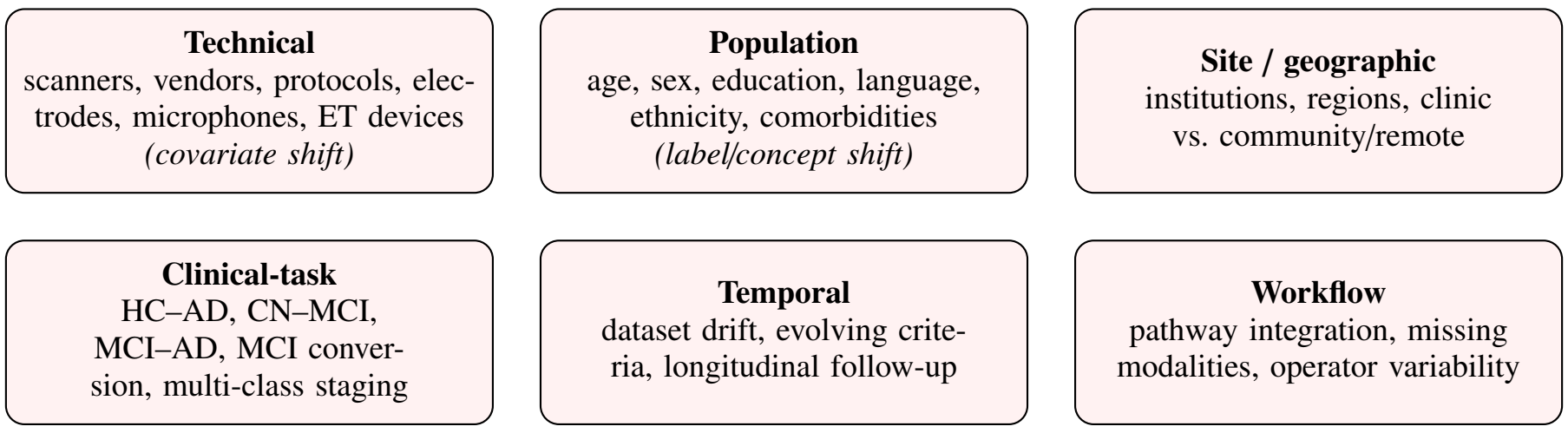

These sources of domain shift ultimately shape what we refer to, in the title of this review, as \emph{clinical generalizability}, which is not a single property but a set of distinct dimensions that are often conflated. To make our contribution explicit and to organize the modality-specific evidence that follows, we group these dimensions into a structured framework (Figure~\ref{fig:generalizability}). \emph{Technical generalization} concerns robustness to differences in scanners, vendors, acquisition protocols, electrode montages, microphones, and ET devices (primarily covariate shift). \emph{Population generalization} concerns robustness across age, sex, education, language, ethnicity, socioeconomic status, and comorbidities. \emph{Site/geographic generalization} concerns transfer across institutions, regions, and care settings (e.g., specialist clinic versus community or remote assessment). \emph{Clinical-task generalization} concerns whether a model transfers across diagnostic endpoints (e.g., HC-vs-AD, CN-vs-MCI, MCI-vs-AD, MCI conversion, multi-class staging), which may exhibit very different shift structures. \emph{Temporal generalization} concerns robustness to dataset drift, evolving diagnostic criteria, and longitudinal follow-up. \emph{Workflow generalization} concerns integration into real clinical pathways, including missing modalities, operator variability, and changes in clinical practice. Throughout the review, we use this framework to characterize which dimension(s) of generalizability a given DA/DG study actually evaluates, and we note that most studies address only technical and, less often, site/geographic generalization, while population, clinical-task, temporal, and workflow generalization remain comparatively underexplored. Framing clinical generalizability in this way also clarifies how the present review is positioned relative to prior surveys.

To the best of our knowledge, as of 2026, this work is among the first comprehensive scoping reviews of recent research on DA/DG techniques for AD detection across diverse data modalities, including neuroimaging, EEG, ET, speech, and multimodal. Although numerous review articles have explored ML applications in AD detection, none have specifically focused on addressing DA/DG across these different modalities. As summarized in Table~\ref{literature_table}, there has been a growing number of reviews on ML and deep learning (DL) in AD detection. However, the majority of these works are narrowly scoped, typically concentrating on a single modality, most often structural MRI, and overlook the critical challenge of domain shift in real-world deployment. For example, reviews by Dahiya et al. \cite{DahiyaSavita2021ADDu} and Dara et al. \cite{app13148298} focus primarily on MRI-based ML methods, while Ebrahimighahnavieh et al. \cite{EBRAHIMIGHAHNAVIEH2020105242} and Khojaste et al. \cite{KHOJASTESARAKHSI2022102332} summarize DL techniques applied solely to the MRI modality. A very recent systematic review of multimodal AI for AD diagnosis \cite{yu2026multimodal} likewise catalogues datasets, models, and modalities. Although these reviews offer valuable insights into model architectures and performance metrics, the absence of DA/DG considerations limits the practical applicability of the reviewed models in diverse clinical settings, where data variability across sites and devices is unavoidable. 

Although a few existing reviews have touched on DA/DG in the context of AD, their coverage remains limited in scope and modality. For example, Li et al. \cite{doi:10.15212/RADSCI-2022-0013} and Reddy et al. \cite{reddy2022applicationunsuperviseddomainadaptation} focus exclusively on DA/DG techniques for MRI data, and Aghdam et al. \cite{Aghdam2025} concentrate on the evaluation of DG methods for MRI and PET data. Broader reviews, such as Guan et al. \cite{9557808}, address DA strategies in medical imaging generally but only briefly address AD-specific challenges and again concentrate mainly on MRI. Gita et al. \cite{10.1145/3654664} examine DA/DG methods across EEG and MRI modalities but do not extend their analysis to other data types. Modality-specific reviews, like those by Vrindha et al. \cite{k2023alzheimersdiseasedetectionspontaneous} on speech and Shah et al. \cite{SHAH2025103202} on ET, provide focused insights but largely overlook domain shift issues and lack discussion on adaptation strategies. 

These limitations underscore a critical gap in the current literature: the lack of a comprehensive, modality-inclusive review that examines DA/DG techniques across diverse data types used in AD detection. Such a unified review is essential because each modality presents distinct domain-specific challenges. Without a consolidated understanding, it is difficult to identify methodological shortcomings and chart a clear path for developing adaptable and generalizable ML models. Existing reviews tend to focus narrowly on single modalities, particularly neuroimaging, and often neglect the broader context of multimodal data integration, which is increasingly central to real-world practice. Variations in acquisition protocols, hardware, demographics, and usage environments introduce significant domain shifts that these models must overcome. To address this gap, our review offers a structured and comprehensive scoping survey of DA/DG techniques across key modalities, including imaging (e.g., MRI, PET, ET), signal (e.g., EEG, speech), and text (e.g., transcripts, clinical notes). We present a taxonomy of adaptation strategies and analyze how domain shifts are addressed in each modality, aiming to support the development of robust, generalizable ML models for deployment in real-world settings. The goal is therefore not to provide an exhaustive catalogue of all AD biomarkers or software tools, but to identify how modality-specific evidence can be connected through the shared problem of domain shift and how multimodal models may combine complementary biomarkers while remaining robust to missing, noisy, or site-specific inputs. MRI, PET, EEG, speech, eye-tracking, and multimodal combinations were selected as the primary scope because they represent the most active areas of DA/DG research for AD detection and together cover complementary clinical, neurophysiological, behavioral, and scalable screening pathways. Adjacent biomarker types such as plasma amyloid-$\beta$, tau, gait, and sleep metrics are acknowledged as important but are not primary review categories because fewer studies have evaluated them under explicit DA/DG or cross-domain validation frameworks.

The remainder of this review is organized as follows: Section~\ref{data} introduces datasets spanning various modalities; Section~\ref{da_dg} surveys DA/DG methods for AD detection across diverse modalities; Section~\ref{challenge_future} outlines key challenges and future directions; and Section~\ref{conclusion} concludes with a summary of insights and findings.

\begin{table}[htbp]
\centering
\caption{A comparison with other review papers in the field.}
\label{literature_table}
\small 
\centering
\begin{tabular}{|C{8cm}|C{1cm}|C{2cm}|C{2cm}|C{2cm}|}
\hline
\textbf{Papers} & \textbf{Year} & \textbf{Review focus} & \textbf{Modality} & \textbf{Model focus} \\ \hline
DL to detect AD from neuroimaging: A systematic literature review \cite{EBRAHIMIGHAHNAVIEH2020105242} & 2020 & AD & MRI & DL \\ \hline
Transfer Learning for AD through Neuroimaging Biomarkers: A Systematic Review \cite{agarwal2021transfer} & 2021 & AD & MRI \& PET & ML \& DL \\ \hline
AD Detection using ML: A Review \cite{DahiyaSavita2021ADDu} & 2021 & AD & MRI & ML \\ \hline
\textbf{DA for Medical Image Analysis: A Survey \cite{9557808}} & 2022 & \textbf{Medical image} & \textbf{MRI} & \textbf{DA} \\ \hline
\textbf{Application of Unsupervised DA for Structural MRI Analysis \cite{reddy2022applicationunsuperviseddomainadaptation}} & 2022 & \textbf{AD} & \textbf{MRI} & \textbf{DA} \\ \hline
AD Detection from Spontaneous Speech and Text: A review \cite{k2023alzheimersdiseasedetectionspontaneous} & 2022 & AD & Speech \& Text & DL \\ \hline
DL for AD diagnosis: A survey \cite{KHOJASTESARAKHSI2022102332} & 2022 & AD & MRI & DL \\ \hline
Transfer Learning Approaches for Neuroimaging Analysis: A Scoping Review \cite{10.3389/frai.2022.780405} & 2022 & AD & MRI \& PET & DL \\ \hline
\textbf{ML for automatic Alzheimer's disease detection: addressing domain shift issues for building robust models \cite{doi:10.15212/RADSCI-2022-0013}} & 2023 & \textbf{AD} & \textbf{MRI} & \textbf{DA} \\ \hline
Alzheimer's Disease Diagnosis Using ML: A Survey \cite{app13148298} & 2023 & AD & MRI & DL \\ \hline
A Survey of DL for Alzheimer's Disease \cite{make5020035} & 2023 & AD & MRI \& PET & DL \\ \hline
A systematic literature review on the significance of DL and ML in predicting AD \cite{KAUR2024102928} & 2024 & AD & MRI \& PET & DL \\ \hline
\textbf{DA and Generalization of Functional Medical Data: A Systematic Survey of Brain Data \cite{10.1145/3654664}} & 2024 & \textbf{Dementia} & \textbf{EEG \& MRI} & \textbf{DA} \\ \hline
Neuropsychological and electrophysiological measurements for diagnosis and prediction of dementia: a review on ML approach \cite{CARRARINI2024102417} & 2024 & AD & EEG & DL \\ \hline
A Review of Datasets, Optimization Strategies, and Learning Algorithms for Analyzing Alzheimer's Dementia Detection\cite{Thulasimani31122024} & 2024 & AD & EEG, MRI, PET & ML, TL \\ \hline
A Review of ML Approaches for Non-Invasive Cognitive Impairment Detection \cite{AlsuhaibaniMuath2025ARoM} & 2025 & AD & Speech \& Text & ML \\ \hline
Alzheimer's Disease Detection Using DL and ML: A Review \cite{mohsen2025alzheimers} & 2025 & AD & MRI, PET, \& CT & DL \\ \hline
A comprehensive review on early detection of AD using various DL techniques \cite{10.3389/fcomp.2024.1404494} & 2025 & AD & MRI \& PET & ML \\ \hline
Eye tracking based detection of mild cognitive impairment: A review \cite{SHAH2025103202} & 2025 & AD & ET & ML \\ \hline
Unlocking the potential of EEG in AD research: Current status and pathways to precision detection \cite{AKBAR2025111281} & 2025 & AD & MRI \& PET & ML \\ \hline
\textbf{Machine-learning models for Alzheimer's disease diagnosis using neuroimaging data: survey, reproducibility, and generalizability evaluation \cite{Aghdam2025}} & 2025 & \textbf{AD} & \textbf{MRI \& PET} & \textbf{DG} \\ \hline
Multimodal AI for Alzheimer Disease Diagnosis: A Systematic Review of Datasets, Models, and Modalities \cite{yu2026multimodal} & 2026 & AD & Multimodal & DL \\ \hline
\textbf{Ours: DA for Alzheimer's Disease Detection: a survey} & 2026 & \textbf{AD} & \textbf{MRI, PET, EEG, ET, Speech} & \textbf{DA/DG} \\ \hline
\end{tabular}
\end{table}

\section{Review Methodology}
\label{sec:methodology}

This review is conducted as a comprehensive scoping review. Our goal is to map the breadth of DA/DG research for AD detection across modalities and to synthesize methodological trends. Accordingly, we report our search and selection process transparently while acknowledging that it was carried out by a single reviewer.

\noindent\textit{Information sources and search strategy}: Relevant literature was identified through structured keyword searches of six bibliographic sources: Google Scholar, PubMed, IEEE Xplore, Scopus, the ACM Digital Library, and arXiv. These sources were chosen because they collectively index the biomedical, engineering, machine-learning, and preprint venues in which DA/DG work for AD detection appears. Searches combined modality terms (e.g., ``MRI'', ``PET'', ``EEG'', ``eye tracking'', ``speech'', ``text'', ``multimodal'') with methodology terms (``domain adaptation'', ``domain generalization'', ``transfer learning'', ``machine learning'') and disease terms (``Alzheimer'', ``dementia'', ``cognitive impairment''), using Boolean \texttt{AND}/\texttt{OR} combinations. The search covered work published up to 2026, with the included studies concentrated in the period 2016--2026.

\noindent\textit{Eligibility criteria}: We included studies that (i) applied DA, DG, or transfer-learning techniques explicitly addressing domain shift, (ii) targeted AD, MCI, or dementia detection using one of the reviewed modalities, and (iii) reported an identifiable adaptation or generalization setting. Because direct AD-specific evidence is sparse for some modalities (notably EEG and ET), we also included a limited set of methodologically transferable studies from adjacent domains (e.g., general EEG classification and gaze estimation); these are clearly identified as such in the corresponding tables and text. We excluded purely machine-learning or deep-learning studies that did not address domain shift, adaptation, or cross-domain generalization.

\begin{figure}[htbp]
\centering
\begin{tikzpicture}[
  node distance=0.55cm and 0.4cm,
  box/.style={rectangle, draw, rounded corners, align=center, text width=10.5cm, inner sep=5pt, font=\footnotesize},
  small/.style={rectangle, draw, rounded corners, align=center, text width=2cm, inner sep=3pt, font=\scriptsize},
  arr/.style={-{Latex[length=2mm]}, thick}
]
\node[box] (id) {\textbf{Identification.} Structured keyword searches of Google Scholar, PubMed, IEEE Xplore, Scopus, ACM Digital Library, and arXiv \\ (modality $\times$ methodology $\times$ disease terms; published up to 2026)};
\node[box, below=of id] (screen) {\textbf{Screening.} Title/abstract screening for relevance to DA/DG or domain-shift-aware learning for AD/MCI/dementia detection};
\node[box, below=of screen] (elig) {\textbf{Eligibility.} Full-text assessment against inclusion criteria; pure ML/DL studies without any domain-shift component excluded; \\ limited adjacent-domain studies (general EEG classification, gaze estimation) retained where AD-specific evidence is sparse};
\node[box, below=of elig] (incl) {\textbf{Included in qualitative synthesis: N\,=\,83 studies}};
\node[small, below=0.7cm of incl.south, xshift=-4.2cm] (n1) {Neuro-imaging\\ (MRI/PET)\\ \textbf{31}};
\node[small, right=of n1] (n2) {EEG\\ \textbf{10}};
\node[small, right=of n2] (n3) {ET\\ \textbf{15}};
\node[small, right=of n3] (n4) {Speech/\\ text\\ \textbf{14}};
\node[small, right=of n4] (n5) {Multi-\\ modal\\ \textbf{13}};
\draw[arr] (id) -- (screen);
\draw[arr] (screen) -- (elig);
\draw[arr] (elig) -- (incl);
\draw[arr] (incl.south) -- (n1.north);
\draw[arr] (incl.south) -- (n2.north);
\draw[arr] (incl.south) -- (n3.north);
\draw[arr] (incl.south) -- (n4.north);
\draw[arr] (incl.south) -- (n5.north);
\end{tikzpicture}
\caption{ Scoping-review selection flow. As a non-registered single-reviewer search, exact intermediate record counts were not logged; the diagram reports the search process and the final set of 83 studies included in the modality-specific DA/DG synthesis tables. For EEG and ET, the included set contains a limited number of methodologically transferable studies from adjacent domains (general EEG classification and gaze estimation), which are identified as such in the corresponding tables.}
\label{fig:prisma}
\end{figure}
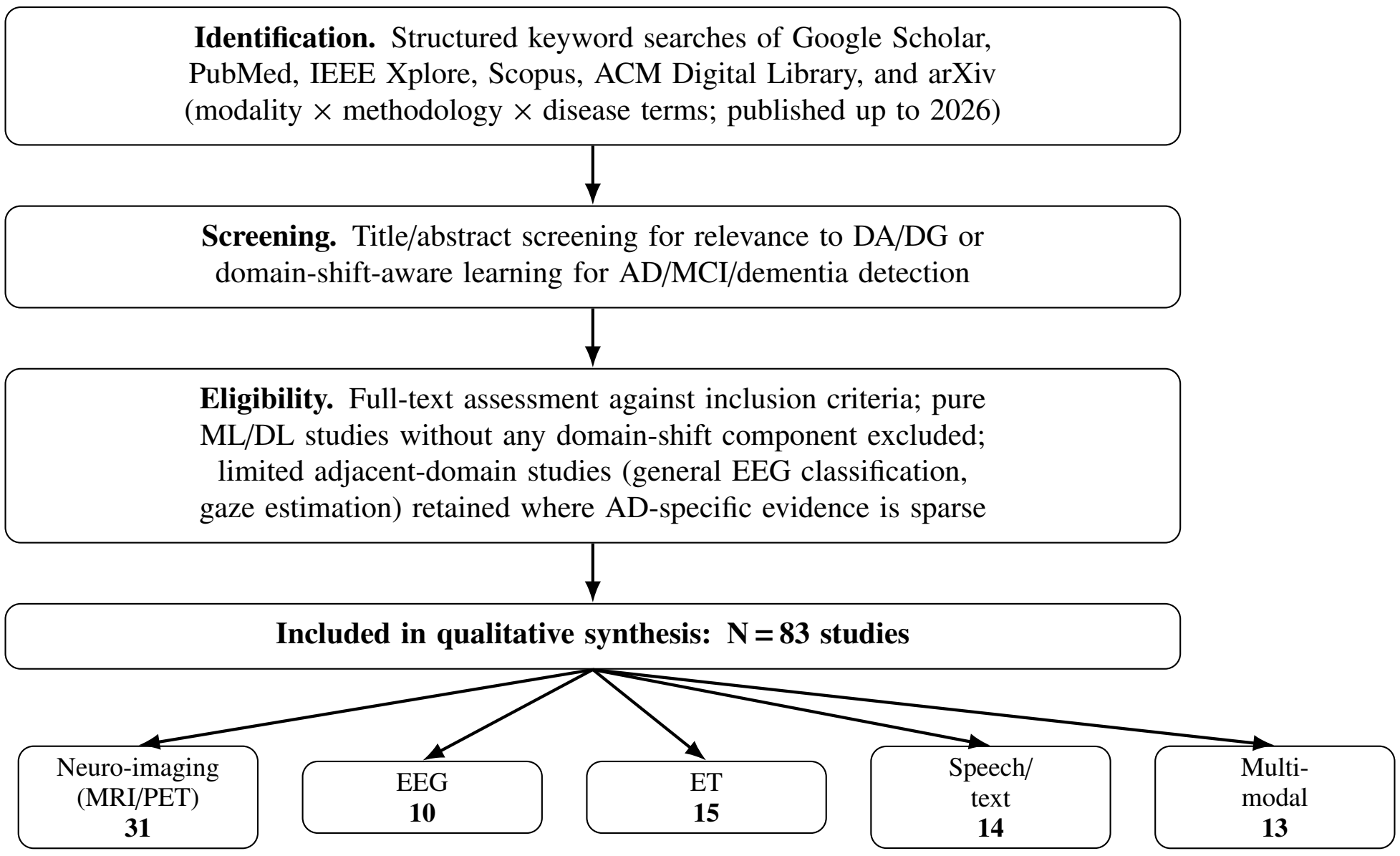

\noindent\textit{Selection outcome}: After screening for relevance and applying the criteria above, 83 studies were retained and organized into the modality-specific DA/DG synthesis tables: 31 for neuroimaging (MRI/PET), 10 for EEG, 15 for ET, 14 for speech and text, and 13 for multimodal integration. Figure~\ref{fig:prisma} therefore summarizes the selection flow and the final included set.

\section{Data Modalities and Sources}
\label{data}

This section reviews widely used datasets for AD detection, organized by data modality as summarized in Table~\ref{dataset_table}: (1) neuroimaging (MRI and PET), (2) EEG signals, (3) ET datasets for gaze and head movement, (4) speech-based datasets including audio recordings and/or transcripts, and (5) multimodal datasets that include more than one modality.

These modalities were selected because they currently form the main body of DA/DG research for AD detection and represent different deployment profiles. MRI and PET provide clinically established structural and molecular information but require specialized infrastructure. EEG provides portable neurophysiological measurements. Speech and ET provide lower-cost behavioral signals that can be collected in clinics, community settings, or remotely. Biomarkers such as amyloid-$\beta$, tau, plasma markers, gait, sleep, and broader digital phenotyping are highly relevant for AD detection; however, they are discussed as complementary or future multimodal inputs rather than primary review categories because fewer studies have evaluated them through explicit DA/DG or cross-domain validation frameworks.

\begin{table}[]
\centering
\small
\caption{Datasets used for AD detection.}
\label{dataset_table}
\begin{tabular}{|C{1cm}C{3cm}|C{4cm}|C{6cm}|}
\hline
\multicolumn{2}{|c|}{\textbf{Data   Modality}}                                       & \textbf{Dataset}                                              & \textbf{Conditions Distribution}   \\ \hline
\multicolumn{1}{|c|}{\multirow{7}{*}{\textbf{Neuroimaging}}} & MRI                   & ADNI -1\cite{mueller2005adni}                                 & 200 AD, 400 MCI, 200 CN        \\ \cline{2-4} 
\multicolumn{1}{|c|}{}                                       & MRI                   & ADNI -GO\cite{mueller2005adni}                                & 200 AD, 600 MCI, 200 CN        \\ \cline{2-4} 
\multicolumn{1}{|c|}{}                                       & PET                   & ADNI -2\cite{mueller2005adni}                                 & 160 AD, 460 MCI, 211 CN        \\ \cline{2-4} 
\multicolumn{1}{|c|}{}                                       & PET                   & ADNI -3\cite{mueller2005adni}                                 & 117 AD, 197 MCI, 211 CN        \\ \cline{2-4} 
\multicolumn{1}{|c|}{}                                       & MRI                   & AIBL\cite{ellis2009aibl}                                      & 211 AD, 133 MCI, 768 CN        \\ \cline{2-4} 
\multicolumn{1}{|c|}{}                                       & MRI                   & IXI \cite{ixi2017dataset}                                     & 600 CN                             \\ \cline{2-4} 
\multicolumn{1}{|c|}{}                                       & MRI                   & PND \cite{aalten2014parelsnoer}                                       & 198 AD, 48 MCI converters, 91 MCI non-converters, 138 SCD                             \\ \cline{2-4}
\multicolumn{1}{|c|}{}                                       & MRI                   & Kaggle dataset \cite{alzheimers_dataset_kaggle}                                       & 64 Moderated Dementia, 896 Mild Dementia, 3200 CN, 2240 very mild dementia                             \\ \cline{2-4}
\multicolumn{1}{|c|}{}                                       & MRI                   & CADDementia challenge \cite{BRON2015562}                                     & 150 AD, 77 MCI, 127 CN                             \\ \cline{2-4} 
\multicolumn{1}{|c|}{}                                       & MRI                   & MIRIAD \cite{malone2013miriad}                                & 46 AD, 23 HC                    \\ \hline
\multicolumn{2}{|c|}{\multirow{3}{*}{\textbf{EEG}}}                                  & Alzheimer's EEG Dataset\cite{alzheimers_eeg_github}           & 10 AD, 10 CN                    \\ \cline{3-4} 
\multicolumn{2}{|c|}{}                                  & Healthy-MCI-AD Dataset\cite{sedghizadeh2023eeg}           & 13 AD, 7 MCI, 15 CN                    \\ \cline{3-4} 
\multicolumn{2}{|c|}{}                                  & CAUEEG dataset \cite{KIM2023120054}          &  311 AD, 417 MCI, 459 CN                   \\ \cline{3-4} 
\multicolumn{2}{|c|}{}                                  & GENEEG dataset \cite{mishra2025challengesdetectingmciusing}         & 28 MCI, 35 HC                    \\ \cline{3-4} 
\multicolumn{2}{|c|}{}                                  & AD-Auditory \cite{lahijanian2024auditory}      & 17 AD, 6 MCI, 10 CN                    \\ \cline{3-4} 
\multicolumn{2}{|c|}{}                                  & ADFSU \cite{vicchietti2023computational}        & 160 AD, 24 CN                    \\ \cline{3-4} 
\multicolumn{2}{|c|}{}                                  & ADFTD \cite{data8060095}        & 36 AD, 23 FTD, 29 CN                   \\ \cline{3-4} 
\multicolumn{2}{|c|}{}                                  & ADSZ \cite{Alves_2022}        & 24 AD, 24 CN                    \\ \cline{3-4} 
\multicolumn{2}{|c|}{}                                  & APAVA \cite{Escudero_2006}       & 11 AD, 11 HC                    \\ \cline{3-4} 
\multicolumn{2}{|c|}{}                                  & DMS Dataset \cite{li2017cognitive}         &  13 AD, 16 MCI, 18 CN                  \\ \cline{3-4} 
\multicolumn{2}{|c|}{}                                                               & OpenNeuro ds004504 \cite{li2024resting}                       & 36 AD, 23 FTD, 29 CN           \\ \cline{3-4} 
\multicolumn{2}{|c|}{}                                                               & PEARL-Neuro \cite{dzianok2024pearl}                               & 79 AD, 192 CN                   \\ \hline
\multicolumn{2}{|c|}{\multirow{3}{*}{\textbf{ET}}}                       & GazeMap \cite{math13111867}                                   & 51 AD, 61 CN, and 23 AD, 21 CN \\ \cline{3-4} 
\multicolumn{2}{|c|}{}                                                               & Eye Focus Time \cite{eyefocustime2025}                        & 16 AD, 16 CN                    \\ \cline{3-4} 
\multicolumn{2}{|c|}{}                                                               & EM-COGLOAD \cite{MILES2024264}                                & 75 CN                              \\ \hline
\multicolumn{1}{|c|}{\multirow{6}{*}{\textbf{Speech-based}}} & Speech, Text          & ADReSS \cite{luz2020alzheimersdementiarecognitionspontaneous} & 78 AD, 78 HC                    \\ \cline{2-4} 
\multicolumn{1}{|c|}{}                                       & Speech, Text          & Pitt Corpus \cite{becker1994natural}                          & 181 AD, 101 HC                 \\ \cline{2-4} 
\multicolumn{1}{|c|}{}                                       & Speech, Text          & CCC \cite{pope2011finding}                                    & 125 AD,  125 HC                    \\ \cline{2-4} 
\multicolumn{1}{|c|}{}                                       & Text          & Alzheimer's Disease Blog Corpus (ADBC)  \cite{masrani-etal-2017-detecting}                               & 1654 AD, 1151 CN                    \\ \cline{2-4} 
\multicolumn{1}{|c|}{}                                       & Speech                & WLS Dataset \cite{herd2014cohort}                             & 115 AD, 839 CN                  \\ \cline{2-4} 
\multicolumn{1}{|c|}{}                                       & Speech                & ADReSSo \cite{luz2021detecting}                               & 122 AD, 115 CN                  \\ \cline{2-4} 
\multicolumn{1}{|c|}{}                                       & Speech                & ADReSS-M \cite{luz2023multilingualalzheimersdementiarecognition}                               & 148 AD, 143 HC                  \\ \cline{2-4} 
\multicolumn{1}{|c|}{}                                       & Speech                & TAUKADIAL \cite{luz2024connectedspeechbasedcognitiveassessment}                               & N/A                  \\ \cline{2-4} 
\multicolumn{1}{|c|}{}                                       & Speech, Text               & Multimodal Dementia Corpus \cite{gkoumas2024longitudinal}                               & 12 AD, 10 HC                  \\ \cline{2-4} 
\multicolumn{1}{|c|}{}                                       & Speech               & VAS Corpus \cite{10095825}                               & 30 AD, 35 MCI, 36 HC                 \\ \cline{2-4} 
\multicolumn{1}{|c|}{}                                       & Speech               & Delaware corpus \cite{lanzi2023dementiabank}                               & 46 MCI, 24 HC                  \\ \cline{2-4} 
\multicolumn{1}{|c|}{}                                       & Speech                & French Dem@Care corpus \cite{10.1145/3154862.3154915}                               & 68 AD, 47 HC                   \\ \cline{2-4}
\multicolumn{1}{|c|}{}                                       & Speech, Text          & NCMMSC2021\cite{ying2023multimodal}                           & 26 AD, 53 MCI, 44 CN           \\ \hline
\multicolumn{1}{|c|}{\multirow{6}{*}{\textbf{Multimodal}}}   & Speech, Text, Video   & PROMPT\cite{kishimoto2020prompt}                              & 100 AD, 200 CN                \\ \cline{2-4} 
\multicolumn{1}{|c|}{}                                       & Speech, Text, Video   & I-CONECT \cite{dodge2024iconect}                              & 100 MCI, 86 CN                   \\ \cline{2-4} 
\multicolumn{1}{|c|}{}                                       & Speech, Text, Video   & Facial and Speech Dataset \cite{takeshige2024digital}         & 93 AD, 99 CN                    \\ \cline{2-4} 
\multicolumn{1}{|c|}{}                                       & ET,  speech & UBC Webcam \cite{Harisinghani_2023}                           & 68 AD, 73 CN             \\ \cline{2-4} 
\multicolumn{1}{|c|}{}                                       & MRI, PET              & OASIS \cite{marcus2007open}                                   & 489 AD, 609 CN                  \\ \cline{2-4} 
\multicolumn{1}{|c|}{}                                       & MRI, EEG              &    BrainLat \cite{prado2023brainlat}                                 & 280 AD, 250 CN                  \\ \cline{2-4} 
\multicolumn{1}{|c|}{}                                       & MRI, speech              & COMPASS-ND \cite{CCNA_COMPASSND_2025}                                 & 113 AD, 280 MCI                  \\ \cline{2-4} 
\multicolumn{1}{|c|}{}                                       & MRI, PET               & NACC \cite{Beekly2004NACC}                                 & N/A                  \\ \cline{2-4}
\multicolumn{1}{|c|}{}                                       & MRI, PET               & J-ADNI \cite{JADNI}                                 &  149 AD, 234 MCI, 154 CN                  \\ \cline{2-4}
\multicolumn{1}{|c|}{}                                       & MRI, PET & FHS \cite{framingham2020dataset}                                 & 400-500 AD, 800-900 MCI, 1000+ CN, \\ \cline{2-4}
\multicolumn{1}{|c|}{}                                       & ET, EEG     & Mendeley \cite{sedghizadeh2023eeg}                          & 13 AD, 7 MCI, 15 CN            \\ \hline
\end{tabular}
\end{table}

A critical limitation across the dataset landscape is the heavy reliance on a small number of benchmark cohorts, especially ADNI for neuroimaging and ADReSS/Pitt-style datasets for speech. These resources are valuable because they enable reproducible experimentation, but they can also narrow the population, geographic, language, and acquisition diversity represented in model development. Dataset imbalance is another recurring concern, because CN samples may outnumber MCI or AD cases, while real-world screening populations include many people who do not fit neatly into binary HC-versus-AD labels. In addition, cross-domain studies must carefully avoid data leakage through subject overlap, preprocessing using full-dataset statistics, slice-level splitting of MRI scans, or reuse of derived features across train and test folds. For this reason, the tables below should be interpreted not only as dataset inventories, but also as evidence of the current limits of reproducibility and external validity.

\subsection{Neuroimaging datasets}

Numerous publicly available neuroimaging datasets have played a pivotal role in advancing AD detection by enabling the development and evaluation of ML models across various modalities. Among these, the Alzheimer's Disease Neuroimaging Initiative (ADNI) \cite{mueller2005adni} remains the most extensively used benchmark. It provides comprehensive data, including structural MRI, PET scans, and clinical assessments, covering individuals at different stages of cognitive decline. Complementing ADNI, the Australian Imaging, Biomarkers and Lifestyle (AIBL) study \cite{ellis2009aibl} emphasizes amyloid imaging and lifestyle factors in an Australian cohort, making it particularly valuable for research on cross-cohort generalization.

Several other neuroimaging datasets support specific benchmarking or baseline analysis. For instance, the Information eXtraction from Images (IXI) dataset \cite{ixi2017dataset} consists of MRI scans from healthy individuals and is commonly used for baseline modeling or as a control dataset in AD research. The CADDementia challenge dataset \cite{BRON2015562} was specifically designed to benchmark automated AD diagnosis systems, using T1-weighted MRI data collected from multiple centers, with a hidden test set enabling rigorous validation of ML model generalization and performance. 

For longitudinal studies, the Minimal Interval Resonance Imaging in Alzheimer's Disease (MIRIAD) dataset \cite{malone2013miriad} provides serial MRI scans collected at varying intervals, allowing researchers to investigate disease progression and the utility of imaging biomarkers over time. Similarly, the Parelsnoer Neurodegenerative Diseases Biobank (PND) \cite{aalten2014parelsnoer} offers a rich multimodal dataset with MRI and clinical assessments from people with cognitive complaints. It includes subgroups such as AD, MCI converters and non-converters, and individuals with subjective cognitive decline (SCD), making it highly suitable for longitudinal modeling despite the absence of healthy controls.

In addition, the Kaggle AD dataset \cite{alzheimers_dataset_kaggle}, which contains 2D structural brain MRI images labeled across four diagnostic categories, is widely used for training and benchmarking convolutional neural networks (CNNs), especially in lightweight or multiclass classification tasks.

The neuroimaging datasets reviewed provide some of the strongest resources for AD model development because they include structural MRI, PET, clinical assessments, longitudinal follow-up, and multi-center benchmarking settings. However, the same dataset landscape also introduces important limitations for DA/DG research. ADNI remains the dominant benchmark, which improves comparability across studies but can also lead to overfitting of methods to ADNI-specific scanner protocols, recruitment criteria, preprocessing pipelines, and demographic composition. Although datasets such as AIBL, OASIS, J-ADNI, NACC, PND, CADDementia, and MIRIAD broaden the available imaging landscape, many studies still evaluate models using random splits or limited cross-cohort settings rather than rigorous leave-one-site-out or external validation. Neuroimaging datasets also differ in diagnostic criteria, disease-stage distribution, MRI/PET acquisition protocols, and availability of longitudinal follow-up. These differences create substantial domain shift and make direct comparison difficult. Reproducibility concerns are especially important for imaging studies because data leakage can occur through slice-level splitting, subject overlap across derived datasets, preprocessing statistics computed on the full dataset, or repeated scans from the same participant appearing in both training and testing sets. Therefore, future neuroimaging DA/DG studies should report subject-level splits, site-level validation, preprocessing procedures, and demographic subgroup performance more transparently.

\subsection{EEG datasets}

EEG offers a non-invasive, cost-effective, and portable method for recording real-time brain activity, making it particularly valuable for early AD detection. Unlike structural neuroimaging, EEG provides high temporal resolution, enabling the capture of dynamic neural oscillations that reflect subtle cognitive impairments, even at preclinical stages. These properties, combined with its compatibility with lightweight and ML models, have made EEG an increasingly popular modality in AD research. This section provides an overview of publicly available EEG datasets commonly used in ML studies, especially those addressing DA/DG challenges.

Several EEG datasets center on resting-state recordings and involve cognitive tasks targeting memory and sensory processing. For instance, the DMS dataset \cite{li2017cognitive} and the Healthy-MCI-AD dataset \cite{sedghizadeh2023eeg} include experimental paradigms such as working memory retrieval and olfactory stimulation. These datasets enable the analysis of cognitive decline through event-related potentials and frequency-based features, typically across diagnostic categories like CN, MCI, and AD. Similarly, ADFSU \cite{vicchietti2023computational}, ADFTD \cite{data8060095}, and ADSZ \cite{Alves_2022}, focus on high-quality resting-state EEG collected using the standard 10–20 system and supplemented with clinical information such as Mini-Mental State Examination (MMSE) scores. Their consistent protocols and structured annotations make them particularly well-suited for DA/DG research, especially in multi-site model evaluation.

Large-scale clinical datasets have also emerged, offering broader demographic coverage and increased statistical power. For example, the CAUEEG dataset \cite{KIM2023120054} includes EEG recordings from over 1,100 participants across AD, MCI, and CN groups, with multimodal signals and standardized clinical acquisition protocols. Similarly, the large-scale PEARL-Neuro dataset \cite{dzianok2024pearl} provides rich synchronized EEG and MRI recordings during both resting-state and cognitive tasks. The inclusion of extensive behavioural and demographic data supports research on the development of robust, generalizable models under varied experimental conditions.

In contrast, smaller-scale datasets such as the Alzheimer's EEG collection on GitHub \cite{alzheimers_eeg_github}, GENEEG \cite{mishra2025challengesdetectingmciusing}, and APAVA \cite{Escudero_2006} are frequently used for early-stage experimentation and algorithm prototyping. While limited in sample size, they are often preprocessed and formatted to facilitate binary classification tasks, facilitating accessible platforms for new methodology testing. Specialized small datasets like AD-Auditory \cite{lahijanian2024auditory}, which capture EEG responses to 40-Hz auditory stimulation, provide unique insights into AD-related deficits in neural entrainment and rhythmic processing.

Finally, research challenge datasets such as OpenNeuro's ds004504 \cite{li2024resting} provide well-curated and publicly accessible EEG recordings, including data from people with AD and frontotemporal dementia (FTD). These datasets are particularly valuable for benchmarking classification algorithms, supporting differential diagnosis, and fostering reproducibility through open science initiatives.

The contrast between large clinical resources such as CAUEEG and PEARL-Neuro, smaller public datasets such as GENEEG and APAVA, and task-specific datasets such as AD-Auditory highlights both the promise and the limitations of EEG-based AD research. EEG provides a more portable and lower-cost alternative to MRI and PET, but many publicly available EEG datasets for AD detection remain small, sometimes including only tens of participants. This increases the risk of overfitting and limits the reliability of reported performance. Across datasets, differences in electrode montage, sampling rate, recording condition, resting-state or task design, artifact removal procedure, and clinical annotation can create strong domain shift across laboratories and make cross-dataset comparison difficult. Class imbalance is another concern, especially when MCI or non-AD dementia groups are underrepresented. Reproducibility can also be affected by trial-level or segment-level splitting, where multiple windows from the same participant may appear in both training and testing sets. For DA/DG research, EEG studies should therefore prioritize subject-level and site-level splitting, transparent preprocessing, artifact handling, and external validation across different acquisition protocols.

\subsection{ET datasets}

As summarized in Table~\ref{dataset_table}, the number of publicly available ET datasets for AD detection remains limited compared to the more abundant neuroimaging and EEG datasets. Nevertheless, ET technologies become promising in AD research due to their ability to capture valuable non-invasive biomarkers for early diagnosis and ongoing cognitive monitoring. Moreover, the increasing availability and portability of ET hardware, including affordable, webcam-based systems has enhanced the feasibility of integrating ET into both clinical and real-world screening environments.

A notable ET dataset for AD detection is the GazeMap dataset, introduced by Jung et al. \cite{math13111867}, which captures gaze and head movement time-series data from 112 participants (61 CN and 51 AD)  in a clinical setting. Additionally, an out-of-distribution cohort was collected from nursing homes and senior centers, including 23 people with AD and 21 CN.

The Eye Focus Time Dataset \cite{eyefocustime2025} offers a more focused examination of visual attention, capturing fixation durations across 10 trials for each of 32 participants (16 AD and 16 CN). While primarily designed for educational use, this dataset is annotated with clinical labels and provides a useful benchmark for binary classification tasks and early exploration of behavioral biomarkers in AD.

In addition, the EM-COGLOAD dataset \cite{MILES2024264}, available through the Open Science Framework, contains detailed eye movement data from 75 participants engaged in a variety of cognitive tasks. Although not exclusive to AD, its inclusion of age-related and cognitive-load-related gaze behavior makes it relevant for understanding how cognitive impairments affect visual dynamics. The dataset's structured task design and open availability make it valuable for investigating age-related cognitive decline and its manifestation in eye movement patterns.

The three ET resources described above illustrate the current state of ET-based AD research: GazeMap provides clinically collected gaze and head-movement data with an additional out-of-distribution cohort, the Eye Focus Time dataset offers a small controlled benchmark for visual-attention analysis, and EM-COGLOAD supports broader investigation of gaze behavior under cognitive-load conditions. However, these resources also show that ET datasets remain limited in number and size compared with neuroimaging and speech datasets, which restricts the strength of current conclusions about generalizability. Existing ET datasets often involve small cohorts, binary AD-versus-CN labels, and controlled tasks, which may not represent population-level screening conditions. ET features are also sensitive to hardware type, calibration quality, camera resolution, lighting, visual stimulus design, screen size, head movement, and participant familiarity with digital tasks. These factors can introduce device- and environment-specific domain shift. Geographic and demographic diversity is also limited in many ET datasets, and the influence of education, vision impairment, cultural viewing patterns, and motor symptoms is often underexplored. Future ET-based DA/DG studies should therefore include larger and more diverse cohorts, report device and calibration details, evaluate cross-device and cross-environment robustness, and avoid trial-level leakage from repeated gaze sequences of the same participant.

\subsection{Speech-based datasets}

With the advancement of NLP, speech-based datasets have become promising for early AD detection. Speech provides rich linguistic and acoustic markers, such as fluency, prosody, coherence, and lexical diversity, that can reveal subtle signs of cognitive decline. As a result, numerous datasets have been curated to support the development of ML models for identifying cognitive impairments based on spontaneous and task-driven speech.

Several foundational datasets focus on structured clinical interviews or neuropsychological tasks, offering both audio recordings and manual transcripts. Among the most widely used is the Pitt Corpus, also known as the DementiaBank dataset \cite{becker1994natural}, which includes recordings of tasks such as the Cookie Theft picture description, word fluency, and sentence construction. This dataset has been instrumental in detailed linguistic and cognitive analysis. Complementing this, the Delaware Corpus \cite{lanzi2023dementiabank}, as part of the expanded DementiaBank project, emphasizes early detection and includes a broader range of elicited language tasks from participants with MCI and non-impaired individuals. Similarly, the Wisconsin Longitudinal Study (WLS) \cite{herd2014cohort} further contributes large-scale responses to cognitive tasks and the Cookie Theft picture description. 

Benchmark datasets designed for shared tasks have also played a crucial role in standardizing evaluation. The ADReSS \cite{luz2020alzheimersdementiarecognitionspontaneous}, ADReSSo \cite{luz2021detecting}, and ADReSS-M \cite{luz2023multilingualalzheimersdementiarecognition} datasets provide balanced diagnostic categories, MMSE scores, and standardized training-test splits. These datasets enable both classification and regression tasks and serve as strong baselines for model comparison across research studies.

Conversational speech datasets offer another valuable dimension by capturing natural, spontaneous language use in real-world settings. The Carolinas Conversation Collection (CCC) \cite{pope2011finding} contains longitudinal interviews with AD and HC participants across multiple sessions, making it a rich resource for studying temporal changes in pragmatic language use. Similarly, the Voice-Assistant Systems (VAS) dataset \cite{10095825} captures user interactions with voice assistants, offering unique insights into how cognitive decline influences everyday conversational behavior. The Multimodal Dementia Corpus \cite{gkoumas2024longitudinal} further enhances this research landscape by offering a fine-grained longitudinal dataset that captures cognitive and language changes in both AD and HC participants through natural spoken conversations, as well as additional input modalities such as writing and typing.

Multilingual and cross-cultural datasets are also gaining attention for their role in enhancing generalization and cross-linguistic transfer. The NCMMSC2021 dataset \cite{ying2023multimodal} includes Mandarin speech samples from AD, MCI, and HC participants performing various tasks, while the French Dem@Care corpus \cite{10.1145/3154862.3154915} contains verbal and semantic fluency, picture description, and story-telling tasks from French-speaking older adults with and without dementia. The TAUKADIAL dataset \cite{luz2024connectedspeechbasedcognitiveassessment} broadens this multilingual perspective in both Chinese and English by collecting 507 clinical speech samples during cognitive-linguistic assessments. These datasets allow for the development and validation of models across diverse linguistic and cultural contexts, promoting inclusivity and robustness in AD detection tools.

Finally, expanding beyond spoken language, the Alzheimer's Disease Blog Corpus (ADBC) \cite{masrani-etal-2017-detecting} offers a valuable source of written language data. Consisting of blog posts authored by individuals with dementia and their caregivers, ADBC supports the analysis of linguistic deterioration in unscripted, natural text. Although not speech-based, it enables multimodal investigations of language-related cognitive decline and complements spoken corpora by capturing everyday communication patterns in written form.

The diversity of speech and language resources described above shows why speech-based AD detection is attractive for scalable screening, but it also highlights several dataset-level limitations. Foundational corpora such as Pitt/DementiaBank, Delaware, and WLS support detailed analysis of elicited language tasks, while ADReSS-style shared-task datasets improve benchmarking through balanced labels and standardized splits. Conversational datasets such as CCC, VAS, and the Multimodal Dementia Corpus capture more natural communication, and multilingual resources such as NCMMSC2021, French Dem@Care, and TAUKADIAL broaden cross-linguistic coverage. However, these datasets differ substantially in task design, language, recording setting, microphone quality, transcription procedure, interviewer behavior, and participant demographics. As a result, models trained on one speech corpus may learn task-specific or dataset-specific patterns rather than disease-general speech markers.

Heavy reliance on English-language datasets such as Pitt/DementiaBank and ADReSS-style corpora also limits generalization to other languages, dialects, cultures, and education levels. Although balanced challenge datasets improve controlled benchmarking, they may not reflect real-world screening populations where MCI, depression, low literacy, hearing impairment, multilingual speakers, and other neurological conditions are common. Data leakage can occur when multiple speech samples or transcripts from the same participant are split across training and testing sets, or when text normalization and feature extraction use information from the full dataset. Future speech-based DA/DG studies should therefore emphasize participant-level splits, cross-lingual and cross-corpus validation, transparent transcription protocols, and evaluation on broader clinically realistic populations.

\subsection{Multimodal datasets}

Multimodal datasets, which integrate multiple data modalities such as speech, video, EEG, ET, and neuroimaging, have become increasingly valuable in AD research. By capturing complementary aspects of cognitive function, these datasets enable the development of more robust and generalizable ML models for early detection and monitoring of AD. This section reviews key publicly available multimodal datasets, categorized based on the types of modalities included.

The most widely available multimodal resources are neuroimaging-focused datasets that include MRI and PET imaging. For instance, the OASIS dataset \cite{marcus2007open} provides over 2,100 MRI and 1,600 PET sessions from 1,098 participants, covering both cognitively normal individuals and those with varying degrees of cognitive impairment. Similarly, the Japanese J-ADNI dataset \cite{JADNI} mirrors the U.S.-based ADNI project by offering multimodal imaging and clinical data tailored to Japanese cohorts, supporting culturally relevant research in AD and MCI. The Framingham Heart Study (FHS) \cite{framingham2020dataset}, a long-running cohort study, includes MRI and PET imaging and cognitive assessments from participants with AD, MCI, and normal cognition, while the large-scale NACC dataset \cite{Beekly2004NACC} provides PET, MRI, and longitudinal clinical data from over 50,000 individuals across 36 AD Research Centers, enabling extensive epidemiological and diagnostic research.

Beyond traditional neuroimaging, some multimodal datasets combine MRI with other non-invasive signals, facilitating more comprehensive assessment in clinical environments. The BrainLat dataset \cite{prado2023brainlat}, collected across five Latin American countries, includes MRI, EEG, and clinical assessments from 530 participants. By emphasizing diversity and inclusion from underrepresented populations, it offers a valuable resource for studying AD across global contexts. Likewise, the COMPASS-ND dataset \cite{CCNA_COMPASSND_2025}, developed under Canada's national CCNA initiative, includes 393 participants spanning a broad spectrum of neurodegenerative conditions, and integrates MRI, recorded speech, and comprehensive clinical assessments, making it well-suited for both multimodal and longitudinal studies.

Another group of datasets emphasizes audio-visual and behavioral interactions in natural or semi-structured environments. The PROMPT dataset \cite{kishimoto2020prompt}, developed in Japan, contains 447 video recordings of participants performing conversational and cognitive tasks. It captures speech, facial expressions, and body movements from individuals with dementia, depression, bipolar disorder, and healthy controls. Similarly, the I-CONECT dataset \cite{ dodge2024iconect} offers longitudinal video recordings from 186 older adults to evaluate changes in engagement and cognition over time by repeated cognitive assessments and MoCA scores. The dataset introduced by Amano et al. \cite{takeshige2024digital} captures interactions between 192 individuals and a conversational chatbot, recording both audio and facial cues. With 99 healthy controls and 93 participants with MCI or AD, this resource enables exploration of spontaneous communication and multimodal symptom markers in real-world settings.

Smaller but notable datasets combine ET with other modalities, offering additional insight into attention and sensory processing. In the UBC Webcam dataset \cite{Harisinghani_2023}, speech and ET were recorded with a webcam while 141 participants were performing the Cookie Theft picture description task. The combination of gaze and speech features significantly improved classification performance (AUC increased from 0.73 to 0.80), demonstrating the synergy of multimodal signals. Sedghizadeh et al. \cite{sedghizadeh2023eeg} released another multimodal dataset that integrates ET and EEG recordings from 35 participants (15 HC, 7 MCI, and 13 AD) during an olfactory oddball perception task. While originally focused on EEG, the inclusion of gaze data enhances the understanding of sensory and attentional changes in AD.

The multimodal resources reviewed above show two major directions in current AD dataset development: large neuroimaging-centered cohorts such as OASIS, J-ADNI, FHS, NACC, BrainLat, and COMPASS-ND, and more behavior-centered datasets such as PROMPT, I-CONECT, the chatbot interaction dataset, UBC Webcam, and ET--EEG recordings. These datasets are valuable because they can combine complementary biological, cognitive, linguistic, and behavioral information. However, they also introduce additional reproducibility and bias concerns. Many multimodal resources remain centered on large benchmark cohorts or specific regional initiatives, which may limit geographic and demographic representativeness. Missing modalities are common because not all participants complete MRI, PET, EEG, speech, ET, or clinical assessments, and excluding incomplete cases can introduce selection bias toward healthier or more compliant participants. Different modalities also have different acquisition schedules, preprocessing pipelines, noise characteristics, and site-specific effects, making alignment and synchronization difficult. In multimodal DA/DG studies, leakage may occur if modality-specific features are extracted or normalized using the full dataset, if repeated visits from the same participant are split across folds, or if one modality indirectly encodes site or diagnostic information. Therefore, future multimodal dataset design should report missingness patterns, harmonize acquisition metadata, evaluate robustness when one or more modalities are absent, and use subject-level, site-level, and longitudinal splits to support reproducible and clinically meaningful generalization.

\section{DA/DG Techniques for AD Detection}
\label{da_dg}

\begin{figure*}[htbp]
\centerline{\includegraphics[width=\textwidth]{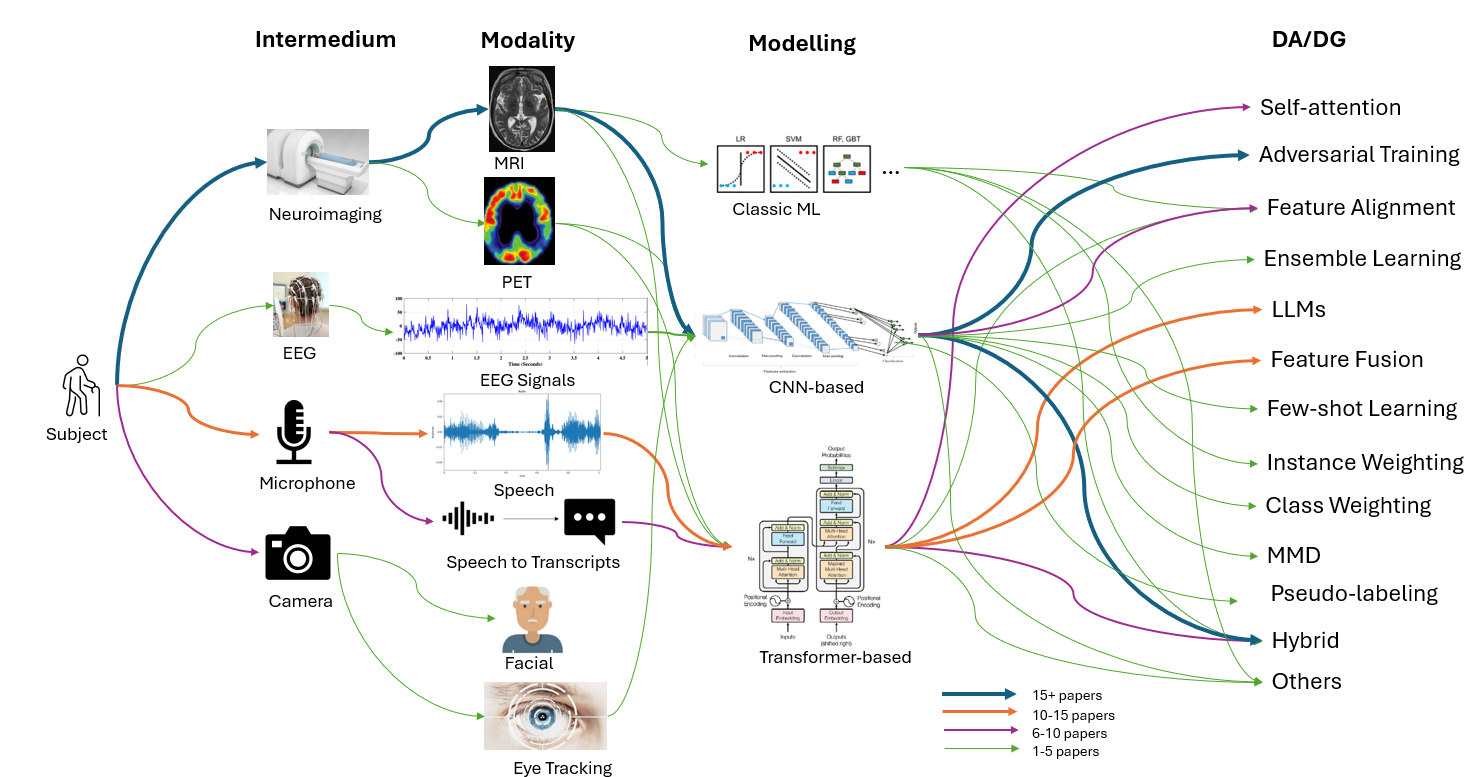}}
\captionsetup{justification=centering,margin=2cm}
\caption{An overview of the reviewed papers on different modalities and DA/DG approaches.}
\label{scheme_fig}
\end{figure*}

This section provides an overview of DA/DG techniques employed across different modalities for AD detection. As illustrated in Figure~\ref{scheme_fig}, the primary data collection intermediaries include neuroimaging (used to generate MRI and PET scans), EEG devices (for brain signals), microphones (for capturing speech), and cameras (for recording facial expressions and eye movement). These modalities serve as the foundation for various data types used in AD detection: structural images, brain activity signals, speech, facial features, and eye gaze patterns.

In terms of modelling, three major categories are utilized: classical ML algorithms such as Logistic Regression (LR) and Support Vector Machines (SVM), CNNs, and transformer-based architectures, such as vision transformer networks and LLMs. Among these, CNNs are the most widely adopted across reviewed studies due to their strong performance in processing structured spatial data such as images and signals.

To classify studies consistently, we apply the following criteria throughout this review. A study is labelled \emph{domain adaptation (DA)} if it accesses some information from the target domain during model development, and \emph{domain generalization (DG)} if it trains only on one or more source domains and is evaluated on an unseen target domain without using target data. DA studies are further distinguished as \emph{supervised} (labelled target data used), \emph{semi-supervised} (a small amount of labelled plus unlabelled target data), or \emph{unsupervised} (only unlabelled target data); DG studies are distinguished as \emph{single-source} or \emph{multi-source}. We reserve the DA/DG label for methods that explicitly target domain shift; techniques that improve robustness more generally (e.g., data augmentation, class or instance weighting, attention modules, generic transfer learning, few-shot learning, multimodal fusion, and LLM-based inference) are noted where they co-occur with DA/DG but are not treated as DA/DG in themselves unless used to align or generalize across domains; we mark such strategies as \emph{dual-use} in Table~\ref{critical_synthesis_table}. To make the experimental domain structure explicit, each modality section is accompanied by a companion table (Tables~\ref{mripet_companion_table}, \ref{eeg_companion_table}, \ref{et_companion_table}, \ref{speech_companion_table}, and \ref{multimodal_companion_table}) reporting, for every study, the source and target domains, whether target labels were available during training, the DA/DG setting, the clinical endpoint, and the reported train/test split and leakage-control practices.  The "Split \& leakage control" column explicitly marks whether a study was evaluated internally (within a single cohort) or cross-cohort (on a separate dataset/site), so that a model adapted to a target site using that site's data is now clearly separated from a DG model tested at a new institution. Cells marked ``NR'' indicate that the corresponding information was not reported in the study, which we treat as a reporting limitation rather than evidence of a particular practice. Beyond internal versus cross-cohort evaluation, this column also records, where reported, whether splitting was performed at the subject level rather than at the slice, trial, or segment level, whether preprocessing was confined to the training data, and whether model selection or hyperparameter tuning avoided the test set, since tuning on test data is itself a form of leakage. Several further quality dimensions raised during review, in particular sample size, the selection of control groups, and statistical uncertainty (for example confidence intervals, variability across folds or random seeds, and significance testing), were considered but are not tabulated as separate per-study columns because they are inconsistently or rarely reported. We address sample size and control-group selection qualitatively in the data and open-problem sections; and we treat the frequent absence of statistical uncertainty as a reporting limitation in the quantitative synthesis, reinforced in the calibration and uncertainty subsection, while reporting confidence intervals and other available uncertainty measures in the evaluation columns wherever the primary studies provide them.

MRI remains the most widely used modality for AD detection due to its rich structural and functional information. Speech-based approaches have emerged as the second most popular modality, driven by advancements in LLMs that enable more effective analysis of linguistic and acoustic biomarkers. Other modalities, such as PET, EEG, facial analysis, and ET, have received comparatively less research attention but demonstrate strong potential for future exploration, particularly facial analysis, which offers a non-invasive and accessible means of assessment. However, progress in AD detection with facial modality has been constrained by the limited availability of open-source datasets, hindering large-scale model development and validation.

We further categorize the DA/DG and related robustness strategies for AD detection used in the literature into techniques such as self-attention, adversarial training, feature alignment, ensemble learning, few-shot learning, and others. The connections in Figure~\ref{scheme_fig} are color-coded to indicate the number of publications employing each method; adversarial training is the most frequently published approach, which reflects research interest rather than demonstrating, by itself, superior effectiveness in domain-invariant feature learning. Additionally, LLMs for speech and transcript data and feature fusion across modalities are promising strategies to mitigate domain shift and improve generalizability. 

Across modalities, DA/DG methods for AD detection should not be treated as interchangeable solutions because they address different types of domain shift and require different assumptions about target-domain data. Adversarial learning is useful when the goal is to suppress site-, scanner-, device-, language-, or cohort-specific features, but it can remove clinically meaningful variation if the domain label is correlated with disease stage, age, sex, or acquisition protocol. Discrepancy-based methods such as maximum mean discrepancy (MMD), feature alignment, or distribution matching are usually more stable and easier to optimize, but they may align marginal feature distributions without preserving class boundaries. Self-training and pseudo-labeling can exploit unlabeled target data, but their performance depends strongly on the calibration of the source model and may degrade through error accumulation. Few-shot and meta-learning methods are attractive when only a small number of labeled target samples are available, but their reliability depends on whether the support samples represent the full heterogeneity of the target population. Weighting-based strategies, including class weighting and instance weighting, can reduce class imbalance or source-target distribution bias, but they cannot fully correct feature-level or label-shift problems when the weighting model is inaccurate. 

Self-attention and transformer-based methods can model long-range dependencies in imaging, speech, EEG, or multimodal sequences, but they often require large datasets and may overfit to site-specific patterns without explicit domain regularization. Ensemble learning can improve robustness by combining complementary models or domains, but it may increase computational cost and reduce interpretability. LLM-based approaches may support text-based cognitive assessment, report summarization, and knowledge-guided reasoning, but they introduce additional challenges related to hallucination, privacy, prompt sensitivity, and clinical validation. Hybrid methods combine two or more of these strategies and are often more flexible, but their benefit depends on whether each component addresses a distinct source of shift rather than simply increasing model complexity. Multimodal feature fusion can improve robustness by integrating structural, functional, linguistic, and behavioral evidence, but it also introduces missing-modality, synchronization, and modality-specific bias problems. These trade-offs suggest that the best DA/DG strategy should be selected according to the expected shift type, available target data, modality characteristics, and deployment setting rather than only according to overall accuracy. The critical synthesis of these common DA/DG-related strategies are summarized in Table~\ref{critical_synthesis_table}.

\begin{table}[htbp]
\centering
\small
\caption{Critical synthesis of common DA/DG-related strategies for AD detection across modalities. Following the classification criteria in Section~\ref{da_dg}, only adversarial alignment, discrepancy-based alignment, self-training/pseudo-labeling, and hybrid combinations of these act on domain shift directly and are treated as DA/DG techniques in themselves. The remaining strategies (few-shot/meta-learning, class/instance weighting, attention/transformer architectures, ensemble learning, LLM-based methods, and multimodal feature fusion) are marked \emph{dual-use}: they are general modeling or robustness techniques that address domain shift only when explicitly combined with a domain-alignment or generalization objective (e.g., importance weighting used to correct source--target shift, or few-shot meta-learning used for cross-cohort generalization), and are included here because they frequently co-occur with DA/DG in the reviewed studies.}
\label{critical_synthesis_table}
\begin{tabular}{|p{2.5cm}|p{3.0cm}|p{3.3cm}|p{3.5cm}|p{3.2cm}|}
\hline
\textbf{Strategy} & \textbf{Typical shift addressed} & \textbf{Main strengths} & \textbf{Main limitations} & \textbf{Most suitable settings} \\ \hline

Adversarial alignment & Site, scanner, device, language, or cohort differences & Learns domain-invariant features and is flexible across MRI, PET, EEG, speech, and multimodal representations & Can be unstable to train and may suppress disease-relevant variation when domain and diagnosis are correlated & Multi-site imaging, speech, EEG, and multimodal tasks with clear domain labels \\ \hline

Discrepancy-based alignment & Marginal or feature-distribution mismatch & Transparent objective, comparatively stable optimization, and useful with limited target labels & May align global distributions while leaving class-conditional mismatch unresolved & Cross-dataset MRI, PET, EEG, eye-tracking, and multimodal feature spaces \\ \hline

Self-training and pseudo-labeling & Limited labeled target data with available unlabeled target samples & Uses unlabeled target samples and can improve adaptation to local data & Sensitive to noisy pseudo-labels, poor calibration, and class imbalance & Deployment sites that can provide unlabeled local data \\ \hline

Few-shot and meta-learning (dual-use) & Scarce labeled target data or new cohorts & Requires only a small number of labeled examples and supports rapid adaptation & Performance depends on support-set quality and may be unstable in heterogeneous populations & New clinics, devices, languages, or underrepresented cohorts with limited annotation resources \\ \hline

Class weighting and instance weighting (dual-use) & Class imbalance, sample-selection bias, or source-target sampling mismatch & Simple to implement and can reduce the influence of overrepresented classes or less transferable source samples & Cannot fully correct complex feature shift or label shift if the weights are poorly estimated & Imbalanced AD/MCI/control datasets or multi-source settings with unequal source relevance \\ \hline

Attention / transformer architectures (dual-use) & Long-range temporal, spatial, or cross-modal dependencies; address domain shift only when paired with a domain-alignment objective & Can capture global relationships in imaging, EEG, speech, language, and multimodal sequences & Not domain-adaptive on their own; often data-hungry and may overfit site-specific or cohort-specific patterns without domain regularization & Large-scale imaging, speech, EEG, text, or multimodal datasets, combined with a DA/DG objective \\ \hline

Ensemble learning (dual-use) & Model variance, source heterogeneity, or modality-specific uncertainty & Improves robustness by combining complementary models, domains, or modalities & Increases computational cost and may reduce interpretability & Multi-source, multi-model, or multimodal AD detection tasks \\ \hline

LLM-based methods (dual-use) & Language, clinical-text, report, or knowledge-reasoning variability & Useful for text-based cognitive assessment, clinical report analysis, explanation, and knowledge-guided support & Requires careful validation because of hallucination, privacy, prompt sensitivity, and possible demographic or linguistic bias & Speech-language assessment, clinical note analysis, caregiver support, and knowledge-assisted screening \\ \hline

Hybrid DA/DG methods & Multiple simultaneous shifts across site, modality, label distribution, and cohort & Can combine complementary strengths, such as adversarial alignment with pseudo-labeling or attention with weighting & More complex to train and harder to interpret; gains may come from added capacity rather than true domain robustness & Realistic clinical deployment where several shift sources occur together \\ \hline

Multimodal feature fusion (dual-use) & Heterogeneous but complementary biomarkers & Combines structural, functional, linguistic, cognitive, and behavioral evidence & Vulnerable to missing modalities, synchronization issues, and modality-specific bias & Clinical decision support where multiple data streams are available \\ \hline

\end{tabular}
\end{table}

Quantitative comparison across studies remains challenging because evaluation protocols differ in cohort definitions, diagnostic labels, preprocessing, train-test splitting, and whether target-domain labels are used. Nevertheless, reported cross-domain gains suggest that DA/DG can improve robustness when compared with non-adapted baselines. The figures cited throughout this review should be read as illustrative examples from individual studies against their own baselines, not as representative or expected gains, because the baselines, datasets, diagnostic tasks, and evaluation settings differ across studies. Unless otherwise stated, reported improvements are expressed in absolute percentage points relative to a study-specific baseline. For example, one MRI harmonization study reported an absolute target-domain AD-classification accuracy increase of about 8 percentage points using attention-guided GAN-based harmonization across ADNI, AIBL, and OASIS \cite{sinha2021alzheimers}, and a speech study reported an increase in AD-classification accuracy from about 82.6\% using manual transcripts to about 88\% using ASR transcripts refined with domain-adaptive beam-search decoding on the same corpus \cite{Li2024UsefulBlunders}. In contrast, an ET study reported a drop from 91.1\% in-distribution to 83.3\% out-of-distribution accuracy \cite{math13111867}; we cite this not as a DA/DG gain but as an illustration of the domain-shift problem that motivates DA/DG. A further caveat is that most studies report only accuracy or AUC; sensitivity, specificity, balanced accuracy, macro-F1, AUPRC, calibration, and confidence intervals are frequently absent, which we regard as a reporting limitation given the class imbalance typical of AD cohorts. For these reasons we avoid ranking methods or modalities on the basis of these heterogeneous point estimates, and we treat publication frequency of a technique as an indicator of research interest rather than of demonstrated effectiveness.

\begin{figure*}[htbp]
\centerline{\includegraphics[width=0.6\textwidth]{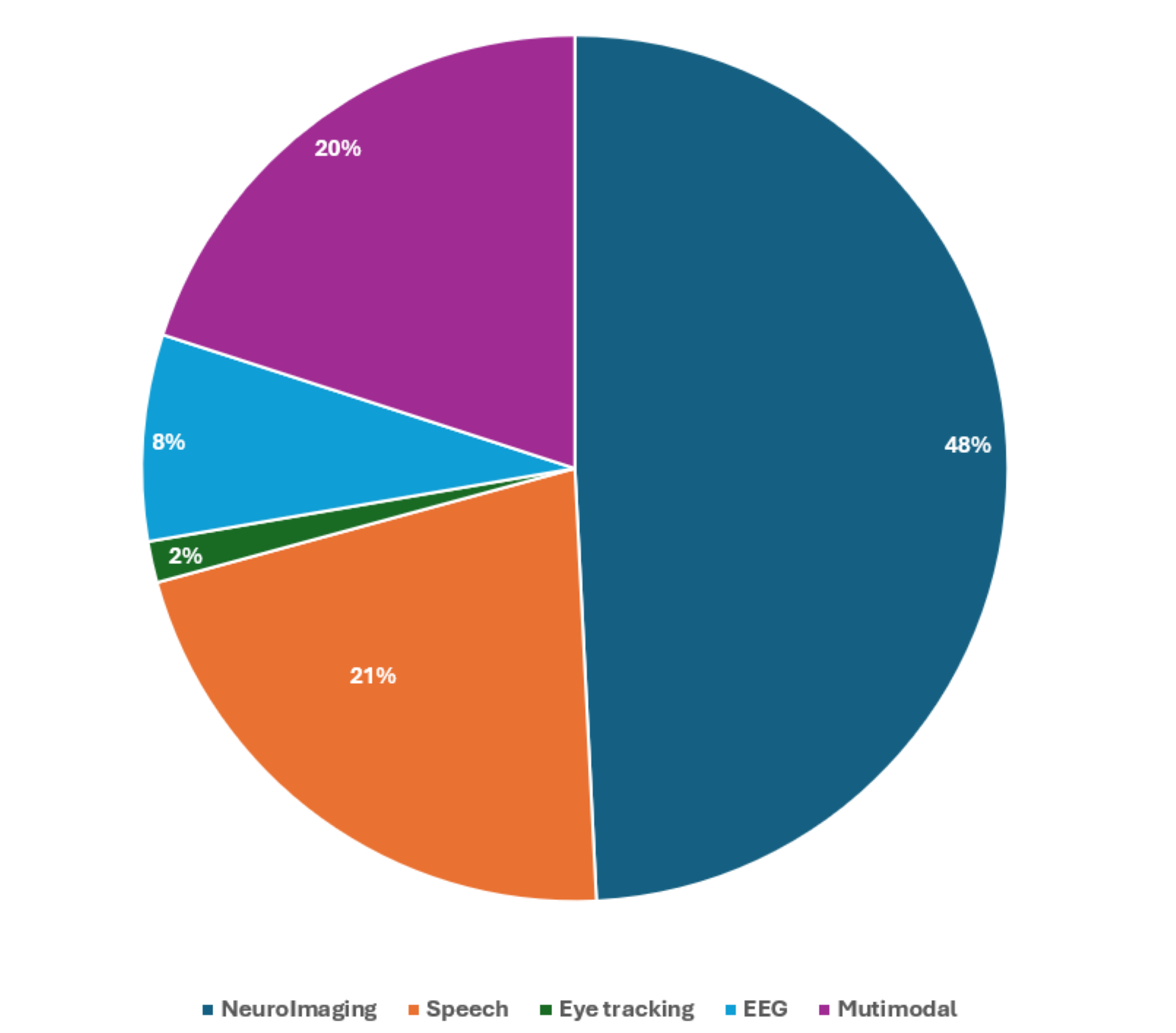}}
\captionsetup{justification=centering,margin=2cm}
\caption{The reviewed research works with different modalities using DA/DG on AD detection. This chart shows the distribution of AD-specific DA/DG studies by primary modality. Unlike the scoping-flow diagram (Figure 2) and the synthesis tables, which additionally retain methodologically transferable non-AD works (gaze estimation for ET, general EEG classification), these are excluded here; the ET and EEG shares therefore appear smaller than their table counts.}
\label{pie_fig}
\end{figure*}

To illustrate the distribution of research efforts applying DA/DG techniques across different data modalities for AD detection, we summarized the reviewed articles by modality, as shown in Figure~\ref{pie_fig}. Neuroimaging data, particularly MRI and PET, dominate the field, accounting for 48\% of the studies, reflecting its long-standing clinical use and availability of standardized datasets. Speech-based approaches represent 21\% of the research, indicating growing interest in non-invasive, cost-effective, and scalable screening tools. ET (2\%) and EEG-based (8\%) modalities are also being explored, though to a lesser extent. Multimodal approaches, combining two or more data types, make up 20\% of the studies and show strong potential for improving generalizability and robustness in real-world deployment. 

\subsection{DA/DG approaches for AD detection with neuroimaging data}

Neuroimaging data, particularly MRI and PET, play a central role in clinical AD detection. However, models trained on a single domain, such as data from a specific scanner, acquisition protocol, or clinical population, often fail to generalize across different clinical settings due to domain shifts \cite{fu2025prediction}. Addressing this challenge has driven the development of DA/DG techniques for neuroimaging-based AD detection.

Tables~\ref{mripet_table} and~\ref{mripetmethod_table} summarize the recent literature in this area. Table~\ref{mripet_table} outlines the datasets, imaging modalities (MRI or PET), and backbone architectures used, while Table~\ref{mripetmethod_table} details methodological strategies employed to enhance cross-domain performance.

From the literature in Table~\ref{mripet_table}, MRI emerges as the dominant modality for DA/DG studies, with the ADNI dataset serving as the most widely used benchmark. DL methods, especially 2D and 3D CNNs, have largely replaced traditional ML approaches. The main DA/DG strategies summarized in Table~\ref{mripetmethod_table} include transfer learning, data augmentation, self-attention mechanisms, ensemble learning, adversarial training, discrepancy minimization, pseudo-labeling, and few-shot learning.

Generalization-oriented strategies such as transfer learning, augmentation, ensemble learning, and self-attention are frequently adopted to handle variations in imaging protocols and limited sample sizes. Transfer learning leverages pre-trained models to reduce inter-domain discrepancies, and augmentation generates diverse samples to expand the existing data distribution. Ensemble learning aggregates multiple model predictions to improve robustness, and self-attention modules capture both local and global features to enhance model generalization and interpretability. For example, Binzagr et al. \cite{10.3389/frai.2025.1540646} combined a pre-trained InceptionV3 \cite{Szegedy2016Rethinking} backbone with multi-head self-attention and augmentation to achieve robust performance across varied MRI domains, while Tang et al. \cite{10.1145/3584371.3613000} improved interpretability and generalization by enhancing a ResNet-10 \cite{He2016Deep} with attention modules. Similarly, Batool et al. \cite{batool2025singledomaingeneralizationalzheimers} employed a pre-trained 3D U-Net \cite{Cicek2016U} with contrastive learning and augmentation, achieving strong cross-dataset performance when training on NACC and testing on unseen ADNI and AIBL datasets.

Adversarial learning is one of the most frequently published DA techniques in neuroimaging; we note, however, that its frequency of use reflects research interest rather than establishing, by itself, superior effectiveness relative to other strategies. By introducing a domain discriminator, adversarial methods encourage the extraction of domain-invariant features. Sinha et al. \cite{sinha2021alzheimers}, Reddy et al. \cite{reddy2022applicationunsuperviseddomainadaptation}, and Guan et al. \cite{GUAN2021102076} all demonstrated promising results using this approach. Generative adversarial networks (GANs) extends this idea by generating neuroimaging data to expand training sets, balance demographics, and model disease progression. Sinha et al. \cite{sinha2021alzheimers} used attention-guided GAN-based MRI harmonization across cohorts to improve target-domain classification accuracy by over 8\%. Other techniques include pseudo-labeling, where high-confidence predictions are used to label unlabeled target data \cite{10135388, CAI2023106570}, and discrepancy-based methods such as MMD, which minimize feature distribution differences between source and target domains \cite{CAI2023106570}, has also proven effective to improve the model adaptation across diverse domains.

Few-shot learning offers a practical solution when the target domain has only a small amount of labeled data. By adapting to new domains with minimal supervision, few-shot approaches reduce the need for extensive retraining. Atitallah et al. \cite{10704046} and Orouskhani et al. \cite{OROUSKHANI2022100066} used pre-trained 2D CNNs within few-shot frameworks and reported competitive accuracy, highlighting their applicability in low-resource clinical environments.

Lightweight domain alignment methods have also been explored as alternatives to DL approaches. Li et al. \cite{8362917} applied Principal Component Analysis (PCA) to align MRI-derived features, while Park et al. \cite{PARK2024122253} minimized inter-domain divergence by matching feature distributions directly. These techniques are computationally efficient and offer greater interpretability, making them suitable for scenarios with limited resources.

Multi-source training is another effective strategy to enhance model adaptations across different domains. Exposing models to a variety of domains during training encourages the learning of more robust, domain-invariant representations. Kushol et al. \cite{10.1007/978-3-031-45857-6_1} and Guan et al. \cite{GUAN2021102076} demonstrated that combining unsupervised DA with multi-site MRI data improved performance on unseen target domains.

Finally, class imbalance remains a significant challenge in DA/DG for AD detection, as CN participants often outnumber AD and MCI cases in neuroimaging datasets. This imbalance can bias models toward the majority class, lowering diagnostic accuracy for minority classes and hindering effective cross-domain adaptation. To address this, Wachinger et al. \cite{wachinger2016domain} and Guan et al. \cite{9433971} applied instance weighting to balance source and target contributions, while Batool et al. \cite{batool2025singledomaingeneralizationalzheimers} used class weighting to counteract skewed distributions. These strategies improved diagnostic accuracy across all adaptation scenarios, with balanced performance between classes in the target domain.

The methods summarized above show that neuroimaging DA/DG research has moved from simple transfer learning and feature alignment toward adversarial learning, pseudo-labeling, few-shot learning, multi-source training, self-attention, and class- or instance-weighting strategies. These approaches address different aspects of imaging domain shift. Feature alignment and MMD-based methods are useful when source and target distributions differ in scanner protocol or cohort composition, because they directly reduce distributional discrepancy. Adversarial learning can learn more flexible domain-invariant representations, but it may be harder to optimize and may remove disease-relevant variation if site and diagnosis are correlated. Few-shot learning is more suitable when a small number of labeled target samples are available, whereas multi-source training is more useful when several source cohorts are accessible. These methodological differences are reflected in the reported quantitative gains. Several MRI-based studies reported target-domain accuracy improvements of approximately 5--22\% compared with non-adapted baselines, with improvements observed for transfer learning, feature alignment, instance weighting, and adversarial adaptation. Multi-source training and cross-cohort evaluation across ADNI, AIBL, OASIS, NACC, MIRIAD, PND, and J-ADNI further suggest that exposing models to more diverse imaging domains can improve robustness. However, direct ranking of neuroimaging DA/DG methods remains difficult because studies differ in 2D versus 3D input representation, diagnostic labels, preprocessing, target-label availability, and whether evaluation is performed with random splits, cross-cohort splits, or external validation. Therefore, the neuroimaging literature supports the value of DA/DG, but also highlights the need for standardized cross-dataset benchmarks and consistent reporting of target-domain performance.

Reading the companion Table~\ref{mripet_companion_table} as a whole, rather than study by study, exposes three patterns that are directly relevant to clinical generalizability. First, in terms of the experimental domain structure, the reviewed studies concentrate on \emph{technical} and \emph{site/geographic} generalization, adapting or generalizing across scanners, acquisition protocols, and cohorts, most often from ADNI to AIBL, OASIS, NACC, MIRIAD, PND, or J-ADNI, whereas \emph{population}, \emph{temporal}, and \emph{workflow} generalization are rarely evaluated. Second, in terms of clinical endpoint, binary AD-versus-CN discrimination dominates, and comparatively few neuroimaging studies target the clinically harder MCI-conversion or multi-class staging tasks, pointing to limited clinical-task generalization. Third, and most important for weighing the evidence, the ``Split \& leakage control'' column is frequently marked ``NR'': many studies do not explicitly report subject-level (rather than slice-level) splitting, external or site-level testing, or training-only preprocessing. Genuine cross-cohort evaluation is therefore documented in only a subset of studies, and internal-split results, some reporting near-ceiling accuracy, should be weighed more cautiously than externally validated ones. The companion table indicates that neuroimaging DA/DG has progressed furthest on technical robustness while still under-reporting methodological safeguards and addressing comparatively narrow diagnostic endpoints, which tempers how far current results can be assumed to transfer to real clinical deployment. A further caveat, made explicit in the companion table, is that several entries grouped under DA/DG in the summary tables are, on closer inspection, generic transfer-learning, ensemble, or architectural-robustness methods evaluated within a single cohort rather than explicit cross-domain adaptation or generalization; we flag these so that they are not over-read as evidence of domain transfer.

\begin{table}[]
\centering
\scriptsize
\caption{Summary of literature on DA/DG approaches for AD detection with neuroimaging data.}
\label{mripet_table}
\begin{tabular}{|C{2.2cm}|C{3cm}|C{0.9cm}|C{0.3cm}|C{0.3cm}|C{2.6cm}|C{4.8cm}|}
\hline
\textbf{Paper}                                                        & \textbf{Dataset}                        & \textbf{Modality}  & \textbf{2D} & \textbf{3D} & \textbf{Backbone} & \textbf{Evaluation}                                                \\ \hline
Tanveer et al. \cite{9440810}                                         & ADNI\cite{mueller2005adni}            & MRI                       & \ding{51}          &             & VGG16          & Accuracy of 99.09\% for CN vs AD and 98.71\% for MCI vs AD.                                                   \\ \hline
Guan et al. \cite{9433971}                                            & ADNI-1\cite{mueller2005adni}, ADNI-2\cite{mueller2005adni}, ADNI-3\cite{mueller2005adni}                  & MRI                          &             &             & SVM                    & All adaptation scenarios exceed the literature methods.                                             \\ \hline
Sinha et al. \cite{sinha2021alzheimers}                               & ADNI\cite{mueller2005adni}, AIBL\cite{ellis2009aibl}, OASIS\cite{marcus2007open}                       & MRI                           & \ding{51}            &             & AlexNet   & The target domain classification accuracy was increased by 8\%.                                                        \\ \hline
Cheng et al. \cite{cheng2017multi}                                    & ADNI\cite{mueller2005adni}                                    & MRI                           &             & \ding{51}            & Deep neural networks with stacked   autoencoders         & Accuracy of target domains was increased by 5-10\%.           \\ \hline
Wang et al. \cite{8856726}                                            & ADNI, private Tongji dataset            & MRI                          & \ding{51}            &             & Kendall correlation   coefficient  + Linear classifier  & The target domain accuracy was greatly improved by 22\%.          \\ \hline
Li et al. \cite{8362917}                                              & ADNI\cite{mueller2005adni}, private Tongji dataset            & MRI                           & \ding{51}            &             & PCA + Linear classifer     & The target domain accuracy was improved by 20\%.                                       \\ \hline
Wachinger et al.   \cite{wachinger2016domain}                         & ADNI\cite{mueller2005adni}, AIBL\cite{ellis2009aibl},CADDementia challenge\cite{BRON2015562}        & MRI                           & \ding{51}           &             & Elastic Net multinomial   regression    & The target domain accuracy was improved by up to 10\%.                            \\ \hline
Cai et al. \cite{CAI2023106570}                                       & ADNI\cite{mueller2005adni}                                    & MRI                           &             & \ding{51}            & 3D CNN       & Outperformed the literature methods on the target domain. Accuracy 92.11\% (AD/CN), 82.37\% (MCI/CN), and 76.01\% (AD/MCI).                                                         \\ \hline
Turrisi et al. \cite{TURRISI2025121016}                               & ADNI\cite{mueller2005adni}                                    & MRI                           &             & \ding{51}            & ADnet \cite{turrisi2024deep},   ResNet18, ResNet50, and ResNet101 & Improved the target domain accuracy to 99\%.\\ \hline
Reddy    \cite{reddy2022applicationunsuperviseddomainadaptation}      & OASIS\cite{marcus2007open}                                   & MRI                           & \ding{51}           & \ding{51}            & EfficientNet,   ResNeXt-50,3D-ResNet-18     & Consistent improvements in all adaptation scenarios.                      \\ \hline
Yu et   al. \cite{10316553}                                           & OASIS\cite{marcus2007open}, ADNI\cite{mueller2005adni}                             & MRI                           & \ding{51}           &             & Siamese networks with VGG16   backbone     & Improved the target domain accuracy by 10.9\%.                       \\ \hline
Guan   et al. \cite{GUAN2021102076}                                   & ADNI-1\cite{mueller2005adni},ADNI-2\cite{mueller2005adni},ADNI-3\cite{mueller2005adni}, AIBL\cite{ellis2009aibl}               & MRI                           &             & \ding{51}            & 3D CNN                      & Outperformed the literature in all adaptation scenarios.                                       \\ \hline
Batool   et al. \cite{batool2025singledomaingeneralizationalzheimers} & ADNI\cite{mueller2005adni}, AIBL\cite{ellis2009aibl} , NACC\cite{Beekly2004NACC}                        & MRI                           &             & \ding{51}            & 3D U-Net        & Exceeds the literature methods on all target domains.                                                   \\ \hline
Zhang   et al. \cite{10135388}                                        & ADNI\cite{mueller2005adni}             & MRI                           &             & \ding{51}            & 3DCNN               & Outperforms the literature methods on the target domain.                                                \\ \hline
Park   et al. \cite{PARK2024122253}                                   & ADNI\cite{mueller2005adni}, AIBL\cite{ellis2009aibl}, OASIS\cite{marcus2007open}                       & MRI                           & \ding{51}            &             & classifiers (NN, SVM, LDA, LR,   DT) & The performance was greatly improved on average with all adaptation scenarios.                               \\ \hline
Kushol   et al. \cite{10.1007/978-3-031-45857-6_1}                &  ADNI\cite{mueller2005adni}, AIBL\cite{ellis2009aibl}, MIRIAD\cite{malone2013miriad} & MRI                           & \ding{51}           &             & ResNet-50           & Improved the target domain accuracy from 80\% to 89\%.                                                \\ \hline
Tobari   et al. \cite{tobari2025domain}                               & ADNI\cite{mueller2005adni}, OASIS\cite{marcus2007open}                             & MRI                           &             & \ding{51}            & 3DCNN      & Outperforms the literature methods in the target domains.                                                        \\ \hline
Rehman   et al. \cite{10570178}                                       & ADNI\cite{mueller2005adni}                                    & PET                          & \ding{51}            &             & MobileViTv3      & Achieved an accuracy of 92.75\% in target domain.                                                   \\ \hline
Duan   et al. \cite{duan2023broad}                                    & ADNI\cite{mueller2005adni}                                    & PET                           & \ding{51}           &             & 2D CNN            & Outperforms the literature even with class-imbalanced and small datasets. Per-class sensitivity/specificity/AUC: AD 92.16\%/97.56\%/0.97, MCI 89.34\%/95.58\%/0.98, CN 95.16\%/95.09\%/0.99; EMCI-vs-LMCI sensitivity 81.63\%, specificity 85.19\%, F1 82.47\%.                                               \\ \hline
Wang   et al. \cite{wang2022high}                                     & ADNI\cite{mueller2005adni}, J-ADNI\cite{JADNI}                   & MRI                           &             & \ding{51}            & 3DCNN              &  Achieved accuracy 0.84 on J-ADNI with trained on ADNI. Cross-cohort J-ADNI AUC 0.91 (internal NA-ADNI accuracy 0.88, AUC 0.95).                                                  \\ \hline
Lteif   et al. \cite{Lteif2023.09.22.23295984}                        & ADNI\cite{mueller2005adni}, AIBL\cite{ellis2009aibl}, NACC\cite{Beekly2004NACC}, FHS\cite{framingham2020dataset}                    & MRI                          &             & \ding{51}            & UNet3D                 & Outperforms the literature on all adaptation scenarios.                                             \\ \hline
Bron   et al. \cite{bron2021cross}                                    & ADNI\cite{mueller2005adni}, PND\cite{aalten2014parelsnoer}                               & MRI                           &             & \ding{51}            & 3DCNN          & Consistently performs better on target domains. External-validation AD--CN AUC 0.876 (95\% CI 0.836--0.913) for the CNN and 0.896 (0.855--0.932) for the SVM.                                                   \\ \hline
Tang   et al. \cite{10.1145/3584371.3613000}                          & ADNI\cite{mueller2005adni}                                    & MRI                           & \ding{51}            &             & ResNet-10        & 2\% drop in F1 compared to a ~29\% drop for baseline models.                                                   \\ \hline
Choi   et al. \cite{8950078}                                          & ADNI\cite{mueller2005adni}                                    & MRI                           & \ding{51}            &             & VGG-16, GoogLeNet,AlexNet     & Outperforms the literature in all test cases.                                        \\ \hline
Majee   et al. \cite{10650462}                                        & ADNI\cite{mueller2005adni}                                     & MRI                         &             & \ding{51}            & 3DCNN, ViT    & Outperformed baseline models, achieving accuracy of ~96.06\%.                                                    \\ \hline
Rawashdeh   et al. \cite{alrawashdeh2024ensemble}                     & OASIS                                   & MRI                           & \ding{51}            &             & YOLOv7,EfficientNet B3   & Reached an average accuracy of 99.8\% on all test cases.                                         \\ \hline
Binzagr   et al. \cite{10.3389/frai.2025.1540646}                      & ADNI\cite{mueller2005adni}                                    & MRI                          & \ding{51}            &             & InceptionV3    & Reached an average accuracy of 96.67\% for all test cases.                            \\ \hline
Xin et   al. \cite{XinJiaming2023Casb}                                & ADNI\cite{mueller2005adni}, AIBL\cite{ellis2009aibl}                              & MRI                          & \ding{51}            &             & 2DCNN, ViT     & Achieved 92.8\% balance accuracy and 91.1\% sensitivity on the target domain.                                                    \\ \hline
Wang   et al. \cite{WangRongguang2022Etdi}                            & ADNI\cite{mueller2005adni}, AIBL\cite{ellis2009aibl}                              & MRI                          &             & \ding{51}            & 3DCNN     & The target performance was greatly improved with age and sex target data.                                                         \\ \hline
Atitallah   et al. \cite{10704046}                                    & ADNI\cite{mueller2005adni}, Kaggle dataset\cite{alzheimers_dataset_kaggle}                    & MRI                           & \ding{51}            &             & ResNet18, ResNet34, MobileNetV2,   VGG16    & Achieved over 99\% accuracy on all test cases.           \\ \hline
Orouskhani   et al.\cite{OROUSKHANI2022100066}                        & OASIS\cite{marcus2007open}                                   & MRI                           & \ding{51}           &             &  VGG16     & Outperforms the literature in accuracy.                                  \\ \hline
\end{tabular}
\end{table}

\begin{table}[]
\centering
\small
\caption{Summary of DA/DG approaches for AD detection using neuroimaging data. Following the classification criteria in Section~4, the final two columns mark only studies with an explicit domain-adaptation or domain-generalization objective; studies that apply generic transfer learning, ensembles, few-shot metric learning, or architectural-robustness techniques without such an objective are listed for completeness but left unmarked in the DA and DG columns, consistent with the companion Table~\ref{mripet_companion_table}.}
\label{mripetmethod_table}
\begin{tabular}
{|C{3cm}|C{0.2cm}|C{0.2cm}|C{0.2cm}|C{0.2cm}|C{0.2cm}|C{0.2cm}|C{0.2cm}|C{0.2cm}|C{0.2cm}|C{0.2cm}|C{0.2cm}|C{0.2cm}|C{0.2cm}|C{0.2cm}|C{0.2cm}|C{0.2cm}|C{0.2cm}|C{0.2cm}|C{0.2cm}|}
\hline
\textbf{Paper} & 
\rotatebox{90}{Multi-source} & 
\rotatebox{90}{Pre-trained} & 
\rotatebox{90}{Customized} & 
\rotatebox{90}{Ensemble learning} & 
\rotatebox{90}{Adversarial learning} & 
\rotatebox{90}{Generative} & 
\rotatebox{90}{Feature space alignment} & 
\rotatebox{90}{MMD} & 
\rotatebox{90}{Augmentation} & 
\rotatebox{90}{Self-attention} & 
\rotatebox{90}{Pseudo-label training} & 
\rotatebox{90}{Instance Weighting} & 
\rotatebox{90}{Class weighting} & 
\rotatebox{90}{Few-shot learning} & 
\rotatebox{90}{Supervised DA} & 
\rotatebox{90}{Unsupervised DA} & 
\rotatebox{90}{Code available} & 
\rotatebox{90}{Domain Adaptation (DA)} &
\rotatebox{90}{Domain Generalization (DG)}\\
\hline
Tanveer et al. \cite{9440810}                                         &                       & \ding{51}                    &                     & \ding{51}                          &                               &                     &                                     &              &                       &                         &                                &                             &                          &                            &                        &                         & \ding{51}                       & &                                                                                                             \\ \hline
Guan et al. \cite{9433971}                                            &                       &                      &                     &                            &                               &                     &                                     &              &                       &                         &                                & \ding{51}                           &                          &                            &                        & \ding{51}                        &                         & \ding{51}&                                                                                                                                \\ \hline
Sinha et al. \cite{sinha2021alzheimers}                               &                       & \ding{51}                    &                     &                            & \ding{51}                             & \ding{51}                   &                                     &              &                       &                         &                                &                             &                          &                            &                        & \ding{51}                        &                         &\ding{51} &                                                                                                                                                                                          \\ \hline
Cheng et al. \cite{cheng2017multi}                                    & \ding{51}                     &                      & \ding{51}                   &                            &                               &                     & \ding{51}                                   &              &                       &                         &                                &                             &                          &                            & \ding{51} &                         &                         &\ding{51} &                                                                                                                                        \\ \hline
Wang et al. \cite{8856726}                                            &                       &                      &                     &                            &                               &                     & \ding{51}                                   &              &                       &                         &                                & \ding{51}                           &                          &                            &                        & \ding{51}                        &                         &\ding{51} &           \\ \hline
Li et al. \cite{8362917}                                              &                       &                      &                     &                            &                               &                     & \ding{51}                                   &              &                       &                         &                                & \ding{51}                           &                          &                            & \ding{51} &                         &                         &\ding{51} &                                                                                                                                                                                     \\ \hline
Wachinger et al.   \cite{wachinger2016domain}                         &                       &                      &                     &                            &                               &                     &                                     &              &                       &                         &                                & \ding{51}                           &                          &                            & \ding{51}                      &                          &                         &\ding{51} &                            \\ \hline
Cai et al. \cite{CAI2023106570}                                       &                       &                      & \ding{51}                   &                            & \ding{51}                             &                     &                                     & \ding{51}            &                       & \ding{51}                       & \ding{51}                              &                             &                          &                            &                        & \ding{51}                        &                         &\ding{51} &                                                                                                                                                                                                      \\ \hline 
Turrisi et al. \cite{TURRISI2025121016}                               &                       & \ding{51}                    &                     &                            &                               &                     &                                     &              & \ding{51}                     &                         &                                &                             &                          &                            &                        &                          & \ding{51}                       &\ding{51} &                                                                                                                                                                  \\ \hline
Reddy    \cite{reddy2022applicationunsuperviseddomainadaptation}      &                       & \ding{51}                    &                     &                            & \ding{51}                             &                     &                                     &              &                       &                         &                                &                             &                          &                            &                        & \ding{51}                        &                         & \ding{51}&                                                                                                                           \\ \hline
Yu et   al. \cite{10316553}                                           &                       & \ding{51}                    &                     &                            &                               &                     &                                     &              &                       &                         &                                &                             &                          &                            &                        & \ding{51}                        &                         &\ding{51} &                                                                                                                                           \\ \hline
Guan   et al. \cite{GUAN2021102076}                                   & \ding{51}                     &                      & \ding{51}                   &                            & \ding{51}                             &                     &                                     &              &                       & \ding{51}                       &                                &                             &                          &                            &                        & \ding{51}                        &                         &\ding{51} &                                                                                                                                     \\ \hline
Batool   et al. \cite{batool2025singledomaingeneralizationalzheimers} &                       & \ding{51}                    &                     &                            &                               &                     &                                     &              & \ding{51}                     &                         &                                &                             & \ding{51}                        &                            &                        &                          &                         & &   \ding{51}                                                                                                                                              \\ \hline
Zhang   et al. \cite{10135388}                                        &                       &                      & \ding{51}                   &                            &                               &                     &                                     &              &                       & \ding{51}                       & \ding{51}                              &                             &                          &                            &                        & \ding{51}                        &                         &\ding{51} & \\ \hline
Park   et al. \cite{PARK2024122253}                                   &                       &                      &                     &                            &                               &                     & \ding{51}                                   &              &                       &                         &                                &                             &                          &                            &                       & \ding{51}                        &                         &\ding{51} &                       \\ \hline
Kushol   et al. \cite{10.1007/978-3-031-45857-6_1}                & \ding{51}                     &                      &                     &                            &                               &                     &                                     & \ding{51}            &                       &                         &                                &                             &                          &                            &                        &                          & \ding{51}                       & \ding{51}&                                                                                                                                                               \\ \hline
Tobari   et al. \cite{tobari2025domain}                               &                       &                      & \ding{51}                   &                            & \ding{51}                             &                     &                                     &              &                       & \ding{51}                       &                                &                             &                          &                            &                        & \ding{51}                        &                         &\ding{51} &                                                                                                                                                                                                      \\ \hline
Rehman   et al. \cite{10570178}                                       &                       & \ding{51}                    & \ding{51}                   &                            &                               &                     &                                     &              &                       & \ding{51}                       &                                &                             &                          &                            &                        &                          &                         & &                                                                                              \\ \hline
Duan   et al. \cite{duan2023broad}                                    &                       &                      & \ding{51}                   &                            &                               &                     &                                     &              &                       &                         &                                &                             &                          &                            &                        &                          & \ding{51}                       & &                                                                               \\ \hline
Wang   et al. \cite{wang2022high}                                     &                       &                      & \ding{51}                   &                            &                               &                     &                                     &              & \ding{51}                     & \ding{51}                       &                                &                             &                          &                            &                        &                          &                         &  &  \ding{51}                                                                        \\ \hline
Lteif   et al. \cite{Lteif2023.09.22.23295984}                        &                       &                      & \ding{51}                   &                            &                               &                     &                                     &              &                       & \ding{51}                       &                                & \ding{51}                           &                          &                            &                        &                         & \ding{51}                       & & \ding{51}                                                                                                                            \\ \hline
Bron   et al. \cite{bron2021cross}                                    &                       &                      &                     &                            &                               &                     &                                     &              &                       &                         &                                &                             &                          &                            &                        &                         & \ding{51}                       & & \ding{51}                 \\ \hline
Tang   et al. \cite{10.1145/3584371.3613000}                          &                       &                      &                     &                            &                               &                     &                                     &              &                       & \ding{51}                       &                                &                             &                          &                            &                        &                         &                         & &\ding{51}                                                                                                \\ \hline
Choi   et al. \cite{8950078}                                          &                       & \ding{51}                    &                     & \ding{51}                          &                               &                     &                                     &              &                       &                         &                                &                             &                          &                            &                        &                         &                         & &                                                                                                  \\ \hline
Majee   et al. \cite{10650462}                                        &                       &                      & \ding{51}                   &                            &                               &                     &                                     &              &                       & \ding{51}                       &                                &                             &                          &                            &                        &                         &                         & &                                                              \\ \hline
Rawashdeh   et al. \cite{alrawashdeh2024ensemble}                     &                       &                      &                     & \ding{51}                          &                               &                     &                                     &              &                       &                         &                                &                             &                          &                            &                        &                          &                         & &                                                                                                                                                                                          \\ \hline
Binzagr   et al. \cite{10.3389/frai.2025.1540646}                      &                       & \ding{51}                    &  \ding{51}                   &                            &                               &                     &                                     &              & \ding{51}                     & \ding{51}                       &                                &                             &                          &                            &                        &                          &                         & &                                                                                                                            \\ \hline
Xin et   al. \cite{XinJiaming2023Casb}                                &                       &                      &                     &                            &                               &                     &                                     &              &                       & \ding{51}                       &                                &                             &                          &                            &                        &                          &                         & & \ding{51}                         \\ \hline
Wang   et al. \cite{WangRongguang2022Etdi}                            &                       &                      & \ding{51}                   &                            &                               &                     &                                     &              &                       &                         &                                &                             &                          &                            &                       \ding{51}  & \ding{51}                        &                         & \ding{51}&                                                                \\ \hline
Atitallah   et al. \cite{10704046}                                    &                       & \ding{51}                    &                     & \ding{51}                          &                               &                     &                                     &              &                       &                         &                                &                             &                          & \ding{51}                          &                        &                          & \ding{51}                       & \ding{51}&                                                                            \\ \hline
Orouskhani   et al.\cite{OROUSKHANI2022100066}                        &                       & \ding{51}                    &                     &                            &                               &                     &                                     &              &                       &                         &                                &                             &                          & \ding{51}                          &                        &                          &                         & &                                                                                                                                                                          \\ \hline
\end{tabular}
\end{table}

\begin{table}[]
\centering

\tiny
\caption{ Companion table for neuroimaging DA/DG studies: experimental domain structure, adaptation/generalization setting, clinical endpoint, and split/leakage reporting. ``NR'' = not reported. The studies that apply generic transfer-learning, ensemble, or architectural-robustness methods without an explicit cross-domain objective are labelled accordingly. ``Internal'' denotes evaluation within a single cohort, ``cross-cohort'' denotes testing on a separate dataset/site.}
\label{mripet_companion_table}
\begin{tabular}{|C{2.1cm}|C{3cm}|C{1.5cm}|C{2.3cm}|C{1.8cm}|C{2.6cm}|}
\hline
\textbf{Paper} & \textbf{Source $\rightarrow$ Target domain} & \textbf{Target labels in training} & \textbf{DA/DG setting} & \textbf{Clinical endpoint} & \textbf{Split \& leakage control} \\ \hline
Tanveer et al. \cite{9440810} & ADNI (internal); ImageNet pre-training & No & Transfer learning (ImageNet); & CN--AD; MCI--AD & Internal; subject-level NR \\ \hline
Guan et al. \cite{9433971} & ADNI-1 $\rightarrow$ ADNI-2/3 & No & Unsupervised DA (cross-phase; mechanism NR) & AD vs CN & Cross-phase; leakage NR \\ \hline
Sinha et al. \cite{sinha2021alzheimers} & ADNI $\leftrightarrow$ AIBL, OASIS (mutual harmonization) & No & Unsupervised DA (adversarial/GAN) & AD vs CN & Cross-cohort (external); NR \\ \hline
Cheng et al. \cite{cheng2017multi} & Within-ADNI multi-domain (auxiliary tasks) & Yes & Supervised DA (multi-domain feature selection) & AD vs CN + MCI conversion (early diagnosis) & Within-cohort (multi-task); NR \\ \hline
Wang et al. \cite{8856726} & ADNI $\rightarrow$ Tongji (private) & Yes & DA (Joint Distribution Adaptation \& Target-Emphasized Loss Weighting) & AD vs CN & Cross-cohort (external); NR \\ \hline
Li et al. \cite{8362917} & ADNI $\rightarrow$ Tongji (private, fMRI) & Yes (few labeled) & Supervised/semi-supervised DA (PCA align.) & AD vs CN & Cross-cohort (external); NR \\ \hline
Wachinger et al. \cite{wachinger2016domain} & ADNI $\rightarrow$ AIBL (main); +CADDementia & Yes & Supervised DA (instance weighting) & AD vs MCI vs CN & Cross-cohort (external); NR \\ \hline
Cai et al. \cite{CAI2023106570} & ADNI 1.5T $\leftrightarrow$ 3T (cross-protocol) & No (pseudo-label) & Unsupervised DA (max-density div.\ + adversarial + prototype) & AD/CN, AD/MCI, MCI/CN & Cross-protocol (within-ADNI); NR \\ \hline
Turrisi et al. \cite{TURRISI2025121016} & ADNI 1.5T $\rightarrow$ 3T (cross-generation) & Yes & Supervised transfer learning (fine-tuning) & AD vs CN & Cross-protocol (within-ADNI); NR \\ \hline
Reddy \cite{reddy2022applicationunsuperviseddomainadaptation} & OASIS (internal) & No & Unsupervised DA (adversarial) & AD vs CN & Internal; NR \\ \hline
Yu et al. \cite{10316553} & OASIS $\rightarrow$ ADNI & No & Unsupervised DA (optimal transport + contrastive) & AD vs CN & Cross-cohort (external); NR \\ \hline
Guan et al. \cite{GUAN2021102076} & ADNI-1/2/3 $\rightarrow$ AIBL & No & Multi-source unsupervised DA (attention-guided) & AD vs CN & Cross-cohort (external); NR \\ \hline
Batool et al. \cite{batool2025singledomaingeneralizationalzheimers} & NACC $\rightarrow$ ADNI, AIBL (unseen) & None & Single-source DG & AD vs CN & Cross-cohort (external, DG); NR \\ \hline
Zhang et al. \cite{10135388} & ADNI (internal) & No & Source-free unsupervised DA & AD vs CN & Internal; NR \\ \hline
Park et al. \cite{PARK2024122253} & In-house $\rightarrow$ public (ADNI/AIBL/OASIS) & No & Unsupervised DA (feature align.) & MCI $\rightarrow$ AD conversion & Cross-cohort (external); NR \\ \hline
Kushol et al. \cite{10.1007/978-3-031-45857-6_1} & ADNI-1/2, AIBL, MIRIAD (domain-shift analysis) & No & Multi-source DA / domain-shift analysis & AD vs CN & Cross-cohort (external); NR \\ \hline
Tobari et al. \cite{tobari2025domain} & Multi-dataset (8 sets, incl.\ non-AD/PPMI) & No & Unsupervised DA (adversarial) & Content-based image retrieval (not AD-vs-CN) & Cross-cohort (external); NR \\ \hline
Rehman et al. \cite{10570178} & ADNI PET (internal) & None & Single-cohort regularization (no target domain; DG loose) & Early/MCI diagnosis & Internal; NR \\ \hline
Duan et al. \cite{duan2023broad} & ADNI PET (internal) & None & Internal classifier on large test dataset & EMCI/LMCI classification & Internal; class-imbalanced \\ \hline
Wang et al. \cite{wang2022high} & ADNI $\rightarrow$ J-ADNI & None & Single-source DG & MCI $\rightarrow$ AD conversion / progression & Cross-cohort (external, DG); NR \\ \hline
Lteif et al. \cite{Lteif2023.09.22.23295984} & NACC $\rightarrow$ ADNI, AIBL, FHS & None & Single-source DG & CN vs MCI vs AD (3-class) & Cross-cohort (external, DG); NR \\ \hline
Bron et al. \cite{bron2021cross} & ADNI $\rightarrow$ PND & None & External validation / cross-cohort generalization (no target adaptation) & AD vs CN + MCI conversion & Cross-cohort (external); NR \\ \hline
Tang et al. \cite{10.1145/3584371.3613000} & ADNI 1.5T $\rightarrow$ 3T (cross-scanner) & None & Single-source DG (privileged info / ROIs) & AD vs CN & Cross-protocol; robustness (F1 drop) reported \\ \hline
Choi et al. \cite{8950078} & ADNI (internal) & None & Internal ensemble (fusion-weight objective) & AD vs CN & Internal; NR \\ \hline
Majee et al. \cite{10650462} & ADNI (internal) & None & Internal classifier on the held-out test set across the diverse sub-studies within ADNI & AD vs CN & Internal; NR \\ \hline
Rawashdeh et al. \cite{alrawashdeh2024ensemble} & Kaggle, OASIS (internal) & None & Ensemble classifier & Multi-class dementia stages & Internal; very high acc.\ (leakage risk, NR) \\ \hline
Binzagr et al. \cite{10.3389/frai.2025.1540646} & ADNI (internal) & None & Inductive transfer learning & AD vs CN & Internal; NR \\ \hline
Xin et al. \cite{XinJiaming2023Casb} & ADNI $\rightarrow$ AIBL & None & Single-source DG & AD vs CN (bal.\ acc., sens.\ reported) & Cross-cohort (external, DG); NR \\ \hline
Wang et al. \cite{WangRongguang2022Etdi} & ADNI, PENN, AIBL (iSTAGING sites) & Yes & DA across sites (both supervised and unsupervised DA ) & AD vs CN & Cross-cohort (external); NR \\ \hline
Atitallah et al. \cite{10704046} & ADNI / Kaggle (evaluated separately) & No (few-shot) & Transfer learning combined with metric-based few-shot learning & AD severity/progression (multi-class) & Cross-source; NR \\ \hline
Orouskhani et al. \cite{OROUSKHANI2022100066} & OASIS (internal) & No (few-shot) & Few-shot metric learning (internal) & AD vs CN & Internal; NR \\ \hline
\end{tabular}
\end{table}

\subsection{DA/DG approaches for AD detection with EEG}

EEG differs fundamentally from neuroimaging or ET modalities in that it captures multi-channel time-series signals, making it better suited for signal processing. The dimensionality of EEG data is influenced by the number of electrodes (channels), the sampling rate, and the recording duration. To harness the power of DL, many studies have converted raw EEG into time-frequency representations—such as spectrograms—to enable the use of CNNs. For example, Jain et al. \cite{JAIN2025105} achieved over 98\% accuracy in AD detection using CNN-based models like SqueezeNet \cite{iandola2016squeezenet}, MobileNet \cite{howard2017mobilenets}, and ResNet \cite{He2016Deep} on these transformed data. However, these models did not address domain shifts caused by variability in EEG acquisition hardware or protocols. The authors acknowledged this limitation and emphasized the need for DA and few-shot learning to improve generalizability.

Effective DA/DG for EEG-based AD detection must account for variability in electrode setups, sampling rates, and recording environments across datasets. These discrepancies introduce distributional shifts that challenge conventional DA/DG methods, particularly those designed for image-based data. As summarized in Table~\ref{eeg_table}, although DA/DG research works in AD detection remains limited, its significance is increasingly recognized. For instance, Mishra et al. \cite{mishra2025challengesdetectingmciusing} demonstrated that models trained on open EEG datasets underperform in clinical settings, highlighting the need for broader, more diverse datasets and robust DA/DG strategies.

Several recent studies have explored ways to improve generalization across heterogeneous EEG recording conditions. Tran et al. \cite{tran2024eegssmleveragingstatespacemodel} enhanced cross-dataset generalization by integrating spectral and temporal features across various segment lengths and temporal resolutions. Likewise, Wang et al. \cite{wang2025leadlargefoundationmodel} developed a foundation model trained on eleven heterogeneous EEG datasets using self-supervised contrastive learning at both sample and subject levels. Their model, capturing both temporal dynamics and spatial patterns, was fine-tuned on five AD datasets with channel-aligned setups, yielding a 9-point gain in F1 score, showcasing the value of diverse pretraining for AD generalization.

When EEG signals are transformed into 3D tensors, CNN-based DA/DG techniques originally developed for vision tasks become applicable. For example, Jiao et al. \cite{JiaoJie2024AADL} and Fan et al. \cite{9560928} used adversarial domain adaptation on CNN networks to mitigate cross-dataset discrepancies. Going beyond image transformation, recent studies have combined 1D EEG signals with features from 2D representations using transformers and adversarial DA (ADA), enriching feature representations and improving cross-dataset generalization performance \cite{nirmaladevi2025novel}.

Due to limited DA/DG studies specifically focused on AD, insights from general EEG classification tasks are often leveraged. Wu et al. \cite{wu2024unsupervised} addressed DA challenges in general EEG classification, particularly when source and target domains differ in channel configurations and signal characteristics due to hardware differences, unlike typical settings where domains share a feature space but differ in distribution \cite{gao2020hhhflhierarchicalheterogeneoushorizontal, 9413712}. Their method fuses multiple representations via attention mechanisms and aligns domains through MMD to enhance transferability.

A significant portion of EEG-based DA/DG research, around 44\% centers on emotion recognition tasks \cite{10.1145/3654664}, which may offer methodological insights for AD applications. For instance, Bao et al. \cite{bao2021two} proposed a two-stage pipeline involving MMD-based distribution alignment followed by class-level alignment with DANN. Kim et al. \cite{Kim2025DomainGeneralized}  combined four DG techniques, Deep CORAL (CORrelation ALignment), GroupDRO, VREx, and DANN, with three representative DL architectures (ShallowFBCSPNet, EEGNet, and TSception) to enable improved subject-independent EEG-based emotion recognition. While not specific to AD, these frameworks provide useful references for mitigating domain shifts in EEG-based research. However, applying such methods across domains requires caution. The underlying EEG biomarkers differ between tasks: AD detection is characterized by changes in spectral power and functional connectivity in theta and gamma bands \cite{chetty2024eeg}, whereas emotion recognition relies on alpha, beta, and gamma activity, power asymmetry, event-related potentials (ERPs), and time–frequency features \cite{brainsci15030220}. These differences must be carefully accounted for when adapting DA/DG techniques from emotion recognition to AD detection.

The EEG studies reviewed above indicate that DA/DG for AD detection is still less mature than neuroimaging DA/DG, but several methodological directions are emerging. AD-focused studies have begun to explore adversarial alignment, spectral--temporal feature integration, and large-scale self-supervised pretraining, while related EEG classification studies provide additional evidence for MMD-based alignment, multi-source learning, and domain-invariant representation learning. The relative usefulness of these methods depends on the type of EEG shift. Spectral--temporal feature integration is suitable when disease information is distributed across frequency bands and time segments. Adversarial and MMD-based methods are useful when electrode montage, channel configuration, sampling rate, or recording environment differs across datasets. Large-scale self-supervised pretraining is especially promising when labeled AD EEG data are scarce, because it can learn transferable temporal and spatial representations before task-specific fine-tuning. Quantitatively, large-scale EEG pretraining has shown F1-score improvements of approximately 9 percentage points at both sample and subject levels, and related EEG DA/DG studies report improvements from adversarial alignment, MMD-based feature fusion, and multi-source representation learning. However, many EEG datasets remain small, and reported improvements are not always directly comparable because evaluations differ between segment-level, trial-level, subject-level, and cross-dataset settings. Future EEG studies should report subject-level cross-domain metrics and variability across independent runs to make robustness gains easier to compare.

Considering the companion Table~\ref{eeg_companion_table} as a whole clarifies the current state of EEG evidence. A large fraction of the rows are not AD studies but adjacent-domain EEG works (motor-imagery or emotion recognition), retained only for methodological transfer and marked ``Non-AD'' in the endpoint column, so they should not be read as evidence of AD generalization. Among the genuinely AD-focused entries, the recorded domain structure is dominated by technical shift (electrode montage, sampling rate, and recording environment), and the AD-relevant cohorts are few, small, and heterogeneous in their diagnostic groups. The split-and-leakage column is frequently ``NR'', and the central EEG leakage risk, trial- or segment-level windows from the same participant appearing in both training and test sets, is only occasionally controlled by explicit subject-level splitting. The table therefore indicates that EEG DA/DG for AD remains early-stage: useful for methodological transfer, but still short of the subject-level, cross-cohort, AD-specific evaluation needed for clinical claims.

\begin{table}[]
\footnotesize
\centering
\caption{Summary of related literature on DA/DG approaches for AD detection using EEG data.}
\label{eeg_table}
\begin{tabular}{|C{2cm}|C{2.5cm}|C{2cm}|C{3cm}|C{3.5cm}|C{1cm}|}
\hline
\textbf{Paper}                                               & \textbf{Dataset}                                                                                          & \textbf{Research area}     & \textbf{Method}                                                                                                                                                                                 & \textbf{Evaluation}                                                                                                                                                                          & \textbf{Code} \\ \hline
Nirmala   et al. \cite{nirmaladevi2025novel}                 & Healthy-MCI-AD EEG   Dataset\cite{sedghizadeh2023eeg}  ,   OpenNeuro ds004504 \cite{li2024resting}        & CN/AD/FTD classification   & Adversarial Domain   Adaptation (ADA)                                                                   & Reaches over 98\% accuracy on the two datasets                                                                                                      &                           \\ \hline
Tran et   al. \cite{tran2024eegssmleveragingstatespacemodel} & OpenNeuro ds004504   \cite{li2024resting}                                                                 & CN/AD/FTD  classification   & Combining temporal and spectral components.                                                                                              & Achieves 91\% accuracy.                                                                                                                      &                           \\ \hline
Wu et   al. \cite{wu2024unsupervised}                        & BCI-III-Iva   \cite{bcicompIII_IVa}, WCCI \cite{cbcic2020}                                                & General EEG classification & Features fusion, attention mechanism and MMD                                                                      & The average cross-dataset accuracy outperforms literature methods for all adaptations                                                                                           &                           \\ \hline
Fan et   al. \cite{9560928}                                  & DMS Dataset   \cite{li2017cognitive},  REST Dataset   \cite{9560928}                                      & MCI detection              & ADA                                                                                                    & The accuracy outperforms the literature on all adaptations.                                                                                                                       &                           \\ \hline
Wang et   al. \cite{wang2025leadlargefoundationmodel}        & 6 publicly available EEG dataset containing AD subjects (AD-Auditory \cite{lahijanian2024auditory}, ADFSU \cite{vicchietti2023computational}, ADFTD \cite{data8060095}, ADSZ \cite{Alves_2022}, APAVA \cite{Escudero_2006}, BrainLat \cite{prado2023brainlat}), 3 private datasets (Cognision-ERP \cite{cecchi2015clinical}, Cognision-rsEEG \cite{wang2025leadlargefoundationmodel}, CNBPM \cite{AMEZQUITASANCHEZ201988}), 7 non-AD datasets (Depression \cite{cavanagh2019depression}, PEARL-Neuro \cite{dzianok2024pearl}, REEG-BACA \cite{getzmann2024resting}, REEG-PD \cite{singh2023evoked}, REEG-SRM \cite{hatlestad2022bids}, TDBrain \cite{vanDijk2022tdbrain}, TUEP \cite{8257044}) & AD and CN classification       & Pretraining with large datasets by self-supervised contrastive learning on both sample and subject level and channel-aligned unified fine-tuning across multiple sources.                   & Outperformed literature by up to +9.86\% F1 at the sample level and +9.31\% F1 at the subject level                                                           & \ding{51}                         \\ \hline
Jiao et   al. \cite{JiaoJie2024AADL}                         & OpenBMI \cite{lee2019openbmi},   BCIC IV 2a \cite{tangermann2012bci}                                      & General EEG classification & Convert into 3D array to leverage the CNN models for feature extraction and ADA & Accuracy rates of 70.42$\pm$12.44 and 73.91$\pm$11.28 on the OpenBMI and BCIC IV 2a datasets, outperformed the baseline models. &                           \\ \hline

Zhong et   al. \cite{10609514}                         & OpenBMI \cite{lee2019openbmi},   BCIC IV 2a \cite{tangermann2012bci}                                      & General EEG classification & Domain-invariant feature representation with multiple source domains & Demonstrated superiority against literature methods on all test cases. &                           \\ \hline
Liang et   al. \cite{10198467}                         &  BCIC IV 2a \cite{tangermann2012bci}, BCIC IV 2b \cite{tangermann2012bci}, BCIC III 4a \cite{tangermann2012bci}                                       & General EEG classification & Category-relevant and -irrelevant feature decorrelation and cross-view invariant feature learning with multi-source domains &  Outperformed the available state-of-the-art methods on all test cases. &                           \\ \hline
Zheng et   al. \cite{bioengineering12050495}                         &  BCIC IV 2a \cite{tangermann2012bci}, the Korean University dataset            & General EEG classification                            & Domain-invariant feature extraction with DL from multi-source domains &  Achieved 8.93\% and 4.4\% accuracy improvements on two datasets, respectively, compared with current state-of-the-art models. &                           \\ \hline
Ballas et   al. \cite{BallasAristotelis2024TDGf}                         &  SEED \cite{Zheng2015Investigating}, SEED-FRA \cite{Schaefer2010Assessing} and SEEDGER \cite{Schaefer2010Assessing}                                     & General EEG classification & Combination of CNN's intermediate features to represent invariant attributes &  Demonstrated improved effectiveness compared to the baseline and the state-of-the-art methods. &                           \\ \hline

\end{tabular}
\end{table}

\begin{table}[]
\centering

\scriptsize
\caption{ Companion table for EEG DA/DG studies: experimental domain structure, adaptation/generalization setting, clinical endpoint, and split/leakage reporting. ``NR'' = not reported. Studies whose endpoint is marked ``Non-AD'' are adjacent-domain EEG classification works (motor imagery/emotion) retained for methodological transfer, not direct AD evidence.}
\label{eeg_companion_table}
\begin{tabular}{|C{2.1cm}|C{3cm}|C{1.5cm}|C{2.3cm}|C{1.9cm}|C{2.5cm}|}
\hline
\textbf{Paper} & \textbf{Source $\rightarrow$ Target domain} & \textbf{Target labels in training} & \textbf{DA/DG setting} & \textbf{Clinical endpoint} & \textbf{Split \& leakage control} \\ \hline
Nirmala et al. \cite{nirmaladevi2025novel} & Healthy-MCI-AD EEG $\leftrightarrow$ OpenNeuro ds004504 & No & Unsupervised DA (adversarial) & CN/AD/FTD & Cross-dataset; leakage NR \\ \hline
Tran et al. \cite{tran2024eegssmleveragingstatespacemodel} & OpenNeuro ds004504 (within-corpus) & None & Within-corpus model (DG/cross-dataset) & CN/AD/FTD & Cross-segment; subject-level NR \\ \hline
Wu et al. \cite{wu2024unsupervised} & BCI-III-IVa $\rightarrow$ WCCI & No & Unsupervised DA (MMD, attention) & Non-AD (motor imagery) & Cross-dataset; NR \\ \hline
Fan et al. \cite{9560928} & DMS $\rightarrow$ REST & No & Unsupervised DA (adversarial) & MCI vs CN & Cross-dataset; NR \\ \hline
Wang et al. \cite{wang2025leadlargefoundationmodel} & 11 EEG datasets $\rightarrow$ 5 AD datasets & No (self-supervised) Yes (supervised fine-tuning) & Self-
supervised pretraining and supervised fine-tuning & AD vs CN & Cross-dataset; subject-level reported \\ \hline
Jiao et al. \cite{JiaoJie2024AADL} & OpenBMI, BCIC IV 2a & Partial (fine-tuned) & Semi-supervised DA (MMD) & Non-AD (motor imagery) & Cross-subject; NR \\ \hline
Zhong et al. \cite{10609514} & OpenBMI, BCIC IV 2a & None & Multi-source DG & Non-AD (motor imagery) & Cross-subject; NR \\ \hline
Liang et al. \cite{10198467} & BCIC IV 2a/2b, BCIC III 4a & None & Multi-source DG & Non-AD (motor imagery) & Cross-subject; NR \\ \hline
Zheng et al. \cite{bioengineering12050495} & BCIC IV 2a, Korea Univ.\ dataset & None & Multi-source DG & Non-AD (motor imagery) & Cross-subject; NR \\ \hline
Ballas et al. \cite{BallasAristotelis2024TDGf} & SEED, SEED-FRA, SEED-GER & None & Multi-source DG & Non-AD (emotion recognition) & Cross-subject; NR \\ \hline
\end{tabular}
\end{table}

\subsection{DA/DG approaches for AD detection with ET}

ET technology has emerged as a powerful, non-invasive tool for capturing visual attention and cognitive processing patterns. By recording fine-grained eye movement behaviors, ET facilitates assessment of memory, attention, and information processing speed affected by AD \cite{azizi2022video}. Its ease of use and suitability for mobile, clinical, and home environments also make it ideal for longitudinal cognitive monitoring \cite{medicina59081361}.

In recent years, ML approaches have been successfully applied to ET data for AD detection, yielding promising results. Classical models such as decision trees (DT) and support vector machines (SVMs) have achieved classification accuracies exceeding 90\% and AUC scores over 80\% in some studies \cite{TADOKORO2021117529, song2023eye, lage2021distinctive, info:doi/10.2196/48093, 10.1145/3613905.3650731}. DL methods, such as CNNs \cite{BoliaCharu2024ASoN, kastrati2021eegeyenet, MILES2024264, sun2022novel, HaqueRafiU.2021DCNN} and transformer-based architectures \cite{candela2024cognitive, SUN2024121929, ALSUHAIBANI2024124185}, have further advanced the field by automatically extracting complex spatiotemporal features from raw ET signals, reducing the reliance on handcrafted inputs.

Despite this progress, most studies evaluate model performance using data collected under the same conditions on devices, protocols, and populations with the training set. This limits ML models' generalizability to real-world scenarios, where domain shifts can result from differences in devices, environments, and user demographics. To ensure robustness in such diverse settings, DA/DG strategies are essential. However, the application of DA/DG techniques specifically for ET-based AD detection remains limited, as shown in Table~\ref{et_table}, highlighting a critical opportunity for future research.

A notable step toward addressing this gap is the work by Jung et al. \cite{math13111867}, who developed a dual-pathway CNN model based on the GazeMap dataset. Their approach transformed 1D gaze and head movement time-series data into 2D spatial maps, effectively capturing both short- and long-term temporal dynamics. Importantly, the model was evaluated not only on an in-distribution test set (achieving 91.1\% accuracy) but also on an out-of-distribution cohort collected from nursing homes and senior centers (achieving 83.3\% accuracy). This work demonstrates the importance of DA/DG to build robust ML models to improve the application of ET in AD detection in real-world settings. 

While direct applications of DA/DG in ET-based AD diagnosis are still emerging, a substantial body of related work has focused on domain-adaptive gaze estimation, an essential upstream task for many cognitive assessment pipelines. Accurate gaze estimation is crucial for deriving reliable biomarkers of cognitive decline, especially under real-world conditions where data sources may differ. For instance, Lahiri et al. \cite{10.1145/3293353.3293423} proposed an adversarial DA/DG framework using dynamic layer-wise weighting to enhance gaze estimation across domains. Nguyen et al. \cite{10.1145/3654703} introduced a Sim2Real learning strategy that aligns synthetic and real ET data, improving generalization. Similarly, Guo et al. \cite{10.1007/978-3-030-69541-5_18} developed an embedding space consistency method for unsupervised DA in gaze estimation. These studies aim to bridge the gap between controlled experimental data and real-world use cases. Xia et al. \cite{XIA2025111244} proposed the Cross Gaze Generalization (CGaG) framework to enhance gaze estimation performance across different domains and highlight real-world applications in healthcare, attention monitoring, driver safety. The authors specifically point to CGaG's potential for deployment on mobile devices for monitoring gaze behavior in people with AD, reinforcing its relevance to cognitive assessment.

Although most of the studies summarized in Table~\ref{et_table} do not directly apply DA/DG to AD detection, they provide critical foundations for developing generalizable ET-based diagnostic systems. As research in AD detection expands toward practical and non-clinical deployment, leveraging DA/DG strategies from gaze estimation will be key to building robust, adaptive tools to maintain performance across diverse populations and environments.

The ET literature reviewed above shows a clear contrast between the limited number of AD-specific DA/DG studies and the broader progress in domain-adaptive gaze estimation. This distinction is important because ET-based AD detection depends on both the diagnostic model and the reliability of upstream gaze estimation. In AD detection, the GazeMap study provides direct evidence of domain shift, with accuracy decreasing from 91.1\% on the in-distribution test set to 83.3\% on an out-of-distribution cohort. This drop indicates that ET models trained in controlled clinical settings may not transfer directly to nursing homes, senior centers, or home-like environments. In related gaze-estimation studies, DA/DG methods have reported substantial error reductions, including approximately 10--43\% improvements using unsupervised adaptation, adversarial learning, consistency regularization, and Sim2Real alignment. These methods are useful in different scenarios: Sim2Real alignment is suitable when synthetic gaze data are available, adversarial learning is useful for learning device-invariant features, and consistency-based methods can stabilize predictions across camera viewpoints and head poses. Although these gaze-estimation studies are not all AD-specific, they are relevant because reliable ET-based cognitive biomarkers depend on accurate gaze estimation under different devices, lighting conditions, head poses, and environments. Thus, the ET literature suggests that robust AD screening will require both diagnostic-domain validation and upstream gaze-estimation adaptation. More AD-specific cross-device and cross-environment benchmarks are needed before firm conclusions can be drawn.

The companion Table~\ref{et_companion_table} makes the thinness of ET-based AD evidence explicit: only the GazeMap study directly targets AD detection, while the remaining rows are upstream domain-adaptive or domain-generalized gaze-estimation studies, marked ``Non-AD'' and retained solely for methodological transfer. The dominant shift recorded across the table is technical and environmental (device, camera, lighting, head pose, and calibration), and no ET entry evaluates population, temporal, or workflow generalization for AD. Because the single AD entry already shows a clear in-distribution-to-out-of-distribution accuracy drop, and because the leakage column indicates that repeated trial-level gaze sequences from the same participant are a standing risk, the table supports only a cautious reading: ET is a promising but under-evidenced modality whose clinical generalizability for AD cannot yet be established from the current literature.

\begin{table}[]
\centering
\footnotesize
\caption{Summary of related literature on DA/DG approaches for AD detection using ET data. }
\label{et_table}
\begin{tabular}{|C{2cm}|C{3cm}|C{1.8cm}|C{3.2cm}|C{3cm}|C{1cm}|}
\hline
\textbf{Paper}                                               & \textbf{Dataset}                                                                                                                                            & \textbf{Research area}                   & \textbf{Method}                                                                                                                      & \textbf{Evaluation}                                                                                                                         & \textbf{Code} \\ \hline
Jung et al. \cite{math13111867}                         & GazeMap   \cite{math13111867}                                                                 & AD detection   & dual-pathway CNN                                                                                                  & Accuracy of 91.1\% on the in-distribution and 83.3\% on the out-of-distribution set                                                                                                                     &                         \\ \hline
Nguyen et al. \cite{10.1145/3654703}                         & RIT-Eyes   \cite{10.1145/3385955.3407935}, OpenEDS   \cite{garbin2019openedsopeneyedataset}                                                                 & ET performance   enhancement   & combination of CycleGAN, Siamese Network, and domain-adversarial neural   network (DANN)                                                                                                  & 20\% improvement in mIoU on real-world datasets                                                                                                                     & \ding{51}                         \\ \hline
Xia et   al. \cite{XIA2025111244}                            & ETH-Xgaze   \cite{zhang2020eth}, Gaze360 \cite{kellnhofer2019gaze360}, MPIIFaceGaze   \cite{zhang2017s},  EyeDiap   \cite{10.1145/2578153.2578190}                 & Gaze estimation for realworld   settings & Hybrid approach combining   contrastive learning and ADA                                                  & A relative error reductions of   ~15-40\% on real-world datasets                                                                        &                           \\ \hline
Bixler   et al. \cite{10.1145/3448017.3457386}               & Raw gaze   recordings\cite{zhao2021mwdet}                                                                                                                   & Attention analysis using eye gaze        & Instance based and feature level   adaptation techniques                                                                              &  The proposed cross-domain models achieved AUROCs of 0.56 to 0.68.                                                                                            &                           \\ \hline
Guo et   al. \cite{10.1007/978-3-030-69541-5_18}             & MPIIGaze \cite{7299081}, EYEDIAP   \cite{10.1145/2578153.2578190}                                                                                           & ET performance enhancement   & Unsupervised DA with consistency of embedding representation across domains.          & Error reduction of 10\% on MPIIGaze and 19\% on EYEDIAP compared with no adaptation                                                   &                           \\ \hline
Lahiri   et al.\cite{10.1145/3293353.3293423}                & MPIIGaze \cite{7299081}                                                                                                                                     & Gaze estimation                          & ADA with dynamic importance on different layers & 43\% error reduction compared   with no adaptation                                                                                            & \ding{51}                         \\ \hline
Ahn et   al. \cite{AhnByungtae2021Hpge}                      & MPIIGaze \cite{7299081}, SynFace   \cite{AhnByungtae2021Hpge}, Columbia Gaze \cite{smith2013gaze}                                                           & Gaze estimation                          &  DANN                                                                                           & Reduced the mean error of 70\% on Columbia Gaze dataset and 60\% on MPIIGaze dataset                                                      &                           \\ \hline
Soundararajan   et al. \cite{SoundararajanVidivelli2025SoEG} & SBVPI \cite{sbvpi_dataset},   Kaggle dataet \cite{shah2020eye}                                                                                              & Gaze estimation and detection            & Transfer learning with   pre-trained CNN models                                                                                      & Achieved an average of 80\%   accuracy      for eye gaze detection and 0.19 MAE for gaze estimaton  &                           \\ \hline
Yin et   al. \cite{YinPengwei2025TAWS}                       & ETH-Xgaze \cite{zhang2020eth}, d   Gaze360 \cite{kellnhofer2019gaze360}, MPIIFaceGaze \cite{zhang2017s}, RT-GENE   \cite{kellnhofer2019gaze360}, EyeDiap \cite{10.1145/2578153.2578190} & Gaze estimation                          & Transformer based supervised and   self-supervised DA                                                                  & Reduced errors on all adaptation scenarios by an average of 15\% compared to baseline                                                          &                           \\ \hline
Zhang   et al. \cite{ZhangSihui2024DaUN}                     & MPIIFaceGaze \cite{zhang2017s},   Gaze360 \cite{kellnhofer2019gaze360}, ETH-XGaze dataset   \cite{zhang2020eth}, EyeDiap \cite{10.1145/2578153.2578190}             & Gaze estimation                          & Enforce inter and intera domain   consistency with Domain-Consistent and Uncertainty-Aware Network                                   & Reduced errors by an average of   4\% compared with baselines for all adaptation tests                                                        &                           \\ \hline
Cui   et al. \cite{8296894}                     & UT Multiview dataset \cite{6909631}            & Gaze estimation                          & Geodesic flow kernel by deriving a new feature representation                                   & Reduced target estimation errors compared with no DA                                                        &                           \\ \hline
Kothari   et al. \cite{10.1145/3530880}                     & ExCuSe \cite{Fuhl2015Excuse}, ElSe \cite{Fuhl2016ElSe} and PupilNet \cite{Fuhl2017PupilNet2}            & Gaze estimation                          & Jointly optimizing on multiple eye image datasets to learn a generalized feature representation                                  &  Datasets with higher variability generalize better from multiset optimization                                                    &                           \\ \hline
Mo   et al. \cite{10809283}                     & ETH-XGaze \cite{zhang2020eth}, Gaze360 \cite{kellnhofer2019gaze360}, EyeDiap \cite{10.1145/2578153.2578190}, and MPIIGaze \cite{7299081}            & Gaze estimation                          &  Integrates frequency domain data augmentation with a lightweight multilayer perceptron                                  &  Exhibited superior generalization capabilities compared to existing methods.                                                    &                           \\ \hline
Back   et al. \cite{Back_2023}                     & ETH-XGaze \cite{zhang2020eth}, MPIIGaze \cite{7299081}, and GazeCapture \cite{Krafka2016EyeTracking}            & Gaze estimation                          &  Domain-agnostic
feature learning and gaze consistency regularization                                &  Outperformed all the state-ofthe-art methods on the DG tasks                                                     &                           \\ \hline
Kim   et al. \cite{10972620}                     & ETH-XGaze \cite{zhang2020eth}, Gaze360 \cite{kellnhofer2019gaze360}, EyeDiap \cite{10.1145/2578153.2578190}, and MPIIFaceGaze \cite{zhang2017s}            & Gaze estimation                          &  Integrates  adversarial training and an additional penalizing term to extract domain-invariant features                                  &  Achieved state-of-the-art performance in the DG on gaze estimation benchmark.                                                    &                           \\ \hline
\end{tabular}
\end{table}

\begin{table}[]
\centering

\scriptsize
\caption{ Companion table for ET DA/DG studies: experimental domain structure, adaptation/generalization setting, clinical endpoint, and split/leakage reporting. ``NR'' = not reported. Only Jung et al.\ directly targets AD detection; the remaining entries are upstream domain-adaptive gaze-estimation studies (non-AD) retained for methodological transfer.}
\label{et_companion_table}
\begin{tabular}{|C{2.1cm}|C{3cm}|C{1.5cm}|C{2.3cm}|C{2cm}|C{2.4cm}|}
\hline
\textbf{Paper} & \textbf{Source $\rightarrow$ Target domain} & \textbf{Target labels in training} & \textbf{DA/DG setting} & \textbf{Clinical endpoint} & \textbf{Split \& leakage control} \\ \hline
Jung et al. \cite{math13111867} & GazeMap in-distribution $\rightarrow$ OOD (nursing/senior) & None & Single-source DG & AD vs CN & Cross-environment (in vs OOD) \\ \hline
Nguyen et al. \cite{10.1145/3654703} & RIT-Eyes/OpenEDS (synthetic $\rightarrow$ real) & No & Unsupervised DA (Sim2Real) & Non-AD (gaze estimation) & Cross-domain; NR \\ \hline
Xia et al. \cite{XIA2025111244} & ETH-XGaze, Gaze360, MPIIFaceGaze, EyeDiap & None & DG (contrastive + ADA) & Non-AD (gaze estimation) & Cross-dataset; NR \\ \hline
Bixler et al. \cite{10.1145/3448017.3457386} & Raw gaze recordings & No & Instance/feature-level DA & Non-AD (attention/gaze) & Cross-domain; AUROC 0.56--0.68 \\ \hline
Guo et al. \cite{10.1007/978-3-030-69541-5_18} & MPIIGaze $\rightarrow$ EYEDIAP & No & Unsupervised DA (consistency) & Non-AD (gaze estimation) & Cross-dataset; NR \\ \hline
Lahiri et al. \cite{10.1145/3293353.3293423} & MPIIGaze & No & Unsupervised DA (adversarial) & Non-AD (gaze estimation) & Cross-domain; NR \\ \hline
Ahn et al. \cite{AhnByungtae2021Hpge} & MPIIGaze, SynFace, Columbia Gaze & No & Unsupervised DA (DANN) & Non-AD (gaze estimation) & Cross-dataset; NR \\ \hline
Soundararajan et al. \cite{SoundararajanVidivelli2025SoEG} & SBVPI, Kaggle dataset & No & Transfer-learning DA & Non-AD (gaze est./detect.) & Cross-dataset; NR \\ \hline
Yin et al. \cite{YinPengwei2025TAWS} & ETH-XGaze, Gaze360, MPIIFaceGaze, RT-GENE, EyeDiap & Partial & Supervised/self-supervised DA & Non-AD (gaze estimation) & Cross-dataset; NR \\ \hline
Zhang et al. \cite{ZhangSihui2024DaUN} & MPIIFaceGaze, Gaze360, ETH-XGaze, EyeDiap & No & Unsupervised DA (uncertainty) & Non-AD (gaze estimation) & Cross-dataset; NR \\ \hline
Cui et al. \cite{8296894} & UT Multiview & No & Unsupervised DA (geodesic flow) & Non-AD (gaze estimation) & Cross-domain; NR \\ \hline
Kothari et al. \cite{10.1145/3530880} & ExCuSe, ElSe, PupilNet & None & Multi-source DG & Non-AD (gaze estimation) & Cross-dataset; NR \\ \hline
Mo et al. \cite{10809283} & ETH-XGaze, Gaze360, EyeDiap, MPIIGaze & None & Multi-source DG & Non-AD (gaze estimation) & Cross-dataset; NR \\ \hline
Back et al. \cite{Back_2023} & ETH-XGaze, MPIIGaze, GazeCapture & None & DG (domain-agnostic) & Non-AD (gaze estimation) & Cross-dataset; NR \\ \hline
Kim et al. \cite{10972620} & ETH-XGaze, Gaze360, EyeDiap, MPIIFaceGaze & None & DG (adversarial) & Non-AD (gaze estimation) & Cross-dataset; NR \\ \hline
\end{tabular}
\end{table}

\subsection{DA/DG approaches for AD detection with speech-based data}

DA/DG have become essential in speech-based AD detection, due to the persistent challenge of poor model generalization across datasets, modalities, and languages. Chafouleas \cite{chafouleas2021alzheimer} demonstrated this limitation by showing that models trained on the ADReSS dataset failed to generalize to the COMPASS-ND dataset, and vice versa. This poor transferability highlights the need for robust DA/DG strategies to build scalable and reliable AD detection systems. Latif et al. \cite{latif2025deep} further emphasized that integrating DA techniques can enhance demographic generalizability, improving model suitability for real-world applications. A summary of representative DA/DG studies in speech-based AD detection is provided in Table~\ref{speechml_table}.

A foundational step in DA/DG is identifying speech features that remain stable across domains. Petti et al. \cite{doi:10.3233/JAD-220847} analyzed longitudinal speech from public figures and found that lexical diversity and pronoun usage consistently signaled cognitive decline. Similarly, Ahmed et al. \cite{ALKENANI2021103803} and Masrani et al. \cite{masrani-etal-2017-detecting} highlighted reduced grammatical complexity and idea density-derived from blogs and narratives—as early signs of AD. Farzana and Parde \cite{farzana-parde-2023-towards} evaluated both feature-based and instance-based DA across multiple datasets, identifying robust domain-invariant predictors such as token count, phrase length, acoustic quality, named entities, and vocabulary richness, which significantly improved adaptation performance.

Recent advancements in LLMs have further transformed feature extraction and generalization. Mo et al. \cite{Mo2024.08.22.24312463} used LLMs to summarize linguistic markers across diverse speech datasets using few-shot and zero-shot prompting, significantly boosting AD detection performance to 97\% accuracy. Botelho et al. \cite{botelho2024macro} similarly extracted macro descriptors from transcripts with LLMs and classified them using traditional models (e.g., SVM, RF), achieving over 80\% accuracy on ADReSS.

Beyond feature extraction, LLMs have also been fine-tuned to improve generalization. Zheng et al. \cite{10.1007/978-981-96-1151-5_21} applied soft prompt tuning to LLAMA2 \cite{touvron2023llama2openfoundation} on the ADReSS dataset. Their method achieved 81.31\% accuracy, outperforming BERT \cite{devlin2019bert} by 4.46\%. Building on this, Casu et al. \cite{10495546} fine-tuned multiple 7B-scale models (LLAMA2 \cite{touvron2023llama2openfoundation}, MISTRAL \cite{jiang2023mistral7b}, OpenHermes \cite{teknium2023openhermes}) on ADReSSo, reaching over 90\% accuracy, further validating the use of LLMs for cross-domain speech modeling.

DA/DG has also been applied at the speech recognition stage. Kang et al. \cite{kang2024errorsequalinvestigationspeech} found that different types of automatic speech recognition (ASR) errors provide important information for dementia classification. Relative to a baseline using manually produced transcripts (accuracy $\approx$82.6\%, AUC $\approx$0.873), classifying from ASR transcripts refined with domain-adaptive beam-search decoding on the same corpus improved AD classification accuracy to $\approx$88\% and AUC to $\approx$0.90 \cite{Li2024UsefulBlunders}.

To increase model robustness, some studies have integrated both acoustic and textual modalities using DA/DG techniques. Ilias et al. \cite{ILIAS2023101485} converted audio to log-Mel spectrograms and processed them alongside text transcripts using BERT \cite{devlin2019bert} and DeiT \cite{touvron2021training}. Optimal transport-based DA was used to align intra- and inter-modal features, resulting in strong performance on the ADReSS and ADReSSo datasets. Loukas et al. \cite{ILIAS2023110834} expanded this work, proposing a more sophisticated fusion of modalities across text and audio images. Their approach outperformed conventional feature fusion methods across both datasets, demonstrating the effectiveness of cross-modal DA/DG.

Cross-lingual generalization has also gained attention. Skirrow et al. \cite{SkirrowCaroline2024SiAA} applied multilingual DA layers on speech-to-text features in the ADNI4 Storyteller dataset, aligning language representations for AD and MCI classification. Their models achieved AUC scores between 0.71 and 0.84 across language pairs. Fraser et al. \cite{fraser-etal-2019-multilingual} also investigated cross-lingual generalization with English-to-French with feature augmentation \cite{daume-iii-2007-frustratingly} on the small French dataset and transferable features across different languages, improving the French classification AUC from 0.85 to 0.89.

Traditional DA strategies have also proven effective. Liu et al. \cite{Liu2022FeaturePurification} used adversarial training with a Gradient Reversal Layer (GRL) and a transformer-based purification network to separate domain-invariant and discriminative features, achieving 91.6\% accuracy on the Pitt dataset. Masrani et al. \cite{10.1007/978-3-319-57351-9_29} focused on transferring models from AD detection to MCI classification using feature-level augmentation and correlation alignment, reporting a 5\% gain in F-measure for MCI detection.

The speech-based studies reviewed above show that DA/DG can be applied at several levels, including linguistic feature selection, instance-based adaptation, ASR adaptation, cross-lingual transfer, LLM-based representation learning, and audio--text fusion. These approaches are not interchangeable because they address different sources of speech-domain shift. Feature- and instance-based adaptation are useful when datasets differ in task design or participant composition but share similar linguistic feature spaces. ASR adaptation is important when recognition errors differ across recording conditions, accents, or clinical speech patterns. Cross-lingual transfer is suitable when labeled data are limited in the target language, while LLM-based representation learning can capture higher-level discourse and semantic markers but may be sensitive to prompt design, training data bias, and language coverage. Audio--text fusion can improve robustness by combining acoustic and linguistic evidence, but it also introduces synchronization and modality-specific noise. These different adaptation points are reflected in the reported quantitative improvements. Feature- and instance-based DA produced a 22\% accuracy increase across speech datasets, ASR transcripts with domain-adaptive decoding improved AD classification accuracy from approximately 82.6\% (manual transcripts) to 88\% on the same corpus, feature-level augmentation and correlation alignment improved MCI classification F-measure by 5\%, and English-to-French transfer improved AUC from 0.85 to 0.89. These findings indicate that both linguistic-feature alignment and upstream ASR adaptation can improve robustness. However, direct comparison remains difficult because studies differ in whether they use audio, transcripts, ASR outputs, handcrafted linguistic features, LLM-derived descriptors, or multimodal audio--text fusion. Speech DA/DG benchmarks also vary in language, elicitation task, participant demographics, and diagnostic labels. Therefore, future work should report cross-corpus, cross-language, and participant-level evaluation using common metrics to clarify which adaptation strategies generalize beyond specific datasets.

The companion Table~\ref{speech_companion_table} shows that speech is the modality where cross-corpus and cross-language shift is most directly confronted: the source-to-target column is dominated by genuine cross-corpus and cross-language transfers such as ADReSS to ADReSSo and English to French, which is a strength relative to the other modalities. Two limitations are equally visible, however. First, the split-and-leakage column shows that participant-level leakage control, critical when the same speaker contributes multiple samples, is rarely reported explicitly, and several entries pair very high single-corpus accuracy with unreported leakage safeguards, a combination that warrants caution. Second, a growing number of entries rely on LLM-based inference, which is flagged for benchmark-contamination risk because widely available corpora such as DementiaBank/Pitt and ADReSS may overlap with model pretraining data. A third caveat, flagged in the companion table, is that several entries the summary table describes as adaptation are on closer reading external-validation studies (which report that models transfer poorly) or ensemble- and LLM-based inference methods rather than explicit domain adaptation, and should not be counted as evidence that domain shift was actively addressed. The table therefore indicates that speech DA/DG has the most mature cross-domain evaluation practices but also the most acute leakage and contamination risks.


\begin{table}[]
\scriptsize 
\centering
\caption{Summary of literature on DA/DG approaches for AD detection using speech-based data.}
\label{speechml_table}
\begin{tabular}{|C{2cm}|C{2cm}|C{2cm}|C{4cm}|C{4cm}|C{1cm}|}
\hline
\textbf{Paper}                                       & \textbf{Dataset}                                                                                                                                                                                                     & \textbf{Research area} & \textbf{Method}                                                                                                                                                                                                                                 & \textbf{Evaluation}                                                                                                                                                                                                                                                                                         & \textbf{Code} \\ \hline
Fraser   et al. \cite{fraser-etal-2019-multilingual} & Pitt Corpus   \cite{becker1994natural}, French Dem@Care corpus   \cite{10.1145/3154862.3154915}                                                                                                                     & AD/HC classification   & Feature extraction from speech transcripts, feature augmentation and classification with linear models.                                                                                                                           & Improved the AUC on the French dataset from 0.85 to 0.89                                                                                                                                                                                                                                &                           \\ \hline
Loukas   et al. \cite{ILIAS2023110834}               & ADReSS   \cite{luz2020alzheimersdementiarecognitionspontaneous}, ADReSSo   \cite{luz2021detecting}                                                                                                                   & AD/CN classification   & Adapting acorss different modalities with Earth Mover's Distance (EMD) Transport                                                                                                                    & Outperforms the literature methods on all adaptations                                                                                                                                                                                                                         &                           \\ \hline
Li et   al. \cite{Li2024UsefulBlunders}              & ADReSS   \cite{luz2020alzheimersdementiarecognitionspontaneous}, WLS Dataset   \cite{herd2014cohort}                                                                                                                 & AD/CN classification   & Domain-adaptive beam search decoding and fine tuning with target domain data                                                                                                           & Accuracy and AUC were improved for both datasets.                                                                                                                                                        & \ding{51}                         \\ \hline
Liu et   al. \cite{Liu2022FeaturePurification}       & Pitt Corpus   \cite{becker1994natural}                                                                                                                                                                               & AD/CN classification   & Adversarial common-feature extraction and feature purification                                        & Achieved 91.6\% accuracy on the Pitt dataset                                                                                                                                                                                                                               &                           \\ \hline
Ahmed   et al. \cite{ALKENANI2021103803}             & Pitt Corpus   \cite{becker1994natural}, ADBC  \cite{masrani-etal-2017-detecting}                                                                                                        & AD/HC classification   & A hybrid stacked fusion model based on ensemble learning.                          & Achieved an AUC of 99.1\%, accuracy of 96.51, and F1 score of 96.74\%, outperforms single classifiers and literature methods.                                                             &                           \\ \hline
Chafouleas   \cite{chafouleas2021alzheimer}          & ADReSS   \cite{luz2020alzheimersdementiarecognitionspontaneous}, COMPASS-ND   \cite{CCNA_COMPASSND_2025}                                                                                                             & AD/HC classification   & Classification with BERT on common feature representations                                                                                                                                                                   & Achieves a test accuracy of 62.6\% on the ADReSS challenge dataset and 66.7\% on the COMPASS-ND dataset                                                                                                                                                      &                           \\ \hline
Zheng et   al. \cite{10.1007/978-981-96-1151-5_21}   & ADReSS   \cite{luz2020alzheimersdementiarecognitionspontaneous}                                                                                                                                                      & AD/CN classification   & LLAMA2 combined with prompt fine-tuning and conditional learning.                                                                                                             & Achieved an accuracy of 81.31\% with a 4.46\% improvement over the literature BERT method.                                                                                                                                  &                           \\ \hline
Farzana   et al. \cite{farzana-parde-2023-towards}   & Pitt Corpus   \cite{becker1994natural}, CCC \cite{pope2011finding}, ADReSS   \cite{luz2020alzheimersdementiarecognitionspontaneous}, ADRC \cite{farzana-parde-2023-towards} & AD/CN classification   & Feature-based and instance-based DA                                                                                                                                  & Achieves a 22\% increase in accuracy compared to baseline.                                                                          & \ding{51}                         \\ \hline
Ilias   et al. \cite{ILIAS2023101485}                & ADReSS   \cite{luz2020alzheimersdementiarecognitionspontaneous}, ADReSSo   \cite{luz2021detecting}                                                                                                                   & AD/CN classification   & Incorporate BERT, DeiT,   self-attention mechanism, and optimal transport DA.          & Outperforms the baselines in accuracy by 0.84-5.35\%.                                                                                                                                                                                     &                           \\ \hline
Masrani   et al. \cite{10.1007/978-3-319-57351-9_29} & Pitt Corpus   \cite{becker1994natural}                                                                                                                                                                              & CN/MCI classification  & A feature-level data   augmentation combined with a statistical correlation alignment technique                                                                                                                                       & Improves the MCI classification F-measure by 5\% compared with no DA                                                                                                                                                                                                  & \ding{51}                         \\ \hline
Mo et   al. \cite{Mo2024.08.22.24312463}                  & Cambridge 398 Dataset,   Pitt Corpus \cite{becker1994natural},ADReSSo \cite{luz2021detecting}                                                                                                                       & AD/CN classification   & Few-shot and zero-shot prompting strategies with LLMs to extract linguistic markers across datasets. & Reached 0.89 accuracy and 0.88 F1 score on the test ADReSSo dataset                                                                                                                                                                                                                                        &                           \\ \hline
Casu et   al. \cite{10495546}                        & ADReSS   \cite{luz2020alzheimersdementiarecognitionspontaneous}, ADReSSo   \cite{luz2021detecting}                                                                                                                   & AD/CN classification   & Fine-tune the pre-trained LLMs                                                                                                                     & The accuracy reached over 90\%   on the unseen ADReSSo dataset                                                                                                                                                                                                                                               &                           \\ \hline
Skirrow   et al.\cite{SkirrowCaroline2024SiAA}       & AMYPRED, ADNI4 \cite{ADNI4_2025}                                                                                                                                                                                              & AD vs. CN, MCI vs. CN  &   Storyteller speech models (text-similarity, lexical, temporal, and acoustic features) developed on AMYPRED and validated out-of-sample on the more diverse ADNI4 cohorts.                                                                                                                                                            & Speech models generalized to unseen ADNI4 data with AUC 0.71-0.84 across remote and in-clinic cohorts.  &                           \\ \hline
Wang \& Wu \cite{wang2026crosslingual} & ADReSSo (English) \cite{luz2021detecting}, NCMMSC2021 (Chinese) & AD/HC classification & Cross-lingual few-shot domain adaptation correcting a cross-lingual ``polarity inversion'' (HuBERT features). & 86.67\% accuracy, 80\% sensitivity, 90\% specificity, AUC 0.871 on the Chinese blind test. & \\ \hline
\end{tabular}
\end{table}

\begin{table}[]
\centering

\scriptsize
\caption{ Companion table for speech/text DA/DG studies: experimental domain structure, adaptation/generalization setting, clinical endpoint, and split/leakage reporting. ``NR'' = not reported. Cross-corpus and cross-language shifts dominate this modality; participant-level leakage control was rarely reported explicitly.}
\label{speech_companion_table}
\begin{tabular}{|C{2.1cm}|C{3cm}|C{1.5cm}|C{2.3cm}|C{1.8cm}|C{2.6cm}|}
\hline
\textbf{Paper} & \textbf{Source $\rightarrow$ Target domain} & \textbf{Target labels in training} & \textbf{DA/DG setting} & \textbf{Clinical endpoint} & \textbf{Split \& leakage control} \\ \hline
Fraser et al. \cite{fraser-etal-2019-multilingual} & Pitt (English) $\rightarrow$ French Dem@Care & No & Unsupervised DA (cross-language) & AD vs HC & Cross-corpus/language; NR \\ \hline
Loukas et al. \cite{ILIAS2023110834} & ADReSS $\leftrightarrow$ ADReSSo & No & Unsupervised DA (EMD transport) & AD vs CN & Cross-corpus; NR \\ \hline
Li et al. \cite{Li2024UsefulBlunders} & ADReSS, WLS (evaluated on both) & Yes (fine-tune) & ASR-robust decoding + fine-tuning (manual-vs-ASR transcripts; DA framing loose) & AD vs CN & Within-corpus (transcript source); NR \\ \hline
Liu et al. \cite{Liu2022FeaturePurification} & Pitt (internal) & No & Adversarial feature purification (internal; DA loose, no cross-domain transfer) & AD vs CN & Internal; participant-level leakage NR \\ \hline
Ahmed et al. \cite{ALKENANI2021103803} & Pitt (spoken) + ADBC (written), fused & No & Ensemble fusion of spoken+written data & AD vs HC & Single-corpus fusion; very high AUC (leakage risk, NR) \\ \hline
Chafouleas \cite{chafouleas2021alzheimer} & ADReSS $\leftrightarrow$ COMPASS-ND (both directions) & None & External validation / generalization & AD vs HC & Cross-corpus (external); NR \\ \hline
Zheng et al. \cite{10.1007/978-981-96-1151-5_21} & ADReSS (internal) & Yes (prompt fine-tune) & LLM prompt fine-tuning & AD vs CN & Internal; LLM contamination risk NR \\ \hline
Farzana et al. \cite{farzana-parde-2023-towards} & Pitt, CCC, ADReSS, ADRC & No & Feature/instance-based DA & AD vs CN & Multi-corpus; NR \\ \hline
Ilias et al. \cite{ILIAS2023101485} & ADReSS $\leftrightarrow$ ADReSSo & No & Unsupervised DA (optimal transport) & AD vs CN & Cross-corpus; NR \\ \hline
Masrani et al. \cite{10.1007/978-3-319-57351-9_29} & Pitt (internal; across visits/tasks) & No & Unsupervised DA (correlation align.) & CN vs MCI & Internal; NR \\ \hline
Mo et al. \cite{Mo2024.08.22.24312463} & Cambridge-398, Pitt $\rightarrow$ ADReSSo & No (few/zero-shot) & LLM marker extraction + classifier & AD vs CN & Cross-corpus; LLM contamination risk NR \\ \hline
Casu et al. \cite{10495546} & ADReSS $\rightarrow$ ADReSSo (unseen) & Yes (fine-tune) & LLM fine-tuning across corpora  & AD vs CN & Cross-corpus (external); LLM contamination risk NR \\ \hline
Skirrow et al. \cite{SkirrowCaroline2024SiAA} & Prior AMYPRED cohort $\rightarrow$ ADNI4 (out-of-sample) & None & External validation / generalization & AD vs CN; MCI vs CN & Cross-cohort (external); NR \\ \hline
Wang \& Wu \cite{wang2026crosslingual} & ADReSSo (English) $\rightarrow$ NCMMSC2021 (Chinese) & Yes (few-shot, 5/class) & Cross-lingual few-shot DA (polarity-inversion correction) & AD vs HC & Cross-lingual/cross-corpus; blind test (Acc 86.67\%, AUC 0.871) \\ \hline
\end{tabular}
\end{table}

\subsection{DA/DG approaches for AD detection with multimodal data}

Single modalities such as MRI, PET, and speech-based data have demonstrated utility for AD detection. However, due to the brain's complex structure and the multifactorial nature of AD, relying on any single modality often limits the ability to capture the full spectrum of pathological characteristics necessary for accurate diagnosis. Consequently, multimodal approaches that integrate information from diverse sources have consistently outperformed single-modality methods in diagnostic accuracy.

The use of multimodal data inherently contributes to improved generalization by leveraging the strengths of different modalities, helping the model learn diverse representations of brain activity and pathology. For instance, Sheng et al.~\cite{SHENG2024108035} highlighted that multimodal fusion leverages complementarity information to improve the robustness and generalization of AD detection models.

Examples of successful multimodal models for improving AD detection performance include Yuan et al.\cite{yuan2025diagnosis}, who developed a 3D Inception-v4 architecture integrating MRI and clinical scores, and Yuan et al.\cite{yuan2024multi}, demonstrating that multi-probe PET/MR imaging significantly improves early AD detection over single-modality approaches. Similarly, Liu et al.\cite{liu2024hammf} used hierarchical attention-based fusion of MRI and PET data for multi-task classification, and Wang et al.\cite{wang2024semi} proposed a multiomics integration framework (MOSEGCN) that substantially boosts classification performance.

While these advances suggest that multimodal data can enhance a model's generalization, it is critical to recognize that multimodality alone does not fully solve the domain shift problem, where models trained on one acquisition setting perform poorly on others. In practical clinical scenarios, model performance is affected by distributional discrepancies caused by variations in imaging protocols, scanners, and participant and clinical populations. Thus, explicit DA/DG strategies remain essential to ensure robust performance across heterogeneous real-world datasets.

Table~\ref{multimodal_table} summarizes key existing DA/DG studies in multimodal AD detection. Most current work focuses on MRI and PET integration. Cheng et al.\cite{cheng2015multimodal} tackled domain differences with a manifold-regularized transfer learning approach combining sample selection and kernel-based MMD metrics, outperforming baseline transfer and single-modality models. Fu et al.\cite{fu2025prediction} introduced a multi-head attention fusion module alongside ADA to align feature distributions between different domains, achieving superior accuracy across adaptation scenarios. Similar DA benefits were observed by Shakhnoza et al.\cite{diagnostics15060717} using adversarial adaptation and batch-normalization statistics adjustment with multimodal neuroimaging. Odusami et al.\cite{ODUSAMI2024e34402} fused MRI and PET with data augmentation and a Mobile Vision Transformer, demonstrating improved generalization across ADNI and OASIS datasets.

Recent work has expanded multimodal fusion to include neuroimaging combined with textual clinical information, often leveraging LLMs. Feng et al.\cite{10403166} fused MRI with GPT-4-generated embeddings from structured clinical notes via transformer modules, achieving strong generalization (AUC 0.90-0.94) even without explicit domain alignment, benefiting from the pretrained LLM's inherent robustness in few-shot scenarios. Similarly, Song et al.\cite{SONG2025100257} fine-tuned GenBERT on multimodal inputs, reaching over 98\% accuracy across heterogeneous data sources and noting that further diversity and longitudinal data could improve generalizability.

Beyond neuroimaging and text, multimodal DA/DG methods have integrated speech and text data. Zhang et al.~\cite{ZhangZhenglin2025DAHA} proposed a multitask learning framework combining audio, text, and expert knowledge via hybrid attention mechanisms. Tested cross-dataset, the model achieved 89.58\% accuracy and 91.67\% recall on AD detection, alongside an RMSE of 4.31 for cognitive assessment, demonstrating robust domain generalization and detailed modality-specific explainability.

These multimodal studies suggest that the benefit of multimodal AD detection does not arise simply from adding more data streams, but from whether the fused modalities provide complementary and domain-stable information. MRI and PET fusion is the most common setting because the two modalities capture related but distinct aspects of AD pathology: structural neurodegeneration and metabolic or molecular abnormalities. This complementarity explains why MRI--PET models often outperform single-modality imaging models. However, most MRI--PET studies still rely heavily on ADNI-like cohorts and therefore remain vulnerable to scanner, protocol, and cohort shifts. Methods that combine imaging with clinical scores, genetic data, or cognitive assessments can improve prediction by adding non-imaging risk information, but they also introduce new sources of heterogeneity because clinical variables may be collected, coded, and normalized differently across sites.

The emerging use of text, speech, and LLM-derived features changes the role of multimodal fusion. These modalities can capture cognitive, linguistic, and behavioral changes that are not directly represented in neuroimaging, and they may improve accessibility for remote or low-cost screening. At the same time, they introduce language, culture, education, recording-device, and prompt-sensitivity biases that are different from scanner-related imaging shifts. Therefore, the connection across existing studies is that multimodal fusion improves generalization only when the fusion strategy explicitly handles both complementarity and heterogeneity. Attention-based fusion, graph-based fusion, adversarial adaptation, MMD-based transfer, and LLM-assisted fusion each address part of this problem, but none fully resolves missing modalities, inconsistent acquisition protocols, site imbalance, and population-level bias. This indicates that future multimodal DA/DG research should move from performance-driven fusion toward shift-aware fusion, where each modality is evaluated according to its clinical contribution, domain stability, missingness pattern, and vulnerability to site- or population-specific bias.

The quantitative findings reported in the multimodal studies above support the preceding interpretation that multimodal AD detection is effective only when fusion captures complementary and domain-stable information. Different fusion strategies address different sources of multimodal shift. MRI--PET fusion is useful because the modalities provide complementary structural and metabolic information, but it remains vulnerable to scanner, protocol, and cohort differences. MMD-based alignment, adversarial adaptation, and batch-statistics adjustment are therefore useful when the main challenge is cross-cohort imaging shift. Attention-based and graph-based fusion can model cross-modal relationships and identify informative modality interactions, but they may overfit when datasets are small or when modality availability differs across sites. Imaging--clinical and LLM-assisted fusion can add non-imaging context, but their robustness depends on the consistency of clinical coding, text generation, and cohort composition. Speech--text fusion can improve accessibility and capture behavioral markers, but its performance remains affected by language, recording condition, and task design. Quantitatively, MRI--PET studies report improved accuracy and AUC when transfer learning, MMD-based alignment, adversarial adaptation, batch-statistics adjustment, or attention-based fusion is used across cohorts. Imaging--clinical and LLM-assisted multimodal studies report high AUC values, often around 0.90--0.94, while speech--text multimodal work reports accuracy around 89.58\% and recall around 91.67\% in one representative cross-dataset study. However, these results should be interpreted carefully because multimodal studies differ in modality availability, missing-data handling, fusion stage, diagnostic task, and external validation protocol. The benchmark-level pattern is that multimodal fusion is most valuable when modalities provide complementary and domain-stable information, while multimodality alone does not remove scanner, language, cohort, missing-modality, or population-level shift.

The companion Table~\ref{multimodal_companion_table} highlights a structural limitation of the multimodal literature: the great majority of studies fuse MRI and PET within ADNI, so the source-to-target column rarely records genuine external site or scanner shift, and the split-and-leakage column shows that missing-modality robustness, arguably the defining real-world challenge for multimodal models, is almost never evaluated. Several entries described as generalization in fact report only within-cohort fusion performance, which means their domain-generalization claims rest on internal evaluation rather than testing on an unseen domain. The endpoint column is dominated by AD-versus-CN discrimination, with only a few studies addressing MCI conversion. The table suggests that current multimodal gains derive mainly from modality complementarity within a single cohort, and that external-cohort, missing-modality, and population-shift evaluation remain the key gaps before multimodal DA/DG can be considered clinically generalizable.

\begin{table}[]
\scriptsize
\centering
\caption{Summary and cross-study synthesis of DA/DG approaches for AD detection with multimodal data. The table reports not only datasets, modalities, and methods, but also the domain-shift implication or limitation of each study.}
\label{multimodal_table}
\begin{tabular}{|C{1.5cm}|C{2cm}|C{1.8cm}|C{3.5cm}|C{4.8cm}|C{1cm}|}
\hline
\textbf{Papers}               
& \textbf{Dataset} 
& \textbf{Modality} 
& \textbf{Method}                                                                   
& \begin{tabular}[c]{@{}c@{}}\textbf{Evaluation and}\\\textbf{domain-shift implication}\end{tabular}                                                                     
& \textbf{Code} \\
\hline

Cheng et al. \cite{cheng2015multimodal}      
& ADNI\cite{mueller2005adni}                              
& MRI, PET                    
& Kernel-based MMD, manifold regularization                                   
& 80.1\% accuracy, AUC 0.852; supports MRI--PET transfer across related diagnostic tasks, but only within ADNI. 
& \\ \hline

Cheng et al. \cite{cheng2015domain}          
& ADNI\cite{mueller2005adni}                              
& MRI, PET                    
& Domain transfer feature selection (DTFS) and domain transfer sample selection (DTSS)                                                        
& 79.4\% accuracy, AUC 0.848; feature/sample selection reduces task shift, but external site shift was not tested.  
& \\ \hline

Fu et al. \cite{fu2025prediction}              
& ADNI-1\cite{mueller2005adni} \& ADNI-2\cite{mueller2005adni}
& MRI \& PET                          
& Multi-head attention fusion and ADA 
& 92.40\% and 94.73\% cross-cohort accuracy; supports attention fusion with adversarial alignment for imaging shift.     
& \\ \hline

Mashhadi et al. \cite{mashhadi2025multimodal} 
& ADNI\cite{mueller2005adni}, OASIS\cite{marcus2007open}, AIBL\cite{ellis2009aibl}, MIRIAD\cite{malone2013miriad}
& MRI \& PET                          
& Modular, compositional graph structure                                                             
& 92.4\% overall accuracy across datasets; improves multi-cohort evidence but depends on consistent MRI--PET availability.                                                     
& \ding{51} \\ \hline

Feng et al. \cite{10403166} 
& ADNI\cite{mueller2005adni}        
& MRI, PET \& clinical data                         
& LLM-based clinical embeddings fused with image features using transformer modules.                                                             
& AUC 0.90--0.94; clinical-text fusion may improve robustness, but validation beyond ADNI is needed.                                                    
& \ding{51} \\ \hline

Zhang et al. \cite{ZhangZhenglin2025DAHA} 
& ADReSS\cite{luz2020alzheimersdementiarecognitionspontaneous}, Pitt Corpus\cite{becker1994natural}         
& Audio, text, expert knowledge                         
& Multi-task learning with hybrid attention and multimodal representations.                                                             
& 89.58\% accuracy, 91.67\% recall; supports speech--text transfer but remains sensitive to language and recording shift.                                                     
& \ding{51} \\ \hline

Odusami et al. \cite{ODUSAMI2024e34402} 
& ADNI-1\cite{mueller2005adni}, OASIS\cite{marcus2007open}
& MRI \& PET                        
& Pulse-coupled neural network and Laplacian pyramid fusion with MViT augmentation.                                                             
& 94.73\%, 92.98\%, and 89.36\% precision; strong imaging fusion, but mainly evaluated on ADNI.                                                   
& \\ \hline

Song et al. \cite{SONG2025100257} 
& AD genetic data \cite{Gao_AlzGeneExpKaggle}, clinical data \cite{Hebishy_ElzihimerKaggle}, neuroimaging data \cite{Chugh_AlzMRI99Kaggle}        
& Clinical notes and MRI                   
& LLM-based biomarker identification with genetic-algorithm classification.                                                             
& 98.30\% accuracy; promising multimodal reasoning, but heterogeneous public sources require leakage and cohort-shift checks.                                                 
& \\ \hline

Shakhnoza et al. \cite{diagnostics15060717} 
& ADNI\cite{mueller2005adni}, OASIS\cite{marcus2007open}       
& MRI, PET, CT scans                   
& ADA and batch-normalization statistics adaptation.                                                             
& Outperformed literature methods; suggests adaptation and batch-statistics adjustment can reduce imaging-domain shift.                                                
& \\ \hline

El-Sappagh et al. \cite{ElSappagh2021} 
& ADNI\cite{mueller2005adni}       
& MRI, genetic data, and cognitive assessments                   
& Key biomarker feature selection and layered classification for diverse clinical profiles.                                                             
& 93.95\% early AD accuracy; multimodal clinical variables help, but missingness and coding differences may limit transfer.                                                
& \\ \hline

Meng et al. \cite{10.3389/fnagi.2022.911220} 
& ADNI\cite{mueller2005adni}       
& MRI, PET                  
& Multimodal neuroimaging data integration and interpretation.                                                             
& High AD/HC accuracy; supports MRI--PET complementarity, but binary single-cohort testing limits screening relevance.                                                
& \\ \hline

Tang et al. \cite{TANG2024108259} 
& ADNI\cite{mueller2005adni}       
& MRI, PET                  
& Multiscale attention and cross-enhancement fusion.                                                             
& Multitask within-ADNI classification (AD/CN, AD/MCI, CN/MCI); strong internal accuracy, but no external-cohort or missing-modality validation.
& \\ \hline

Xiang et al. \cite{xiang2026crossmodal}
& Miltiadous EEG (OpenNeuro ds004504), NCMMSC2021 speech
& EEG + speech
& Cross-modal alignment of EEG and speech with a conditional rectified-flow model that synthesizes speech-derived latents to compensate for missing EEG (missing-modality robustness, not cross-cohort DA/DG).
& Three-class AD/MCI/HC: 89.08\% accuracy, macro-F1 0.8862, AUC 0.9153; +9.28\% over speech-only. Within-dataset evaluation, no external cohort.
& \\ \hline

\end{tabular}
\end{table}

\begin{table}[]
\centering

\scriptsize
\caption{Companion table for multimodal DA/DG studies: experimental domain structure, adaptation/generalization setting, clinical endpoint, and split/leakage reporting. ``NR'' = not reported. Most studies fuse MRI--PET within ADNI; external site/scanner shift and missing-modality robustness are rarely tested.}
\label{multimodal_companion_table}
\begin{tabular}{|C{2.1cm}|C{3cm}|C{1.5cm}|C{2.3cm}|C{1.8cm}|C{2.6cm}|}
\hline
\textbf{Paper} & \textbf{Source $\rightarrow$ Target domain} & \textbf{Target labels in training} & \textbf{DA/DG setting} & \textbf{Clinical endpoint} & \textbf{Split \& leakage control} \\ \hline
Cheng et al. \cite{cheng2015multimodal} & ADNI (MRI/PET, task transfer) & Yes & Supervised DA (MMD, manifold) & MCI-C vs MCI-NC (conversion; AD/CN auxiliary) & Within-ADNI; external shift not tested \\ \hline
Cheng et al. \cite{cheng2015domain} & ADNI (MRI/PET) & Yes & Supervised DA (feature/sample selection) & MCI-C vs MCI-NC (conversion; AD/CN auxiliary) & Within-ADNI; external site not tested \\ \hline
Fu et al. \cite{fu2025prediction} & ADNI-1 $\rightarrow$ ADNI-2 (MRI/PET) & No & Unsupervised DA (attention fusion + ADA) & AD vs CN & Cross-cohort; NR \\ \hline
Mashhadi et al. \cite{mashhadi2025multimodal} & ADNI $\rightarrow$ AIBL, MIRIAD (MRI/PET) & Partial (fine-tuned on AIBL) & Single-source DA/DG hybrid (fine-tunes on target) & AD vs CN & Multi-cohort; depends on MRI/PET availability \\ \hline
Feng et al. \cite{10403166} & ADNI (MRI/PET/clinical) & No & Internal multimodal fusion (LLM); & AD vs CN & Within-ADNI; external validation NR \\ \hline
Zhang et al. \cite{ZhangZhenglin2025DAHA} & ADReSS, Pitt (audio/text; speech modality) & None & Multi-task audio/text fusion (DG on an unseen external cohort) & AD vs CN & Cross-corpus (speech/text); language/recording shift \\ \hline
Odusami et al. \cite{ODUSAMI2024e34402} & ADNI-1 $\rightarrow$ OASIS (MRI/PET) & No & Fusion classifier with cross-cohort & AD stages (incl.\ AD vs CN) & Cross-cohort; mainly ADNI-evaluated \\ \hline
Song et al. \cite{SONG2025100257} & Public genetic/clinical/MRI sources & None & Biomarker discovery + fusion & AD vs CN (biomarker discovery) & Heterogeneous public; leakage checks needed (NR) \\ \hline
Shakhnoza et al. \cite{diagnostics15060717} & ADNI $\rightarrow$ OASIS (MRI/PET/CT) & No & Batch-statistics/ADA adaptation + GAN augmentation  & AD vs CN & Cross-cohort; NR \\ \hline
El-Sappagh et al. \cite{ElSappagh2021} & ADNI (MRI/genetic/cognitive) & No & Internal multimodal XAI classifier & Early AD (multi-class) + MCI$\rightarrow$AD progression & Within-ADNI; missingness/coding may limit transfer \\ \hline
Meng et al. \cite{10.3389/fnagi.2022.911220} & ADNI (MRI/PET) & No & Internal binary fusion classifier (within-ADNI;) & AD vs HC & Within-ADNI (binary); screening relevance limited \\ \hline
Tang et al. \cite{TANG2024108259} & ADNI (MRI/PET) & No & Internal multimodal fusion & AD vs CN & Within-ADNI; no unseen-domain test \\ \hline
Xiang et al. \cite{xiang2026crossmodal} & EEG (Miltiadous/OpenNeuro ds004504) + speech (NCMMSC2021);  & None  & Cross-modal fusion + rectified-flow missing-modality synthesis (cross-modal alignment) & AD/MCI/HC (3-class) & Within-dataset (5-fold CV / fixed split); no external cohort \\ \hline
\end{tabular}
\end{table}

\section {Open Problems and Research Opportunities}
\label{challenge_future}

While ML models leveraging neuroimaging data combined with DA/DG techniques have demonstrated promising results for AD detection, substantial challenges remain before these methods can be effectively translated into real-world clinical practice. In parallel, emerging non-invasive modalities, such as speech analysis, eye-tracking, and EEG, show potential for accessible AD screening; however, their clinical applicability is still in its early stages. This section highlights key open problems across different modalities and identifies future research directions aimed at advancing the development, generalization, and deployment of robust and practical AD detection systems.

\subsection{Expanding modalities and dataset diversity}

EEG and ET offer promising, non-invasive biomarkers that are easier to deploy outside clinical environments. However, as summarized in Table~\ref{eeg_table} and Table~\ref{et_table}, DA/DG research using EEG and ET data remains limited, suggesting considerable potential for future exploration. The scarcity of open-source EEG and ET datasets for AD detection further constrains progress. With the continued development of mobile EEG devices and high-quality cameras embedded in consumer electronics, future work should focus on collecting larger, more diverse datasets from people with AD across different demographics and environments. Such datasets would provide a stronger foundation for DA/DG research aimed at improving real-world applicability.

\subsection{Benchmark datasets and standardized evaluation}

Many studies rely on open datasets such as ADNI for MRI and ADReSS for speech to develop DA/DG algorithms. However, these datasets are often small, imbalanced, typically with more CN than AD cases, and relatively homogeneous, limiting the ability of models to learn robust, domain-invariant features. Furthermore, differences in dataset selection, preprocessing, and artificial balancing make it difficult to compare results across studies. This issue is consistent across modalities, including single-site EEG datasets and limited multimodal resources.

Future work should prioritize the creation of large-scale, multi-center, multimodal benchmarks with standardized data collection, preprocessing, and evaluation protocols. These benchmarks should include detailed metadata on acquisition devices, demographics, and environmental conditions to better reflect real-world variability. Community challenges with fixed train-test splits could further improve reproducibility and comparability across studies.

\subsection{DA/DG methodologies for real-world variability}

Most existing open-source datasets for AD detection are relatively small and class-imbalanced, often containing substantially more CN subjects than AD or MCI cases. This imbalance skews the learning process, biasing models toward the majority class and reducing sensitivity to early-stage or minority categories. While some studies, such as Batool et al.~\cite{batool2025singledomaingeneralizationalzheimers}, explicitly address this issue through techniques like class weighting, many others either overlook it or rely on artificially balanced datasets during development. Such approaches risk producing models that fail to generalize in real-world clinical settings, where class distributions are inherently skewed. Addressing class imbalance is therefore a critical step toward developing DA/DG models that are both accurate and clinically reliable.

In addition to class imbalance, most DA/DG methods focus on a single source of domain shift, such as scanner differences in MRI or microphone variability in speech, but real-world applications must adapt to multiple, concurrent sources of variability. These include differences in acquisition devices, demographic characteristics, languages, and environmental conditions, all of which can interact in complex ways to degrade performance. Future research should explore unified, modality-agnostic DA/DG frameworks capable of adapting to multiple shift types simultaneously. Promising directions include multi-source DA/DG to integrate diverse training domains, continual adaptation to maintain performance as new data arrives, and few-shot domain generalization to adapt with minimal labeled data from new settings. 

For multimodal systems, it is crucial to design flexible fusion mechanisms that maintain robustness even when certain modalities are missing, degraded, or noisy, thereby ensuring reliability in real-world applications. This is a practical necessity rather than an edge case: in real clinical datasets, few participants have MRI, PET, EEG, speech, ET, and clinical assessments acquired concurrently, so restricting analysis to complete cases introduces selection bias and sharply reduces sample size. Several strategies can improve robustness to missing modalities and should be evaluated more systematically in AD DA/DG work, including modality dropout during training, partial or flexible multimodal learning, late fusion with optional inputs, imputation combined with uncertainty estimation, mixture-of-experts architectures that route available modalities to specialized sub-models, and teacher--student distillation in which a multimodal teacher supervises a unimodal or reduced-modality student. Reporting performance under realistic missing-modality conditions, rather than only on fully complete cohorts, would make multimodal DA/DG results far more clinically representative.

\subsection{Model explainability and clinical trust}

Across modalities, DL-based DA/DG models for AD detection are frequently perceived as “black boxes.” While some incorporate attention mechanisms or saliency maps, these explanations are rarely validated for clinical meaningfulness. Without transparent reasoning linked to established AD biomarkers, such as hippocampal atrophy, speech disfluency, or EEG slowing, clinical adoption will remain limited.

Future research should integrate explainable AI (XAI) methods tailored to DA/DG, ensuring that domain-aligned features remain interpretable. Techniques such as concept activation vectors, attention heatmaps aligned with expert annotations, and explanation faithfulness metrics could enhance trust. Hybrid models combining learned features with known clinical markers may further improve interpretability. Evaluation frameworks should expand beyond accuracy to include measures of explanation validity, robustness to domain shift, and clinician trustworthiness. Open access to trained models and code will also be critical for reproducibility.

\subsection{Privacy, security, and deployment constraints}

Data privacy remains a major challenge for modalities like speech and clinical text, where transcripts can contain identifiable participant information. Many high-performing systems rely on commercial cloud-based models (e.g., GPT-4) hosted externally, raising compliance issues under regulations such as HIPAA and PHIPA. For example, Casu et al.~\cite{10495546} demonstrated that fine-tuning medium-sized open-source LLMs such as LLAMA2-7B and Mistral-7B locally on consumer-grade GPUs preserves privacy while delivering competitive diagnostic accuracy. They also applied model quantization to reduce computational demands, enabling secure, on-premise inference.

Nonetheless, the computational requirements of large-scale LLMs or multimodal DA/DG models remain a barrier can be prohibitively resource-intensive for real-time or resource-constrained settings, such as small clinics, community centers, and home-based screening systems. Future work should develop privacy-preserving DA/DG methods, such as federated learning and differential privacy, and lightweight architecture design to enable DA/DG models to operate effectively on low-power devices without sacrificing accuracy. This is particularly relevant for home-based monitoring and large-scale screening applications.

The reported performance of LLM-based approaches also warrants particular caution. Several speech- and text-based studies report very high accuracies using LLMs, but these results are susceptible to specific validity threats that are easy to overlook. Because the pretraining corpora of large proprietary models are unknown and may already contain benchmark transcripts (e.g., DementiaBank/Pitt or ADReSS material that is widely available online), high benchmark performance may partly reflect training-data contamination rather than genuine generalization. LLM outputs are additionally sensitive to prompt wording, decoding parameters, and model version; proprietary API models are updated or deprecated over time, which undermines reproducibility of API-based experiments. Other concerns include transcript leakage between training and test partitions when the same participants contribute multiple samples, and demographic or linguistic bias inherited from the pretraining data. Consequently, high performance on an established AD benchmark should not be interpreted as evidence of domain generalization unless the study explicitly evaluates an unseen domain (e.g., a different corpus, language, or recording setting) and controls for contamination and participant-level leakage. We therefore recommend that LLM-based AD studies report model versions and access dates, fix and disclose prompts and decoding settings, prefer open-weight models where feasible for reproducibility, and test on genuinely held-out cohorts.

\subsection{Inclusivity applicability}

Most current research focuses on English-language speech datasets and clinically acquired neuroimaging datasets, which may lead models to encode dataset-specific biases rather than disease-general patterns. These biases arise not only from language and acquisition protocols, but also from the way study participants are recruited. In many AD datasets, the HC group is highly selected and may be healthier, more educated, more compliant with research procedures, and less clinically heterogeneous than the general population. As a result, models trained to separate HC from AD may learn differences between selected research groups rather than features that are reliable for real-world screening.

This limitation is especially important for screening applications. Binary HC-versus-AD classification is useful for controlled methodological comparison, but it does not reflect the input distribution expected in real-world deployment, where individuals may present with MCI, subjective cognitive decline, depression, vascular cognitive impairment, Parkinsonian syndromes, mixed dementia, medication effects, low education, sensory impairment, or no cognitive impairment. If such individuals are forced into a binary HC/AD label, the system may produce misleading outputs and amplify the selection bias introduced by narrowly defined HC groups. Therefore, screening-oriented AD models should move beyond simple binary classification and should be evaluated on broader, population-like cohorts that include cognitively normal individuals, MCI, AD, non-AD dementias, psychiatric and vascular conditions, and uncertain cases.

Future research should emphasize inclusive data collection and adaptation across diverse linguistic, cultural, educational, socioeconomic, and clinical contexts. Multilingual and culturally adaptive modelling is especially underexplored in speech-based AD detection. Cross-lingual transfer, affordable device-agnostic designs, and community-based validation could broaden accessibility, particularly in underserved regions. Methodologically, future systems should move toward multi-class or ordinal models, longitudinal risk prediction, uncertainty-aware triage, and open-set recognition, so that uncertain or out-of-distribution cases are referred for clinical assessment rather than forced into a binary diagnosis. These changes are necessary for DA/DG-based AD screening tools to become applicable to real-world populations rather than only to highly controlled research datasets.

Beyond inclusive data collection, generalizability should be assessed at the level of clinically relevant subgroups rather than only in aggregate. A model that generalizes across scanners but performs unevenly across linguistic, educational, cultural, socioeconomic, or demographic subgroups is not fully clinically generalizable. Where the relevant metadata are available, we recommend that studies report performance broken down by age, sex, language, education, hearing or vision impairment, and comorbidity, alongside the corresponding population and clinical-task dimensions of the generalizability framework (Figure~\ref{fig:generalizability}). Subgroup reporting is especially important for speech- and ET-based approaches, which are strongly influenced by language, literacy, culture, device familiarity, and sensory impairment, so that aggregate accuracy does not mask systematic underperformance in vulnerable subgroups.

\subsection{Calibration, predictive uncertainty, and abstention}

To be clinically useful, a model must not only separate diagnostic classes but also produce reliable probability estimates and recognize cases that fall outside its competence. This aspect is largely absent from the reviewed literature, which reports discrimination metrics (accuracy, AUC) far more often than reliability metrics. We therefore highlight several complementary tools that DA/DG studies for AD detection should adopt. Calibration should be assessed using expected calibration error, reliability diagrams, and the Brier score, and recalibration (e.g., temperature scaling) should be checked after domain adaptation, since calibration often degrades under domain shift even when accuracy is maintained. Predictive uncertainty can be quantified through Bayesian approximations, deep ensembles, or Monte-Carlo dropout, and conformal prediction can provide distribution-free coverage guarantees. Selective prediction (abstention) and out-of-distribution detection allow a system to defer uncertain or atypical cases for clinical assessment rather than forcing a binary diagnosis. These mechanisms are directly aligned with our recommendation that uncertain cases be referred to clinicians, and they are essential when models are deployed across new sites, devices, or populations where confident but incorrect predictions are especially harmful.

\subsection{Clinical translation and deployment considerations}

Clinical translation requires more than high cross-dataset accuracy. A deployable AD decision-support system must first define its intended use and fit into an existing clinical workflow. For example, a model may be designed for risk stratification, referral support, community screening, longitudinal monitoring, or specialist diagnostic decision support. These use cases have different requirements for accuracy, interpretability, turnaround time, acceptable uncertainty, and clinician oversight. Current binary HC-versus-AD classifiers should therefore be viewed as early methodological prototypes rather than complete clinical screening tools. Real-world decision support must include MCI and other cognitive or neurological conditions, evaluate predictive uncertainty, avoid data leakage, and be validated prospectively in representative clinical and community populations.

It is also useful to organize the evidence according to the stage of clinical evaluation a study actually reaches, because these stages are not equivalent in the strength of evidence they provide. We distinguish (i) \emph{analytical validity} (does the model measure its inputs reliably and reproducibly?), (ii) \emph{clinical validity} (does it accurately classify the intended clinical condition in the target population?), (iii) \emph{clinical utility} (does it improve clinical decisions or outcomes?), (iv) \emph{workflow integration} (can it operate within real clinical pathways?), and (v) \emph{post-deployment monitoring} (does it remain reliable as data distributions drift?). Relatedly, the evaluation designs used in the reviewed studies form an implicit hierarchy of increasing evidential strength: random internal splits, cross-cohort or leave-one-site-out testing, independent external validation, and finally prospective clinical-impact studies. The overwhelming majority of the studies reviewed here stop at retrospective benchmark evaluation (random or, less often, cross-cohort splits); such results should not be assigned the same evidential weight as models validated prospectively in independent clinical populations. Making this hierarchy explicit clarifies how far each modality has actually progressed toward clinical use.

Clinical feasibility also differs substantially across modalities. MRI and PET provide rich structural and molecular information and may be most appropriate for specialist memory clinics where imaging is already part of the diagnostic pathway. However, their high cost, infrastructure requirements, need for trained personnel, and limited accessibility reduce their suitability for repeated monitoring or large-scale community screening, especially in rural or resource-limited settings. EEG is generally less expensive and more portable than MRI or PET, but it remains sensitive to acquisition protocols, artifacts, electrode placement, and operator expertise. Speech-, language-, eye-tracking-, and digital-behavior-based approaches are comparatively low-cost and scalable, making them attractive for primary-care triage, remote screening, or longitudinal monitoring. However, these modalities are strongly affected by language, literacy, education, hearing, culture, device quality, recording environment, task design, and user engagement. Therefore, modality choice should be aligned with the intended clinical use rather than selected only according to reported model accuracy.

Multi-site deployment introduces further challenges because data heterogeneity arises from scanner vendors, imaging protocols, demographic composition, recruitment criteria, cognitive assessment instruments, language, comorbidities, disease-stage distribution, and local clinical workflows. These differences can cause substantial domain shift between development datasets and real-world deployment settings. DA/DG models should therefore be evaluated not only with random train--test splits, but also with external validation, leave-one-site-out testing, subgroup analysis, calibration assessment, and dataset-drift monitoring after deployment. Regulatory and governance considerations are also essential, including privacy-preserving data handling, transparent reporting of intended use and limitations, human oversight, explainable outputs tied to clinically meaningful biomarkers, and procedures for model updating when local data distributions change. Future work should prioritize standardized multi-site and multimodal benchmarks, lightweight and privacy-preserving adaptation, and workflow-based prospective validation. These steps are necessary to move DA/DG-based AD detection from promising experimental results toward reliable, equitable, and clinically useful decision-support systems.

Regulatory validation should therefore include prospective external validation, predefined clinical endpoints, calibration analysis, subgroup fairness assessment, and workflow studies that measure whether the model changes referral decisions or clinical efficiency. Multi-site deployment also requires data governance, privacy protection, version control, monitoring for dataset drift, and clear human oversight. For DA systems, local adaptation must be documented because model behavior can change after exposure to target-site data; for DG systems, validation should include truly unseen sites and devices. These considerations are especially important because a model that performs well in curated HC-versus-AD experiments may not be safe as a population screening tool without evaluation in realistic cohorts containing MCI, subjective cognitive decline, mixed dementia, psychiatric or neurological comorbidities, and cognitively normal individuals with variable education and cultural backgrounds.

\section{Conclusions}
\label{conclusion}

This review examined DA and DG techniques for ML-based AD detection across neuroimaging, EEG, ET, speech, and multimodal data. Rather than treating these modalities as independent research silos, the review connects them through the shared challenge of domain shift. MRI and PET studies provide the strongest body of DA/DG evidence and demonstrate clear gains in cross-site or cross-dataset robustness, but their clinical use is constrained by cost, infrastructure, and access. EEG, speech, and ET offer more scalable and non-invasive pathways for screening and monitoring, yet they require stronger cross-device, cross-language, and cross-environment validation. Multimodal systems can combine complementary biological, neurophysiological, linguistic, and behavioral information, but they must explicitly address missing modalities and modality-specific bias.

The main methodological insight from the reviewed literature is that DA/DG performance depends on the type of shift and the deployment context. Adversarial methods are powerful for learning domain-invariant representations but can be unstable or remove clinically relevant information. Discrepancy-based methods are more transparent and stable but may not preserve class boundaries. Self-training and few-shot approaches are attractive when limited local target data are available, whereas DG is most relevant when target data are unavailable before deployment. Therefore, future research should report not only accuracy or AUC, but also the domain definition, shift type, target-data availability, class distribution, calibration, subgroup performance, and external validation protocol.

For clinical translation, current binary HC-versus-AD classifiers should be viewed as early methodological prototypes rather than complete screening tools. Real-world decision support must include MCI and other cognitive or neurological conditions, evaluate uncertainty, avoid data leakage, and be validated prospectively in representative clinical and community populations. Future work should prioritize standardized multi-site and multimodal benchmarks, transparent reporting, privacy-preserving and lightweight adaptation, explainable outputs tied to clinically meaningful biomarkers, and workflow-based validation. These steps are necessary to move DA/DG-based AD detection from promising experimental results toward reliable, equitable, and clinically useful decision-support systems.

\section*{Declaration of generative AI and AI-assisted technologies in the manuscript preparation process}

During the preparation of this work, the author(s) used Grammarly in order to check grammar and refine languages. After using this tool/service, the author(s) reviewed and edited the content as needed and take(s) full responsibility for the content of the published article.

\section*{CRediT authorship contribution statement}

\textbf{Yingchao Huang}: writing - review \& editing, writing - original draft, visualization, formal analysis, data curation, conceptualization, funding acquisition.
\textbf{Yuhan Su}: writing - original draft, visualization, validation, formal analysis. \textbf{Xin Wang}: review \& editing, visualization, formal analysis, data curation. \textbf{Shanshan Yao}: investigation, review \& editing,  visualization, validation.

\section*{Consent for publication}
All authors declare consent for publication.

\section*{Funding}

This research was supported by funding from Saskatchewan Polytechnic, Faculty of Digital Innovation, Arts and Sciences and the 2025 Applied Research Release Time (ARRT) Fund from Saskatchewan Polytechnic.

\section*{Declaration of competing interest}
The authors declare that they have no known competing financial interests or personal relationships that could have appeared to influence the work reported in this paper.




\bibliography{dementia_da}

\end{document}